\pdfoutput=1
\documentclass[11pt]{article}

\usepackage[T1]{fontenc}
\usepackage[utf8]{inputenc}
\usepackage[letterpaper,margin=1in]{geometry}
\usepackage{amsmath,amssymb,amsthm,mathtools}
\usepackage{newtxtext,newtxmath}
\usepackage{microtype}
\usepackage{setspace}
\usepackage{ragged2e}
\usepackage{booktabs,threeparttable,tabularx,adjustbox,multirow}
\usepackage{graphicx}
\usepackage{subcaption}
\usepackage{float}
\usepackage{placeins}
\usepackage{enumitem}
\usepackage{natbib}
\usepackage{xcolor}
\definecolor{navyblue}{RGB}{0,0,128}
\usepackage[colorlinks=true,linkcolor=navyblue,citecolor=navyblue,urlcolor=navyblue]{hyperref}
\usepackage[nameinlink,noabbrev]{cleveref}

\setlist[itemize]{leftmargin=1.5em,itemsep=0.2em,topsep=0.3em}
\setlist[enumerate]{leftmargin=1.7em,itemsep=0.2em,topsep=0.3em}

\newtheorem{assumption}{Assumption}
\newtheorem{proposition}{Proposition}

\theoremstyle{remark}

\crefname{assumption}{assumption}{assumptions}
\Crefname{assumption}{Assumption}{Assumptions}
\crefname{proposition}{proposition}{propositions}
\Crefname{proposition}{Proposition}{Propositions}
\crefname{theorem}{theorem}{theorems}
\Crefname{theorem}{Theorem}{Theorems}
\crefname{lemma}{lemma}{lemmas}
\Crefname{lemma}{Lemma}{Lemmas}
\crefname{figure}{figure}{figures}
\Crefname{figure}{Figure}{Figures}
\crefname{table}{table}{tables}
\Crefname{table}{Table}{Tables}

\newcommand{\E}{\mathbb{E}}

\newcommand{\R}{\mathbb{R}}
\newcommand{\ind}{\mathbf{1}}

\hypersetup{
  pdftitle={The Cost of Reorganizing Production: Evidence from Airline Outsourcing},
  pdfauthor={Myongjin Kim, Byoung G Park, and Jangsu Yoon},
  pdfsubject={Airline outsourcing, organizational adjustment, pricing, and route participation}
}

\title{\textbf{The Cost of Reorganizing Production:\\ Evidence from Airline Outsourcing}\thanks{This research did not receive any specific grants from funding agencies in the public, commercial, or nonprofit sectors.}}
\author{Myongjin Kim\thanks{Department of Economics, University of Oklahoma. \href{mailto:mjkim@ou.edu}{mjkim@ou.edu}.}
\and Byoung G Park\thanks{Department of Economics, University at Albany, SUNY. \href{mailto:bpark2@albany.edu}{bpark2@albany.edu}.}
\and Jangsu Yoon\thanks{Department of Economics, University of Kentucky. \href{mailto:jangsu.yoon@uky.edu}{jangsu.yoon@uky.edu}.}}
\date{November 2026}

\begin{document}
\clubpenalty=10000
\widowpenalty=10000
\makeatletter
\@clubpenalty=10000
\makeatother
\setcitestyle{citesep={,}}
\counterwithin{assumption}{section}
\counterwithin{proposition}{section}
\counterwithin{theorem}{section}
\counterwithin{lemma}{section}
\maketitle

\begin{abstract}
We study how the costs of reorganizing production shape entry and competition. We estimate a dynamic game of entry, exit, and outsourcing in the U.S. airline industry. Combining own operations with flights purchased from independent regional carriers lowers incumbents' adjustment costs but raises entry costs, explaining why mixed production is common among incumbents but rare at entry. In counterfactuals, persistently higher costs of purchased regional service reduce entry and continued service, even by airlines operating a route internally, because they may outsource later. Requiring some internal production deters entry but reduces service losses when regional supply is interrupted.
\end{abstract}
\vspace{0.4em}
\noindent\textit{Keywords.} Outsourcing, firm organization, airline competition, dynamic entry.
\par\noindent\textit{JEL classifications.} L13, L22, L93.
\clearpage

\section{Introduction}
\label{sec:introduction}

Firms that rely on outside suppliers may respond to higher purchasing costs or a supplier's withdrawal by purchasing from another supplier, expanding internal production, or leaving a market. The costs of these responses affect which firms continue to compete and which products remain available. They also influence entry. An incumbent that already combines internal operations and outside suppliers may be able to shift production between them at relatively low cost. An entrant choosing the same combination must first incur the associated setup costs. We study how setup and adjustment costs shape market structure when the cost or availability of outside supply changes.

The U.S. airline industry provides a setting in which we can observe both how a service is produced and whether a firm continues to offer it. The carrier selling a ticket need not operate the flight. Our data identify both the ticketing and operating carriers, allowing us to follow changes in the share of service a carrier outsources on the same route over time. Flights operated by the carrier or by regional airlines wholly owned by the carrier or its parent company constitute \emph{internal production}. Flights operated by independent regional providers constitute \emph{outsourcing}. We call the use of both on a route \emph{mixed production}. We observe these production choices together with route entry and exit, allowing us to compare entrants' choices with incumbents' adjustments. Regional airlines operated 40 percent of U.S. departures in 2019 \citep{RAA2020}.\footnote{Regional airlines provided the only scheduled passenger service at 436 U.S. airports in 2019 \citep{RAA2020}. Both figures include independent providers and wholly owned subsidiaries.} Legacy carriers market regional flights under brands such as Delta Connection and United Express, using both regional airlines they own and independent providers \citep{ForbesLederman2009}. Changes in the cost and availability of regional service can therefore affect a substantial part of a carrier's network.

Federal rules introduced in 2013 raised the qualifications required of airline first officers \citep{FAA2013PilotRule,GAO2014Pilots}. Republic Airways Holdings attributed a growing scarcity of qualified new pilots to these requirements and warned that recruitment difficulties could raise labor costs and disrupt regional service \citep{Republic2013Form10K}. In 2014, its subsidiary Chautauqua Airlines withdrew aircraft from its agreements with American and United amid difficulties recruiting pilots. American replaced some of the lost service through its subsidiary Envoy. Among active carrier-route pairs exposed to the withdrawal, outsourcing fell by about 30 percentage points relative to the same carriers' other routes, with the decline concentrated on American's routes. Across eleven provider withdrawals between 2004 and 2018, carriers retained most exposed routes, often by purchasing replacement service from another independent provider.\footnote{Figure~\ref{fig:republic_event_study} traces the production and continuation responses. Republic's announcements describe pilot recruitment difficulties and the withdrawal of 27 regional jets \citep{Republic2014InvestorUpdate,Republic2014Contracts}. The pooled difference-in-differences estimate for outsourcing is $-0.299$, with a standard error of $0.078$. Appendix~\ref{app:republic} reports the comparisons and concurrent changes in the carriers' networks, and Appendix~\ref{app:supplier_events} reports the eleven withdrawals.} Retaining a provider can also become more expensive. During Republic's 2016 restructuring, amendments to its agreements with Delta and United increased reimbursement rates \citep{Republic2016QuarterlyReport}. These episodes motivate our analysis of how changes in the cost and availability of purchased regional service affect entry and continued service.

We estimate a dynamic game of entry, exit, and outsourcing using quarterly data on seven U.S. legacy carriers from 1998 through 2019. The model distinguishes the costs of starting service with a particular production mix from the costs of changing that mix on a route already served. Our measure of participation counts each carrier on each route separately: losing one of two carriers reduces participation even if the other continues to serve the route. The observed production choices differ sharply between entrants and incumbents. Among entrants using independent providers, 88 percent outsource completely. Yet carriers that already combine internal and outside production change their outsourcing shares much more frequently. In the quarter following a production adjustment, the probability of adjusting again is 74 percent under mixed production and 17 percent under fully internal or fully outsourced production. The probability of another adjustment falls as an arrangement remains in place, including within the same production category and carrier-route pair. These patterns also appear when outsourcing is measured by operated seats and on routes with frequent departures.\footnote{Section~\ref{sec:data} describes the samples, and Figure~\ref{fig:fit_duration} reports adjustment rates by duration. Appendix Tables~\ref{tab:state_duration}, \ref{tab:measurement}, and \ref{tab:app_frequency} compare production categories, measures of outsourcing, and flight frequencies.} In each quarter, an incumbent can keep its outsourcing share or change it. Changing the share entails a fixed cost that depends on the production already in place and its duration, as well as a cost that increases with the size of the change. An entrant instead pays a route entry cost, a setup cost if it purchases outside service, and an additional setup cost if it begins service with both internal and outsourced production. Carriers take into account both current profits and future opportunities to change production, enter, or exit. Their decisions also depend on the expected choices of competing airlines.

Repeated observations of the same carrier on the same route allow us to separate adjustment costs from persistent differences in profitability. We jointly estimate the common payoff parameters and 15,035 carrier-by-route fixed effects. Adjustment frequencies inform the fixed cost of changing an arrangement; the size and direction of changes help distinguish adjustment costs from the recurring return to outsourcing. Entrants' choices, compared with those of incumbents, inform the additional setup cost of mixed production. The model also allows the return to continued service to depend on previous passenger volume, distinguishing small entrants from established carriers. A separately estimated demand and pricing system translates changes in the production mix and the set of carriers into fares, traffic, and consumer valuation.\footnote{Appendix~\ref{app:estimation} presents estimation and inference. Appendix~\ref{app:no_route_effects} compares the model with a specification that replaces carrier-by-route effects with carrier intercepts.}

The estimates explain why entrants rarely choose mixed production even though incumbents using it adjust frequently. Once internal operations and independent suppliers are both in place, changing their relative shares entails a lower fixed cost. The additional setup cost of mixed production is almost as large as the basic cost of entering a route in 1998. Adjustment also becomes more costly the longer a production arrangement has remained unchanged.

We first use the model to examine an unexpected, persistent increase in the cost of purchasing regional service, holding internal production costs fixed. Carriers can respond by carrying more passengers through their own operations or leaving routes; any outsourcing share, including complete outsourcing, remains available. The increase reduces the number of carriers serving routes. Its effects extend to airlines initially operating a route entirely internally and to potential entrants, because both may purchase regional service in later quarters. Expected future outsourcing costs therefore affect the decision to enter or continue serving a route today. A carrier's current outsourcing share on a route does not fully measure its exposure to a sustained increase in supplier costs. Support that preserves access to regional service may affect competition even on routes a carrier currently operates internally.

We then ask what happens when a carrier can no longer enter or remain on a route using only independent providers. In this counterfactual, every carrier must transport a specified fraction of its passengers on each route through its own operations or its wholly owned regional subsidiaries. The principal comparison sets this minimum at 25 percent. This is an analytical benchmark for studying the consequences of having to operate some service internally. It is not an airline policy observed in the data, and it does not reproduce a particular provider withdrawal: a carrier may replace a departing provider with another independent airline. Under the 25 percent minimum, the number of active carrier-route pairs falls by 3.6 percent, and by 16.7 percent on the third of routes with the lowest traffic. The entry probability falls by 0.42 percentage points. An airline that would enter using only independent providers must switch to fully internal production, pay the additional setup cost of mixed production, or stay out. Even a minimum internal share of 1 percent removes the fully outsourced entry option and produces 64 percent of the participation loss under the 25 percent minimum. Much of the participation loss thus occurs as soon as fully outsourced entry becomes unavailable. Competing carriers partly replace the lost participation: when one airline withdraws or does not enter, serving the route becomes more attractive to others. Holding the observed set of products fixed, average fares change little.

To compare this response with higher outsourcing costs, we choose a cost increase that produces approximately the same mean outsourcing share among active carrier-route pairs as the 25 percent minimum. The cost increase produces a 31 percent larger decline in participation. The minimum internal share removes some production choices but does not change the return to those that remain available. Higher outsourcing costs reduce profits whenever a carrier continues to purchase regional service, including in future quarters. To distinguish this recurring cost from the incentive to reduce outsourcing, we add a common payment to every carrier-route pair that remains active. The payment compensates the average additional cost and reduces the participation loss to 38 percent below that under the minimum internal share. Because the payment does not depend on a carrier's own outsourcing share, it leaves the incentive to reduce outsourcing in place. This exercise illustrates the role of the return to continued service; it requires outside funding, whose cost is not included.\footnote{Proposition~\ref{prop:menu_charge} separates the loss of production choices from the decrease in recurring returns, holding current states, continuation values, and aggregation weights fixed.}

Reducing the cost of reorganizing existing production helps carriers during the transition but does little to offset the longer-run decline in participation. In a further experiment, we give initially fully outsourced carriers the lower fixed adjustment cost estimated for mixed production. The initial participation loss under the 25 percent minimum falls by nearly three quarters, whereas the reduction in the average loss over the full experiment is much smaller. Existing carriers pay less to introduce internal service. Future entrants still face the same restriction on how they can begin serving a route.

Finally, we examine how prior use of internal production affects service when replacement suppliers are unavailable. We simulate an unanticipated four-quarter interruption of all independent regional production beginning in 2014. This is a broad disruption: unlike the Chautauqua episode, it leaves no other independent provider from which to purchase replacement service. With the 25 percent minimum in place before the interruption, the cumulative loss of active carrier-route pairs through 2019 is 72 percent smaller than when carriers are free to outsource completely. In each case, the loss is measured relative to the corresponding history without an interruption. The avoided loss equals approximately seven years of the reduction in carrier-route service under the minimum when independent providers are available, measured without discounting. The comparison includes changes in both production shares and the set of carriers serving routes before the interruption.\footnote{Sections~\ref{sec:reorganization} and \ref{sec:disruption} report the adjustment-cost and disruption experiments. Appendix~\ref{sec:merger} applies the framework to a hypothetical combination of American and United using conduct estimated from the three mergers among the sampled brands.} The minimum internal share thus reduces service losses during this interruption but also discourages entry in ordinary periods. Adjustment costs faced by incumbents and setup costs faced by entrants therefore have different implications for the choices available to passengers.

Our analysis contributes to the literature on firm organization by connecting the costs of reallocating service between internal operations and independent providers to entry and continued service. Theories of transaction costs and ownership explain how firm boundaries affect coordination and investment \citep{Coase1937,Williamson1985,GrossmanHart1986,HartMoore1990}, and \citet{LafontaineSlade2007} survey the empirical evidence on vertical integration. Studies of trucking and of cement and concrete production examine these forces \citep{BakerHubbard2003,HortacsuSyverson2007}, while \citet{AtalayHortacsuSyverson2014} distinguish vertical ownership from physical input flows. Mixed production relates to concurrent sourcing and the coexistence of company-owned and franchised outlets \citep{Bradach1997,Parmigiani2007,LafontaineShaw2005}. In airlines, \citet{ForbesLederman2009} relate regional carrier ownership to adaptation needs, and \citet{ForbesLederman2010} study the consequences for delays and cancellations. Other work examines outsourcing and price competition \citep{Tan2018}, shared subcontractors and fares \citep{AryalCampbellCilibertoKhmelnitskaya2026}, and the formation of subcontracting links \citep{HeKosmopoulou2021}.\footnote{\citet{GilKimZanarone2022} study relational outsourcing contracts under financial stress, and \citet{ArgyresGilZanarone2025} examine supply concentration and cooperation.} We distinguish the setup cost of combining internal and outside production from the cost of changing their relative use once both are in place. This distinction allows the same production arrangement to facilitate adjustment by incumbents while remaining costly for entrants.

The paper also relates to dynamic models of sourcing and airline competition. \citet{Igami2018} estimates a game of offshoring, entry, and exit in hard disk drives, and \citet{Thurk2024} studies how outside fabrication changes the investment required for entry into semiconductor production. We allow repeated changes in the share of internal and outside production, with fixed and variable adjustment costs as in models of capital adjustment \citep{CooperHaltiwanger2006}. Structural airline studies analyze entry \citep{Berry1992,CilibertoTamer2009}, demand and industry outcomes \citep{BerryJia2010}, dynamic entry and network economies \citep{AguirregabiriaHo2012}, demand estimation with endogenous entry \citep{CilibertoMurryTamer2021}, and service choices after a merger \citep{LiMazurParkRobertsSweetingZhang2022}. We allow carriers to choose both whether to serve a route and how much service to purchase from independent regional providers. Changes in outside supply can then affect passengers through the service offered by continuing carriers and through the entry and withdrawal of carriers.\footnote{The analysis also relates to ownership, product positioning, and variety \citep{Fan2013,Sweeting2013,MazzeoSeimVarela2018}. The merger application in Appendix~\ref{app:common_ownership} estimates conduct from service choices and complements the pricing analysis of \citet{MillerWeinberg2017} and the airline conduct estimates of \citet{CilibertoWilliams2014}.}

Estimation builds on models of industry evolution and conditional choice probabilities \citep{EricsonPakes1995,HotzMiller1993} and methods for persistent heterogeneity, including finite mixtures, grouped effects, and sufficient statistics for dynamic logit models with duration dependence \citep{KasaharaShimotsu2009,ArcidiaconoMiller2011,BonhommeManresa2015,IgamiYang2016,AguirregabiriaGuLuo2021}. We combine the value linearization of \citet{DearingBlevins2025} with the exact block elimination of \citet{KohYoon2026} to estimate common payoff parameters and carrier-by-route effects jointly. The method accommodates keeping an existing outsourcing share, choosing fully internal or fully outsourced production, and choosing a continuous share under mixed production. The fixed effects enter both current payoffs and continuation values.

Finally, the comparison of higher outsourcing costs with a minimum internal share relates to taxes and standards with entry and exit \citep{BuchananTullock1975,Spulber1985}, the input-tax and output-subsidy components of intensity standards \citep{HollandHughesKnittel2009}, and regulation in dynamic oligopoly \citep{Ryan2012,FowlieReguantRyan2016}. We distinguish the effect of removing a production option from the effect of reducing the recurring profits available from that option. The disruption experiment connects this distinction to supply assurance \citep{BoltonWhinston1993} and resilient supply \citep{GrossmanHelpmanLhuillier2023}, including airline sourcing under disruption risk \citep{WangWanJiangGu2025}. Supplier disruptions can propagate through production relationships \citep{BarrotSauvagnat2016,CarvalhoNireiSaitoTahbazSalehi2021}; our analysis shows why their consequences depend on both the production that firms already operate and the setup costs they face when entering a market.

\section{Institutional Setting and Data}
\label{sec:data}

\subsection{Regional Outsourcing in the Airline Industry}
\label{sec:background}

U.S. legacy carriers market regional flights under names such as Delta Connection and United Express. Flights sold under a legacy carrier's brand may be operated by the carrier itself, a regional subsidiary, or an independent regional provider. Independent providers may serve one or several legacy carriers \citep{ForbesLederman2009}. We classify flights operated by the legacy carrier and by regional airlines wholly owned by the carrier or its parent company as internal production. Flights operated by independent regional providers constitute outsourcing. Thus a flight on a regional aircraft is not necessarily outsourced: it counts as internal production when a wholly owned subsidiary operates it.

Under a capacity purchase agreement, the legacy carrier chooses routes and schedules, sets fares, and controls seat inventory. The regional provider operates the flights and receives payments based on aircraft, departures, and block hours, which include taxi time. The legacy carrier also reimburses specified operating expenses \citep{ForbesLederman2013,AirWisconsin2021,SkyWest2025}.\footnote{Under prorate agreements, the regional carrier receives passenger revenue according to a negotiated formula, bears more risk from fares, traffic, and costs, and may control schedules and fares on certain routes \citep{SkyWest2025}.} Contractual rates can change when the parties renegotiate. During Republic's 2016 restructuring, amendments to its agreements with Delta and United increased reimbursement rates \citep{Republic2016QuarterlyReport}. Although these agreements generally commit aircraft for several years, assignments across routes can change during the contract. Changes in schedules and aircraft assignments can therefore alter a route's outsourcing share without a new agreement. The share also changes when the legacy carrier or a wholly owned regional subsidiary expands or reduces service.

If an independent provider cuts its flying, the legacy carrier can operate more flights itself or through a wholly owned subsidiary, purchase replacement service from another provider, or withdraw from the affected routes. Replacement requires suitable aircraft and crews, and reallocating them can affect schedules, maintenance arrangements, and service elsewhere in the network. Expansion may be easier when the carrier or a wholly owned subsidiary already operates flights on the route. Aircraft and crews committed elsewhere, together with restrictions in existing contracts, can nevertheless limit the available alternatives. The carrier weighs the cost of establishing and operating replacement service against the revenue it expects to earn on the route.

Contracts between legacy carriers and their pilots' unions limit the number and size of regional aircraft that can be used in the carrier's network. These limits are set out in \emph{scope clauses} \citep{AirWisconsin2021,SkyWest2025}. Scope clauses can apply to wholly owned regional subsidiaries as well as independent providers. We classify flights operated by wholly owned subsidiaries as internal production even when a scope clause limits their use.

\subsection{Products, Carriers, and Outsourcing}

We define a market as a directional route in a calendar quarter. The ticketing carrier sells the service to passengers, while the operating carrier provides the flight. A product is the ticketing carrier's nonstop service in a market. Several operators may provide flights sold as the same product. We combine the Airline Origin and Destination Survey, T100 Segment records, and On Time Performance files from the U.S. Bureau of Transportation Statistics \citep{BTSOnTime,BTSDB1B,BTST100}. We link these records to identify the operators providing each ticketing carrier's service. The Origin and Destination Survey provides a ten percent sample of tickets for the period we study, so we multiply passenger counts by ten. We sum passengers across operating carriers within each product. The product fare is a passenger-weighted average of the mean ticket fares for flights operated by each carrier, and delay measures use the same passenger weights.

For each product, the outsourcing share is the fraction of sampled tickets for flights operated by independent regional providers. We apply the ownership classification described above and exclude other codeshare operations from the regional outsourcing measure.\footnote{Corporate filings establish provider ownership separately from branded service contracts. Appendix~\ref{app:data} documents the ownership histories.} A share of zero means that the carrier or its wholly owned subsidiaries operate all the service; a share of one means that independent regional providers operate all of it. We refer to these arrangements as complete internal production and complete outsourcing, respectively. Mixed production combines the two on the same route. This ticket measure is available for every observed product. We also measure outsourcing as the fraction of available seats supplied by independent regional providers. The seat measure covers 82.1 percent of product observations, allowing us to compare outsourcing measured by tickets with outsourcing measured by capacity.

The panel spans 88 quarters from 1998 through 2019 and covers seven ticketing carriers: American, Alaska, Continental, Delta, Northwest, United, and US Airways. We account for Southwest and other low-cost carriers through controls for market conditions. We exclude tickets sold by Trans World Airlines, which American acquired in 2001, and America West, which merged with US Airways in 2005. Table~\ref{tab:sample_summary} reports the samples used to estimate demand and passenger costs and to study route entry, exit, and production choices. The static sample contains 351,811 products on 6,744 directional routes in 264,597 route-quarter markets. The participation sample includes carriers serving a route and eligible carriers that do not currently serve it. It contains 818,335 carrier-route-quarter decisions in 260,940 route-quarter markets on 6,159 routes. We retain a decision only if the preceding quarter's state is observed and the participation eligibility rules described below are satisfied.\footnote{Appendix~\ref{app:data} provides the sample definitions, eligibility rules, and measurement details.} We remove Northwest, Continental, and US Airways from the set of available carriers when flights are no longer sold under their brands, distinguishing these brand retirements from withdrawal from an individual route.

\begin{table}[!tp]
\centering
\caption{Sample and Summary Statistics}
\label{tab:sample_summary}
\small
\begin{tabularx}{\textwidth}{@{}Xrrrrrr@{}}
\toprule
& Observations & Mean & Std. dev. & 10th pct. & Median & 90th pct. \\
\midrule
\multicolumn{7}{@{}l}{\textit{Panel A. Product variables}} \\
Outsourcing share & 351,811 & 0.322 & 0.427 & 0.000 & 0.000 & 1.000 \\
Fare, dollars & 351,811 & 282.40 & 110.23 & 164.03 & 270.64 & 402.92 \\
Passengers & 351,811 & 3,701 & 6,492 & 30 & 1,330 & 10,000 \\
Carrier network size & 351,811 & 0.481 & 0.327 & 0.095 & 0.400 & 0.938 \\
Active legacy brands & 351,811 & 1.613 & 0.806 & 1.000 & 1.000 & 3.000 \\
Capacity outsourcing & 288,672 & 0.363 & 0.450 & 0.000 & 0.000 & 1.000 \\
Departure delay, minutes & 241,225 & 8.65 & 8.42 & 0.45 & 7.35 & 17.98 \\
Cancellation fraction & 241,237 & 0.022 & 0.030 & 0.000 & 0.013 & 0.055 \\
\midrule
\multicolumn{7}{@{}l}{\textit{Panel B. Static sample}} \\
Markets & \multicolumn{6}{r}{264,597} \\
Directional routes & \multicolumn{6}{r}{6,744} \\
Quarters & \multicolumn{6}{r}{88} \\
Products with another modeled carrier & \multicolumn{6}{r}{157,007} \\
Independent providers & \multicolumn{6}{r}{23} \\
Candidate carrier and provider pairs & \multicolumn{6}{r}{154} \\
\midrule
\multicolumn{7}{@{}l}{\textit{Panel C. Participation and organizational game}} \\
Carrier, route, and quarter decisions & \multicolumn{6}{r}{818,335} \\
Route-quarter games & \multicolumn{6}{r}{260,940} \\
Directional routes & \multicolumn{6}{r}{6,159} \\
Estimated carrier-by-route effects & \multicolumn{6}{r}{15,035} \\
\bottomrule
\end{tabularx}
\par\smallskip
\begin{minipage}{\textwidth}
\footnotesize\justifying
\noindent\textit{Notes.} A product is a ticketing carrier's nonstop service on a directional route in a quarter. A market is a directional route in a quarter. Departure delay and the cancellation fraction describe the flights that serve the product in the On Time Performance files. Capacity outsourcing is available for 82.1 percent of products. Carrier network size is the share of routes from the product's origin airport on which the ticketing carrier sells nonstop tickets; the denominator consists of routes with sampled nonstop tickets. The game uses eligible decisions of carrier-route pairs that serve the route at least once.
\end{minipage}
\end{table}

\subsection{The Distribution and Persistence of Outsourcing}
\label{sec:persistence}

\begin{figure}[!tp]
\centering
\includegraphics[width=\textwidth]{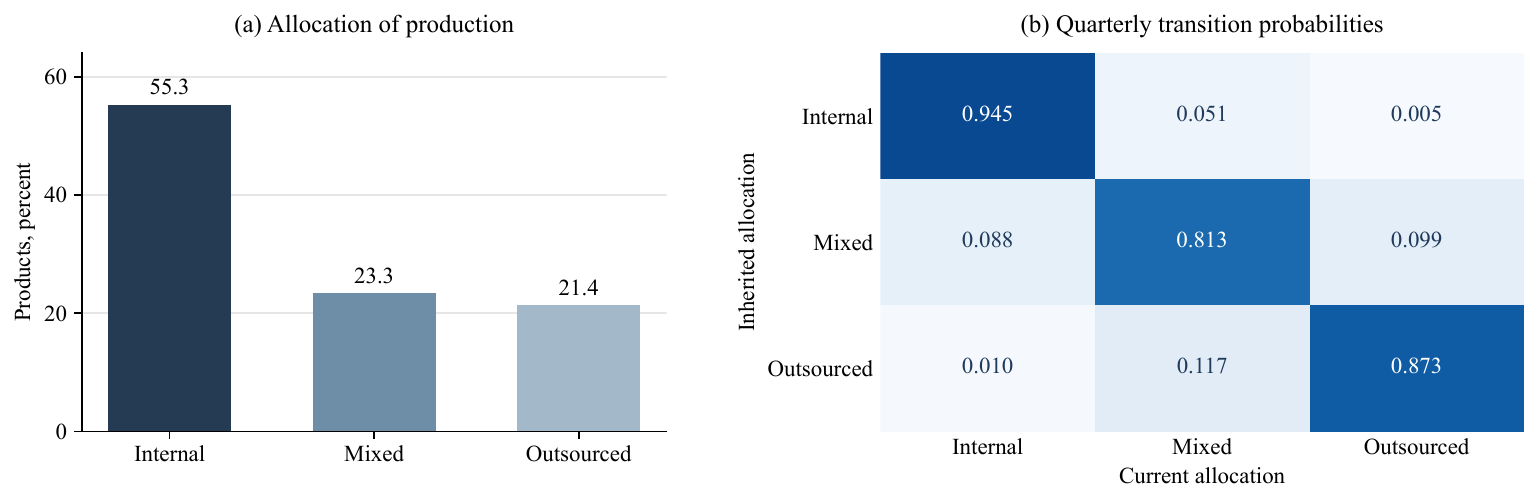}
\caption{The Distribution and Persistence of Outsourcing}
\label{fig:action_distribution}
\par\smallskip
\begin{minipage}{\textwidth}
\footnotesize\justifying
\noindent\textit{Notes.} Panel (a) reports the shares of the 351,811 products under complete internal production, mixed production, and complete outsourcing. Panel (b) reports transition probabilities between these arrangements for adjacent observations of the same carrier on the same route. Rows identify the arrangement in the earlier observation, and columns identify the arrangement in the later observation. Each row sums to one, subject to rounding.
\end{minipage}
\end{figure}

Complete internal production accounts for 55.3 percent of product observations, mixed production for 23.3 percent, and complete outsourcing for 21.4 percent. Figure~\ref{fig:action_distribution} reports this distribution and transitions among the three production regimes. Between adjacent observations of the same carrier on a route, the probability of remaining in the same regime is 0.945 for complete internal production, 0.813 for mixed production, and 0.873 for complete outsourcing. Direct switches between fully internal and fully outsourced production are rare; carriers generally pass through mixed production when moving between them. Persistence also remains after controlling for carrier-by-route and quarter fixed effects. A regression of the outsourcing share on its value in the previous quarter yields a coefficient of 0.868, with a standard error of 0.004, after applying the split-panel jackknife correction of \citet{DhaeneJochmans2015}.

A carrier can change its outsourcing share substantially while remaining in mixed production. For example, raising the share from 40 to 60 percent shifts one fifth of the product's passengers from internal to independent production without changing its classification as mixed production. We define a production adjustment as a change in the outsourcing share of more than five percentage points or any transition between the three regimes. Changes of five percentage points or less count as no adjustment when the regime is unchanged. Under this definition, adjustments occur in 23.1 percent of adjacent observations, with a median absolute change of 12.6 percentage points among adjustments.

A carrier's next adjustment depends on how it currently provides service and how long it has used that production mix. In the quarter after an adjustment, the probability of adjusting again is 74.3 percent under mixed production and 16.5 percent under fully internal or fully outsourced production. Duration is zero in the quarter following entry or an adjustment and increases by one for each subsequent quarter without an adjustment. Following a production adjustment, the pooled adjustment rate falls from 50.4 percent at duration zero to 16.6 percent at durations of two to three quarters and 3.6 percent at durations of eight to fifteen quarters. Fully internal and fully outsourced production are more common among long spells, but adjustment declines with duration within each regime as well. The relationship remains after controlling for carrier-by-route and quarter fixed effects.\footnote{Figure~\ref{fig:fit_duration} compares these patterns with model predictions. Appendix Table~\ref{tab:state_duration} reports the comparisons with fixed effects.}

The seat measure yields similar patterns. Ticket and seat outsourcing shares have a correlation of 0.952 and indicate the same production regime for 92.2 percent of products covered by both measures. They also remain positively associated within route-quarter markets after controlling for carrier-by-quarter effects (Table~\ref{tab:reduced_form_main}, Panel B). For the comparison of adjustment rates, we include observations following entry as well as those following an adjustment. At duration zero, adjustment rates under the seat measure are 72.2 percent for mixed production and 17.6 percent for fully internal or fully outsourced production. The corresponding rates under the ticket measure are 74.3 and 17.8 percent. Adjustment becomes less frequent with duration under both measures.

We also examine how much adjustment would arise from ticket sampling if the underlying outsourcing share were fixed. Under this binomial sampling benchmark, measured shares can change because different tickets are sampled even though the underlying allocation has not changed. Observed adjustment rates for mixed production exceed the benchmark in every group of sampled ticket counts. The gap rises from 24 to 56 percentage points as the number of sampled tickets increases.\footnote{Appendix Tables~\ref{tab:app_action_audit} and \ref{tab:measurement} report the capacity and sampling comparisons.} The decline in adjustment with duration also appears on routes with frequent service, where reallocating a few departures has a relatively small effect on the outsourcing share. Among carrier-route pairs averaging at least five daily nonstop departures in the preceding quarter, adjustment rates under mixed production fall from 64 percent at duration zero to 32 percent at durations of four or more quarters. The corresponding rates for fully internal or fully outsourced production are 27 and 5 percent.\footnote{Appendix Table~\ref{tab:app_frequency} reports the comparisons by lagged T100 departures.}

\subsection{Outsourcing and Fares}
\label{sec:reduced_form}

The relationship between outsourcing and fares differs across comparisons. Panel A of Table~\ref{tab:reduced_form_main} first follows a carrier on a route over time, controlling for carrier-by-route and calendar-quarter effects. The fitted relationship is U-shaped, with the lowest log fare near an outsourcing share of 0.36. Complete outsourcing is associated with fares 0.035 log points above complete internal production. A second specification compares carriers on the same route in the same quarter and includes carrier-by-quarter effects. Here, fares peak at an outsourcing share of 0.16, and complete outsourcing is associated with fares 0.031 log points below internal production. The first comparison removes permanent differences across carrier-route pairs; the estimated cost schedule in Section~\ref{sec:estimates} has the same U shape. The second also reflects differences in market power among carriers serving the same route. Because carriers choose both production and fares in response to demand and costs, neither comparison alone identifies how a change in outsourcing opportunities affects fares. We address this question with the demand, pricing, and production-cost model in Section~\ref{sec:demand_model}.

\begin{table}[!tp]
\centering
\small
\singlespacing
\caption{Outsourcing, Fares, and the Allocation of Production}
\label{tab:reduced_form_main}
\begin{tabularx}{\textwidth}{@{}Xrrrr@{}}
\toprule
\multicolumn{5}{@{}l}{\textit{Panel A. Log fares and outsourcing}} \\
Fixed effects & Outsourcing & Squared share & $o=1$ minus $o=0$ & Observations \\
\midrule
Carrier-by-route and quarter & $-0.0942$ & 0.1292 & 0.0350 & 348,548 \\
 & (0.0182) & (0.0178) & (0.0049) & \\
Route-by-quarter and carrier-by-quarter & 0.0148 & $-0.0455$ & $-0.0307$ & 157,007 \\
 & (0.0271) & (0.0249) & (0.0076) & \\
\end{tabularx}
\begin{tabularx}{\textwidth}{@{}Xrrr@{}}
\midrule
\multicolumn{4}{@{}l}{\textit{Panel B. Persistence and capacity}} \\
Outcome and explanatory variable & Coefficient & Standard error & Observations \\
\midrule
Outsourcing persistence, split-panel jackknife & 0.8681 & 0.0036 & 297,904 \\
Capacity outsourcing on ticket outsourcing & 0.7547 & 0.0400 & 93,280 \\
\bottomrule
\end{tabularx}
\par\smallskip
\begin{minipage}{\textwidth}
\footnotesize\justifying
\noindent\textit{Notes.} Both fare regressions include outsourcing and its square. The final coefficient column sums the linear and quadratic coefficients and compares complete outsourcing with internal production, holding controls fixed. The first regression includes carrier-by-route effects, calendar-quarter effects, the logarithm of the carrier's network size, hub status, and Southwest and other low-cost carrier presence. The second includes route-by-quarter and carrier-by-quarter effects, as does the capacity regression. The persistence estimate uses carrier-by-route and quarter effects and the split-panel jackknife. Parentheses in Panel A report standard errors. The persistence standard error uses 200 airport-pair multiplier draws, and other standard errors are clustered by airport pair.
\end{minipage}
\end{table} 

\subsection{Route Participation and Industry Consolidation}

To study entry and exit, we follow each legacy brand on routes it serves at least once, including quarters when it does not serve the route. A route-quarter enters the sample if at least one legacy carrier served the route in the preceding quarter. We retain the current quarter when all legacy carriers have withdrawn, allowing the panel to record the disappearance of legacy service as well as individual exits. An entry occurs when a brand begins serving a route after a quarter without service; continuation occurs when it serves the route in both quarters. We measure both among brands available in the industry in both quarters. Because each decision requires an observed preceding quarter, the panel contains 87 decision quarters.

We estimate a separate participation effect for each carrier-route pair observed both with and without service. For pairs that serve the route in every observed quarter, we use their production choices and hold their participation fixed. Pairs that never serve the route are omitted. The counterfactuals apply the same treatment to these groups.\footnote{The initial candidate set contains 25,563 never-active pairs and 1,118 always-active pairs. Appendix~\ref{app:data} describes sample construction.}

Quarterly entry rates, averaged across years, fall from 5.6 percent in 1998 through 2015 to 2.1 percent in 2016 through 2019, while continuation rises from 86.2 to 92.7 percent. The later period follows the mergers of Delta and Northwest, United and Continental, and American and US Airways. We allow entry costs to vary by year to account for the decline in entry alongside more persistent incumbent service. Entry is also more frequent on routes inherited through a merger. From the quarter in which the combined carrier receives a single operating certificate through two years after the acquired brand retires, the acquiring brand's quarterly entry rate on the acquired brand's routes is 17 percent, compared with an average entry rate of 5 percent.

\subsection{Responses to Supplier Withdrawals}
\label{sec:republic}

In 2014, Chautauqua Airlines, a subsidiary of Republic Airways Holdings, withdrew aircraft from its agreements with American and United. Republic attributed the decision not to renew these commitments to difficulties recruiting qualified pilots. The withdrawal reduced the capacity available to the two carriers under these agreements. We date the withdrawal period to 2014Q2.\footnote{Republic's Investor Update of February 11, 2014 announced the planned removal of 27 regional jets. The Form 8-K of March 4 specified that 12 ERJ145 aircraft would leave United service on April 1 and 15 ERJ140 aircraft would leave American service between March 1 and August 19. Republic was also adding larger E175 aircraft during this period \citep{Republic2014InvestorUpdate,Republic2014Contracts}.} We classify a carrier-route pair as exposed if Chautauqua carried its passengers during 2013Q2 through 2014Q1.

\begin{figure}[!tp]
\centering
\includegraphics[width=\textwidth]{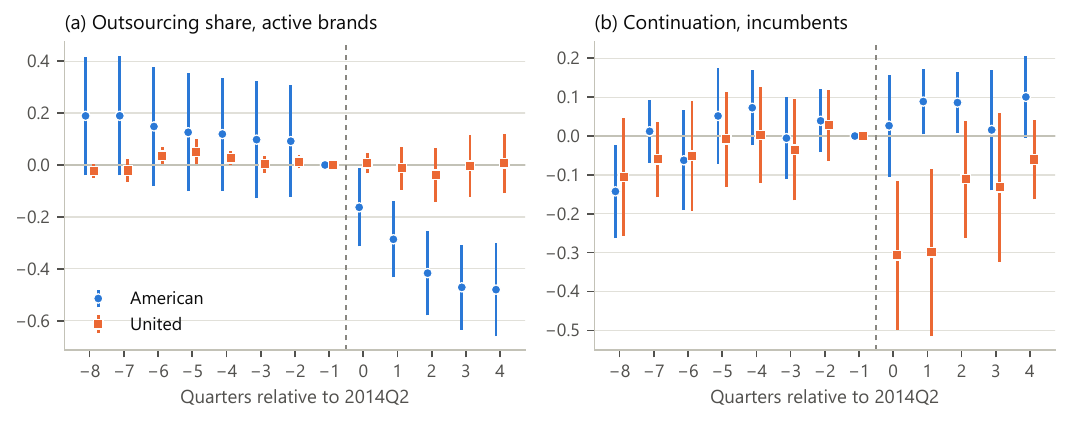}
\caption{Outsourcing and Route Continuation around Chautauqua's Withdrawal}
\label{fig:republic_event_study}
\par\smallskip
\begin{minipage}{\textwidth}
\footnotesize\justifying
\noindent\textit{Notes.} The panels report carrier-specific event-study estimates for routes exposed to Chautauqua's withdrawal. Event time is measured relative to 2014Q2, and 2014Q1 is the omitted quarter. Each regression includes carrier-by-route and carrier-by-quarter fixed effects and excludes the other affected carrier's routes exposed to Chautauqua. The event studies use the difference-in-differences specification. Outsourcing is measured among active observations and continuation among incumbents. Coefficients are in share or probability units. Intervals are pointwise 95 percent confidence intervals based on standard errors clustered by airport pair.
\end{minipage}
\end{figure}

Figure~\ref{fig:republic_event_study} follows outsourcing and route continuation around the withdrawal. Panel (a) shows a gradual decline in American's outsourcing share among active pairs, while United's share changes little. American shifted production toward its subsidiary American Eagle, renamed Envoy in 2014. On American's active exposed routes, the subsidiary's passenger share rose from 62.2 percent in 2013Q2 through 2014Q1 to 81.8 percent in 2014Q3 through 2015Q2. Other independent providers supplied additional replacement service.\footnote{The pooled American and United comparison in Appendix Table~\ref{tab:app_republic_did} gives a 29.9 percentage point decline in outsourcing on exposed routes relative to the carriers' other routes, with a standard error of 7.8 percentage points. Outsourcing is measured among active carrier-route pairs.}

Panel (b) shows different continuation patterns across the two carriers. Continuation on United's exposed routes falls relative to the comparison routes in 2014Q2 and 2014Q3. This decline is concentrated on routes from Cleveland, where United was contracting its hub. American's continuation point estimates are positive after the withdrawal, during its integration with US Airways.\footnote{American and US Airways entered common ownership in 2013Q4 and remained separately observed through 2015Q2. Carrier-by-quarter effects absorb changes common to each carrier's network. A triple difference that adds Delta's Chautauqua routes, where Chautauqua continued flying, as a further comparison gives a continuation estimate for United of $-0.148$, with a standard error of $0.077$. Excluding routes with an endpoint at Cleveland gives $0.236$, with a standard error of $0.081$. Appendix~\ref{app:republic} reports the full comparisons and event timing.}

The broader record contains eleven withdrawals of independent providers from legacy carriers between 2004 and 2018. For each event, we follow the carrier's pairs active before the withdrawal and compare exposed pairs with its other pairs in the same inherited production regime. Carriers continued serving most of the routes exposed to a withdrawal, often by purchasing replacement flights from another independent provider. Among exposed pairs that remained active and initially had outsourcing shares of at least 0.99, the average outsourcing share three quarters after withdrawal exceeded 0.9. Carriers could therefore continue to rely heavily on independent production after a provider withdrew. Service losses varied across events and were concentrated in several withdrawals.\footnote{Appendix Table~\ref{tab:supplier_events} reports all eleven events and the pooled comparisons. A placebo comparison eight quarters before each withdrawal gives a similar difference. A withdrawal occurs when a provider's passengers for a carrier fall below 15 percent of their average over the preceding four quarters and remain below that threshold for four quarters. Ownership changes and mergers of operating certificates are excluded.}

These withdrawals illustrate two responses to a loss of outside supply: carriers can replace production and continue serving a route, or withdraw. The outsourcing share records how continuing carriers divide production between their own operations and independent providers. It cannot record service lost when a carrier leaves a route, because the share is observed only where the carrier still sells tickets. We therefore study production choices together with entry and exit. The dynamic model allows a carrier to weigh the cost of replacing service today against the returns from continuing to serve the route, including the possibility of choosing a different outsourcing share in later quarters.

\section{Model}
\label{sec:model}

\subsection{Environment and Timing}
\label{sec:timing}

A market is a directional route $m$ in quarter $t$. The set $\mathcal{I}_{mt}$ contains the available legacy brands, with at most seven members. Brand $i$ chooses whether to serve the route, $a_{imt}\in\left\{0,1\right\}$, and, when active, the outsourcing share $o_{imt}\in\left[0,1\right]$. This share is the fraction of the brand's sampled tickets for flights operated by independent regional providers on the route. The map $f_t\left(i\right)$ assigns each brand to its corporate owner, and brand availability follows the observed ownership and integration history. Internal production includes flights operated by the carrier and its wholly owned regional subsidiaries. The outsourcing decision determines how much service to purchase from independent regional providers.\footnote{A change of provider that leaves the outsourcing share unchanged does not constitute a production adjustment. Network conditions enter the public state, and participation and production are chosen on each route.}

Decisions occur in two stages within a quarter. Each available brand first observes its private payoff shocks and decides whether to serve the route. A continuing carrier also decides whether to keep the previous quarter's outsourcing share or change it; an entrant chooses how much of its new service to outsource. Brands choose simultaneously, forming expectations about rivals' choices conditional on the public state. The quarter's demand innovation is then realized, and corporate owners set fares for their active brands. A carrier can reorganize production while continuing to serve the route, and the value of future production choices enters the current participation decision.

The dynamic payoff includes the return from ticket sales and the costs of maintaining and reorganizing production. We estimate it from carriers' decisions to enter, continue service, withdraw, or change their outsourcing shares. Carrier-by-route fixed effects allow a route to be persistently more profitable for one carrier than another. Adjustment costs depend on the previous outsourcing share and on how long the carrier has used that arrangement. We estimate demand and passenger costs separately, then use the pricing conditions to calculate the fares, traffic, and consumer valuation associated with counterfactual production and participation choices.\footnote{Appendix Table~\ref{tab:robustness_profit} reports a specification that links the activity payoff directly to predicted variable profit.}

\subsection{Demand, Pricing, and Passenger Cost}
\label{sec:demand_model}

Demand follows a nested logit in which the brands serving a route form one nest \citep{Berry1994}. Let $\mathcal{A}_{mt}$ denote the set of active brands, $q_{jmt}$ the passengers carried by brand $j$, $Q_{mt}=\sum_{j\in\mathcal{A}_{mt}}q_{jmt}$ total route traffic, and $s_{jmt}=q_{jmt}/Q_{mt}$ the brand's share of that traffic. Let $x_{jmt}$ contain observed demand characteristics with coefficient vector $\beta^D$, let $p_{jmt}$ denote the fare in dollars, and let $\xi_{jmt}$ denote the current demand innovation. Carrier-by-route effects $\psi_{jm}$ absorb persistent product differences, and carrier-by-quarter effects $\psi_{jt}$ absorb changes in a brand's appeal common to its routes. The price coefficient is $\alpha>0$, and $\gamma_o$ measures the effect of outsourcing on mean utility. Mean utility is
\[
\delta_{jmt}=x_{jmt}'\beta^D+\psi_{jm}+\psi_{jt}+\gamma_o o_{jmt}-\alpha\log p_{jmt}+\xi_{jmt}.
\]
The outside option includes other modes of travel and not traveling. Let $N_m$ denote the number of potential travelers and let $\sigma\in\left[0,1\right)$ measure the correlation of preferences across brands on the same route. With route traffic small relative to the population of potential travelers, product demand satisfies
\[
\log q_{jmt}=\log N_m+\delta_{jmt}+\sigma\log s_{jmt}.
\]
The term $\log N_m$ is absorbed in the carrier-by-route effects. Logarithmic fares make proportional price changes comparable across routes with different fare levels and allow markups to scale with fares. Differentiating demand at the current production allocation gives
\begin{equation}
\frac{\partial q_{jmt}}{\partial p_{imt}}=-\frac{\alpha}{p_{imt}}\left[\frac{Q_{mt}s_{jmt}\left(\ind\left\{i=j\right\}-s_{imt}\right)}{1-\sigma}+q_{jmt}s_{imt}\right].
\label{eq:coherent_demand_jacobian}
\end{equation}
The first term describes substitution across brands, and the second describes changes in total route traffic. A uniform one percent increase in fares on a route reduces traffic by $\alpha$ percent. For $i\ne j$, equation~\eqref{eq:coherent_demand_jacobian} implies $\partial q_{jmt}/\partial p_{imt}=\alpha\sigma q_{jmt}s_{imt}/\left[\left(1-\sigma\right)p_{imt}\right]$, so brands on the same route are substitutes when $\sigma>0$. When only one brand remains, the response of total traffic determines the markup. Let $D_{mt}$ denote the demand derivative matrix, with entries $D_{ji}=\partial q_{jmt}/\partial p_{imt}$.\footnote{Appendix~\ref{app:demand} derives the demand derivatives and corresponding consumer valuation.}

Each corporate owner sets the fares of its active brands to maximize joint variable profit. Let $mc_{imt}$ denote passenger marginal cost and $r_{imt}^p$ a residual in the pricing condition. The first-order condition is
\begin{equation}
q_{imt}+\sum_{j\in\mathcal{A}_{mt}}\ind\left\{f_t\left(j\right)=f_t\left(i\right)\right\}\left(p_{jmt}-mc_{jmt}\right)\frac{\partial q_{jmt}}{\partial p_{imt}}=r_{imt}^p.
\label{eq:price_foc}
\end{equation}
Within each market, collect fares, marginal costs, passenger quantities, and pricing residuals in $p$, $mc$, $q$, and $r^p$. Define the ownership matrix by $\Omega_{ij}=\ind\left\{f_t\left(i\right)=f_t\left(j\right)\right\}$ and let $\Delta_{ij}=\Omega_{ij}D_{ji}$. When $\Delta$ is nonsingular, equation~\eqref{eq:price_foc} gives $p=mc-\Delta^{-1}q+\tilde r^p$, where $\tilde r^p=\Delta^{-1}r^p$ expresses the pricing residual in dollars. Counterfactual comparisons hold this fitted residual fixed, so baseline fares equal their observed values.

Passenger marginal cost is the cost of carrying an additional passenger, conditional on the production arrangement. We allow this cost to depend on the outsourcing share through a quadratic term added to a positive baseline. Let $K_{im,t-1}$ denote the standardized logarithm of the carrier's passenger volume on the route in the preceding quarter, and let $\mathit{weather}_{imt}$ denote weather conditions. Define the baseline $c_{0,imt}=\exp\left\{\psi_m^c+\psi_{it}^c+\theta_s^cK_{im,t-1}\right\}$ and curvature $c_{2,imt}=\exp\left\{\theta_\kappa^c+\theta_{\kappa W}^c\mathit{weather}_{imt}\right\}$. Route effects $\psi_m^c$ and carrier-by-quarter effects $\psi_{it}^c$ absorb persistent route costs and common operating conditions. For a cost-minimizing outsourcing share $o_{imt}^*$,
\begin{equation}
mc_{imt}=c_{0,imt}+\frac{c_{2,imt}}{2}\left(o_{imt}-o_{imt}^*\right)^2.
\label{eq:marginal_cost}
\end{equation}
The minimizing share varies with inherited passenger volume, route distance $\mathit{dist}_m$, hub status $H_{imt}$, and weather. Define the logistic function $\Lambda\left(v\right)=\exp\left(v\right)/\left[1+\exp\left(v\right)\right]$. The cost-minimizing share is
\[
o_{imt}^*=\Lambda\left(\theta_0^c+K_{im,t-1}\theta_K^c+\mathit{dist}_m\theta_{\mathit{dist}}^c+H_{imt}\theta_H^c+\mathit{weather}_{imt}\theta_W^c\right).
\]

An increase in outsourcing lowers passenger marginal cost when $o_{imt}<o_{imt}^*$ and raises it when $o_{imt}>o_{imt}^*$. Reducing outsourcing can therefore raise passenger costs for one product and lower them for another. The carrier also considers the cost of reorganizing production and the return to future service. Its chosen share need not minimize current passenger costs.

\subsection{Adjustment of the Production Mix}
\label{sec:org_model}
\label{sec:adjustment}

An incumbent begins the quarter with the outsourcing share chosen in the preceding quarter. Let $\ell$ denote this inherited share and let $r\left(o\right)$ identify complete internal production, mixed production, or complete outsourcing. A production adjustment changes the outsourcing share by more than five percentage points or changes the production regime. The feasible new allocations form the set $\mathcal{O}^M\left(\ell\right)=\left\{o\in\left[0,1\right]|\left|o-\ell\right|>0.05\text{ or }r\left(o\right)\ne r\left(\ell\right)\right\}$. An active choice is $\left(o,k\right)$, where $k$ indicates adjustment. We call keeping the inherited share retention. Retention gives $\left(\ell,0\right)$, and adjustment gives $\left(o,1\right)$ with $o\in\mathcal{O}^M\left(\ell\right)$.

The two corner choices are complete internal production and complete outsourcing. Between them is a continuum of shares combining the two. The choice set contains retention and all changes that meet the adjustment definition. Formally, the measure $\mu_\ell$ assigns unit mass to retention and each feasible adjustment to a corner, and Lebesgue measure to the feasible interior intervals. A corner equal to the inherited share appears only as retention. Changes of five percentage points or less within the same regime are treated as retention in the organizational likelihood. An observed interior adjustment contributes a density at the recorded share. Demand and passenger cost estimation use the observed allocation.

Let $X$ collect the public state, and let $W\left(o,k\right)$ be the current payoff plus the continuation value of an active choice. The active inclusive value, $I^A=\log\int\exp\left\{W\left(o,k\right)\right\}d\mu_\ell\left(o,k\right)$, summarizes the return to serving the route when the carrier can choose among these production arrangements. It incorporates current returns and the value of future choices. Conditional on serving the route, the density over production choices is
\begin{equation}
f^A\left(o,k|\ell,X\right)=\exp\left\{W\left(o,k\right)-I^A\right\}.
\label{eq:mixed_continuous_org_probability}
\end{equation}
Equation~\eqref{eq:mixed_continuous_org_probability} is the limit of finite logit models that weight each interior cell by its length.\footnote{Appendix~\ref{app:mixed_choice_proof} derives the limit.} In these models, type-I extreme value shocks have unit scale and mean zero and are independent across brands, dates, and alternatives conditional on public states. Retention and feasible corner adjustments retain unit weights as the cells become finer.

Outsourcing can change because more carriers adjust or because those that adjust choose different shares. The model distinguishes these two responses. Define the adjustment probability $P^R=\Pr\left(k=1|a=1,\ell,X\right)$, mean outsourcing after adjustment $\bar o_R=\E\left[o|k=1,a=1,\ell,X\right]$, and mean outsourcing conditional on activity $m^o=\E\left[o|a=1,\ell,X\right]$. Let $I^R=\log\int_{k=1}\exp\left\{W\left(o,1\right)\right\}d\mu_\ell\left(o,1\right)$ and $W^0=W\left(\ell,0\right)$. A dot denotes a derivative with respect to a scalar shifter $\omega$, holding the inherited state and choice support fixed. Write $\mathcal C_\ell$ for the active choice space with typical element $c$, $U$ for a neighborhood of $\omega$, and $\E_R$ and $\operatorname{Cov}_R$ for expectation and covariance conditional on an active adjustment.

\begin{assumption}[Local organizational regularity]
\label{ass:org_regular}
(i) $0<\mu_\ell\left(k=1\right)<\infty$. (ii) $W$ is measurable in the choice and continuously differentiable in $\omega$. (iii) $\sup_{\omega\in U,c\in\mathcal C_\ell}\left\{\left|W\right|+\left|\dot W\right|\right\}<\infty$.
\end{assumption}

\begin{proposition}[Adjustment and allocation]
\label{prop:revision_allocation}
Under Assumption~\ref{ass:org_regular}, $P^R=\Lambda\left(I^R-W^0\right)$ and $m^o=\ell+P^R\left(\bar o_R-\ell\right)$. Moreover,
\begin{equation}
\dot P^R=P^R\left(1-P^R\right)\left\{\E_R\left[\dot W\right]-\dot W^0\right\},\quad \dot m^o=\dot P^R\left(\bar o_R-\ell\right)+P^R\operatorname{Cov}_R\left(o,\dot W\right).
\label{eq:revision_decomposition}
\end{equation}
\end{proposition}

The first term of $\dot m^o$ captures changes in the frequency of adjustment; the second captures changes in the outsourcing shares chosen by carriers that adjust. A common increase in the payoff to adjustment affects its frequency without changing the relative attractiveness of the new allocations. Now consider an additional current cost $\chi o$ of outsourcing, holding the other components of choice values fixed. The response is $\partial m^o/\partial\chi=-\operatorname{Var}_{f^A}\left(o\right)$, so expected outsourcing falls. Adjustment becomes more frequent when the inherited share exceeds the mean share chosen after adjustment, and less frequent when the ordering is reversed. The dynamic game also allows the cost increase to affect the value of future outsourcing.

Section~\ref{sec:counterfactuals} compares higher outsourcing costs with a counterfactual in which every carrier serving a route must operate part of its service itself or through wholly owned subsidiaries. In this comparison, outsourcing cannot exceed $1-\tau$, where $\tau\in\left[0,1\right]$ is the minimum internal share. We use this constraint to study what happens when complete outsourcing is no longer available, separately from an increase in its cost. The constraint removes every active choice with $o>1-\tau$ from the measure $\mu_\ell$, including retention when $\ell>1-\tau$. The cost increase instead subtracts $\chi o$ from every active choice value and leaves all production choices available. Let $F^A$ denote the probability distribution with density $f^A$ with respect to $\mu_\ell$, and let $\Pi_\tau=F^A\left(o>1-\tau|\ell,X\right)$ denote the probability of the choices removed by the limit. For choice densities $g$ and $f$ with respect to $\mu_\ell$, let $\mathcal D\left(g\Vert f\right)=\int g\log\left(g/f\right)d\mu_\ell$ denote the relative entropy of $g$ with respect to $f$.

For a comparison across brands, let $j=1,\ldots,n$ index brands with weights $\varpi_j\geq0$ summing to one and unrestricted choice densities $f_j^A$. Let $f_{\tau j}$ and $f_{\chi j}$ denote their active choice densities under the outsourcing limit and the common cost increase. Write $m'$ for the weighted mean outsourcing share under either experiment and $W\left(\varnothing\right)$ for the value of inactivity. Proposition~\ref{prop:menu_charge} compares current choices at fixed inherited states, public states, continuation values, and aggregation weights.

\begin{proposition}[Outsourcing limits and cost increases]
\label{prop:menu_charge}
Under Assumption~\ref{ass:org_regular}, (i) the outsourcing limit changes a brand's active inclusive value by $\log\left(1-\Pi_\tau\right)$, with $\Pi_\tau\rightarrow F^A\left(o=1|\ell,X\right)$ as $\tau\downarrow0$. The cost increase changes this value by $\log\E_{f^A}\left[\exp\left(-\chi o\right)\right]\rightarrow0$ as $\chi\downarrow0$. (ii) At the same weighted mean outsourcing share $m'$, the weighted change in active inclusive values under the cost increase minus that under the limit equals $\sum_j\varpi_j\left[\mathcal D\left(f_{\tau j}\Vert f_j^A\right)-\mathcal D\left(f_{\chi j}\Vert f_j^A\right)\right]-\chi m'$, where $\sum_j\varpi_j\mathcal D\left(f_{\tau j}\Vert f_j^A\right)\geq\sum_j\varpi_j\mathcal D\left(f_{\chi j}\Vert f_j^A\right)$. (iii) Under the condition in (ii), a common payment $\chi m'$ to every active choice under the cost increase makes the weighted difference nonnegative.
\end{proposition}

Part (i) shows why the counterfactual minimum internal share can substantially lower the return to continued service for a carrier that previously outsourced all its flights. With the value of inactivity fixed, each activity probability $\Lambda\left(I^A-W\left(\varnothing\right)\right)$ moves in the same direction as its active inclusive value. Any positive $\tau$ removes the probability mass at complete outsourcing. For a carrier that previously outsourced all its service, this mass includes keeping the existing arrangement, a choice with high probability in the data. To continue serving the route, the carrier must now establish some internal production and pay the fixed adjustment cost. Tightening the limit removes additional mixed allocations above $1-\tau$. In the data, 21.4 percent of products use complete outsourcing and 5.1 percent have mixed shares above 0.75, so complete outsourcing accounts for most of the allocations excluded by a 25 percent internal-share floor. The loss in the value of service is therefore discontinuous at $\tau=0$. A carrier that faces a small increase in outsourcing costs can instead pay the higher cost and keep its arrangement. The resulting loss in the value of serving the route is small. The minimum internal share also affects entrants: a carrier that would have entered by outsourcing all its flights must establish internal production or forgo entry.

Part (ii) compares the experiments at the same mean outsourcing share, for either one brand or an average across brands with fixed weights. Among all choice distributions with that mean, those induced by a common cost increase have the smallest weighted relative entropy with respect to the unrestricted distributions \citep{Csiszar1975}. Under this measure, the cost increase changes production choices least: carriers can continue using arrangements for which paying the higher outsourcing cost is preferable to reorganizing. They nevertheless pay more for the outsourced service they retain, whereas the minimum internal share leaves the payoffs to permitted production choices unchanged. The ranking of weighted active inclusive values depends on whether this additional expense exceeds the greater change in production choices caused by the limit. This distinction is related to the comparison of taxes and standards with free entry \citep{BuchananTullock1975,Spulber1985}. Here it arises from the cost of outside production and the persistence of sourcing choices, rather than the cost uncertainty in \citet{Weitzman1974}.

Part (iii) separates the effect of higher expenditure on outsourced service from the effect of changing production. A common payment $\chi m'$ to each brand serving the route offsets the mean added outsourcing cost in the weighted comparison. The remaining difference reflects the greater change in production choices under the limit, so the cost increase with this payment yields at least as high a weighted active inclusive value. The payment is conditional on service, preserves the relative cost of outsourcing, and requires financing separate from the increase in supplier costs. The decomposition is related to the input-tax and output-subsidy representation of an intensity standard in \citet{HollandHughesKnittel2009}. The dynamic game allows continuation values and rivals' participation to respond as well.

\subsection{The Participation and Organization Game}
\label{sec:long_run_mpe}

Index a route-quarter game by $e=\left(m,t\right)$. Brand $i$ inherits the state $s_{ie}=\left(a_{i,e-1},\ell_{ie},d_{ie},z_{ie}\right)$, where $a_{i,e-1}$ indicates whether the brand served the route in the preceding quarter, $\ell_{ie}$ is its previous outsourcing share, $d_{ie}$ counts quarters without a production adjustment, and $z_{ie}$ measures its previous passenger volume. Duration is zero in the quarter following entry or a production adjustment and increases by one after each quarter of retention. For arrangements already in place at the start of the sample, duration begins at zero in the first decision quarter. Inherited scale is the standardized logarithm of the brand's passengers on the route in the preceding quarter, capped at the median among active brands.\footnote{Inherited scale counts sampled passengers, one tenth of the expanded count. The cap equals 135 sampled passengers. An indicator $n^1_{ie}$ for a single sampled passenger enters the payoff with coefficient $b_1$.} An inactive brand has $\ell_{ie}=d_{ie}=z_{ie}=0$. The available actions are inactivity, $\varnothing$, and active production choices $\left(o,k\right)$. Rivals affect payoffs through their expected number on the route, $M^A_{ie}=\sum_{j\ne i}\Pr\left(a_{je}=1|X_e\right)$, and their expected outsourcing, $M^{AO}_{ie}=\sum_{j\ne i}\E\left[a_{je}o_{je}|X_e\right]$. The latter sums rivals' outsourcing shares, counting a rival's share as zero when it does not serve the route.

Entry costs are $c^E_{ie}=\kappa_E+\kappa_{E,y\left(t\right)}+\kappa_Ph_{ie}+\kappa_T\ind\left\{e\in\mathcal{G}_i\right\}$, where $h_{ie}$ indicates that brand $i$ served the route in an earlier quarter and is taken from the observed sequence of public states. Year shifts $\kappa_{E,y\left(t\right)}$ allow entry costs to differ in each calendar year from 1999 to 2019, with 1998 as the reference year. The set $\mathcal{G}_i$ contains the games in which brand $i$ is the acquiring brand of a legacy merger during integration, on a route that the acquired brand served at any time from eight quarters before closing through its last quarter. Integration runs from the quarter in which the combined carrier receives a single operating certificate through eight quarters after retirement of the acquired brand.\footnote{Closing, retirement, and the operating certificates are dated in Appendix~\ref{app:merger_data}.} Membership in $\mathcal{G}_i$ is part of the public state. The component of the quarterly payoff common to all production choices is
\[
A_{ie}=\phi_{im}+\psi_t+x_{ie}'b^A-\ind\left\{a_{i,e-1}=0\right\}c^E_{ie}+\ind\left\{a_{i,e-1}=1\right\}\left(b_zz_{ie}+b_1n^1_{ie}\right)+\rho_{AA}M^A_{ie}+\rho_{AO}M^{AO}_{ie}.
\]
The fixed effect $\phi_{im}$ captures permanent differences in a carrier's return to service on a route, and $\psi_t$ captures calendar-quarter conditions. One calendar effect is set to zero. The vector $x_{ie}$, with coefficient vector $b^A$, contains lagged route traffic, lagged carrier network size, the preceding quarter's rival count, and lagged Southwest and other low-cost carrier presence. The coefficient $b_z$ measures the return to continued service from the brand's own inherited scale, conditional on the carrier-by-route effect.

The recurring return to outsourcing has linear and quadratic components. The coefficient on the outsourcing share is $G_{ie}=\gamma_1+\gamma_{1,i}+\gamma_{1,\mathit{era}\left(t\right)}+w_{ie}'\gamma_w+\rho_{OA}M^A_{ie}+\rho_{OO}M^{AO}_{ie}$, where $w_{ie}$, with coefficient vector $\gamma_w$, contains lagged network size, route traffic, and the number of rivals. The carrier shifts $\gamma_{1,i}$ and the era shifts $\gamma_{1,\mathit{era}\left(t\right)}$ capture differences in the return to outside production across carriers and across the periods 2005 to 2009, 2010 to 2015, and 2016 to 2019, with Delta and 1998 to 2004 as the references.

Changing the outsourcing share incurs a fixed cost that depends on the previous production mix and on how long it has been in place. Taking Delta as the reference carrier, define
\[
R_{ie}=\theta_R+\theta_{R,i}+\theta_I\ind\left\{0<\ell_{ie}<1\right\}+\theta_D\log\left(1+d_{ie}\right)+\theta_\ell\ell_{ie}.
\]
The deterministic payoff from an active choice is
\begin{equation}
\begin{aligned}
u_{ie}\left(o,k\right)&=\begin{cases}A_{ie}+G_{ie}o+\gamma_2o^2-\theta_A\left(o-\ell_{ie}\right)^2-kR_{ie}, & a_{i,e-1}=1,\\A_{ie}+\left(G_{ie}+\gamma_E\right)o+\gamma_2o^2-\kappa_O\ind\left\{o>0\right\}-\kappa_M\ind\left\{0<o<1\right\}, & a_{i,e-1}=0,\end{cases}\\[3pt]
u_{ie}\left(\varnothing\right)&=0.
\end{aligned}
\label{eq:game_payoff}
\end{equation}
An incumbent that changes its outsourcing share pays the fixed component $R_{ie}$ and a quadratic cost that rises with the size of the change. An entrant has no inherited arrangement to reorganize. It pays $\kappa_O$ to establish outside production and an additional $\kappa_M$ if it also establishes internal production. The coefficient $\gamma_E$ allows the recurring return to outsourcing to differ at entry. Once both internal operations and independent service are in place, changing their relative shares may be less costly; $\theta_I$ allows the fixed adjustment cost to differ for these carriers. Duration dependence allows the same change in the outsourcing share to be more costly after a long period without adjustment. The four rival coefficients distinguish competition in route participation from interactions in sourcing choices. Flow payoffs are measured in units of the choice disturbance, whose scale is normalized to one.

In the cost experiment, an increase in the recurring cost of outside production reduces $G_{ie}$ by the common amount $\chi$. This change lowers the route operating payoff in proportion to outsourcing while holding the cost of internal production fixed. It does not directly change passenger marginal cost in equation~\eqref{eq:marginal_cost}. Fares respond when carriers change their production mixes or enter and leave routes. The experiment therefore concerns the cost of maintaining an outsourced production arrangement, measured in normalized payoff units, rather than a specified dollar increase per passenger or block hour.

Write $c$ for an action and $z'$ for scale in the next quarter. The next state $T\left(s,c\right)$ has a stochastic scale component under activity. Inactivity gives $T\left(s,\varnothing\right)=\left(0,0,0,0\right)$. An active incumbent that retains the allocation moves to $\left(1,\ell,d+1,z'\right)$, while an adjustment to share $o$ resets duration and gives $\left(1,o,0,z'\right)$. Entry also starts duration at zero. Scale follows a first-stage Markov process around the brand's mean scale on the route over all sample quarters. For a continuing brand, the deviation from that mean has autoregressive coefficient 0.64. An entrant starts 0.59 log points below the mean.\footnote{Both processes are estimated from consecutive active quarters. Residuals are drawn from their empirical distribution within eight bins of the predicted level, and a logit gives the probability of a single sampled passenger. Appendix~\ref{app:estimation_algorithm} describes the transition and its representation in continuation values.}

Brands know the observed sequence of public states, including brand availability. Let $n\left(ie\right)$ denote the next decision for the same brand and route along that sequence, including the continuation quarters defined below. The quarterly discount factor is $\beta=0.98$. Conditional on rival policies and the sequence of public states, the choice-specific value is $W_{ie}\left(c,s\right)=u_{ie}\left(c,s\right)+\beta\E\left[V_{n\left(ie\right)}\left(T\left(s,c\right)\right)|s,c,X_e\right]$. The expectation integrates over scale in the next quarter using the fixed first-stage transition distribution.

Adding a unit atom at inactivity to the organizational measure $\mu_\ell$ gives the measure $\nu_\ell$. The value function and unconditional choice density satisfy
\begin{equation}
V_{ie}\left(s\right)=\log\int\exp\left\{W_{ie}\left(c,s\right)\right\}d\nu_\ell\left(c\right),\quad f_{ie}\left(c|s,X_e\right)=\exp\left\{W_{ie}\left(c,s\right)-V_{ie}\left(s\right)\right\}.
\label{eq:dynamic_mixed_choice}
\end{equation}
The implied activity probability is $P_{ie}=\Lambda\left(I^A_{ie}-W_{ie}\left(\varnothing\right)\right)$. A carrier's decision to serve the route thus depends on the value of its available production choices. Each observed decision contributes a participation probability and, when the carrier is active, a conditional density for its production choice. Pairs treated as always active have no inactivity option in current or future choice sets. An equilibrium is a set of brand policies such that, in every game, each brand's expected rival activity $M^A_{ie}$ and expected outsourced activity $M^{AO}_{ie}$ equal the values implied by the rivals' choice densities in equation~\eqref{eq:dynamic_mixed_choice}.

To construct continuation values, we extend each observed sequence by twenty quarters, holding public covariates, calendar effects, year shifts of entry costs, era effects, rival policies, brand availability, and merger integration status at their final observed values. Own participation, outsourcing, duration, and scale continue to evolve according to $T$. Values are solved backward from a zero terminal value at the end of the extension. The added quarters affect continuation values, while the likelihood and counterfactual averages use observed decisions.\footnote{Appendix Table~\ref{tab:robustness_game} reports results with forty continuation quarters, estimation excluding 2018 and 2019, and a quarterly discount factor of 0.95.}

\subsection{Prices and Participation}
\label{sec:transmission}

Changes in the cost or availability of independent regional service can affect passengers through fares and through the set of carriers serving their route. A change in a carrier's production mix affects passenger costs and demand. A change in the profitability of serving the route can induce entry or exit, changing the products available to passengers and the competition among the remaining carriers. Let $p=F\left(p,o,\omega\right)$ denote the price policy under a shifter $\omega$, and let $\mathbf{I}$ denote the identity matrix. Subscripts denote partial derivatives, and dots denote derivatives with respect to $\omega$. When $F$ is continuously differentiable and $\mathbf{I}-F_p$ is nonsingular at the solution, the local price response is $\dot p=\left(\mathbf{I}-F_p\right)^{-1}\left(F_o\dot o+F_\omega\right)$. The corresponding activity response is $\dot P_{ie}=P_{ie}\left(1-P_{ie}\right)\left\{\dot I^A_{ie}-\dot W_{ie}\left(\varnothing\right)\right\}$. With the active choice set fixed, the change in the active inclusive value averages changes in current payoffs and continuation values over the choice distribution. A minimum internal share removes the option of complete outsourcing and other shares above the permitted maximum. At given continuation values, it changes the active inclusive value by $\log\left(1-\Pi_\tau\right)<0$, as in Proposition~\ref{prop:menu_charge}(i). The number of carriers serving routes can therefore fall even when offsetting cost and demand responses leave average fares almost unchanged.\footnote{Appendix~\ref{app:transmission_proof} derives the local responses.}

\section{Identification and Estimation}
\label{sec:identification_estimation}

\subsection{Identification}
\label{sec:identification}

Identification draws on prices and quantities, changes in production, and entry and continuation decisions. Demand and pricing identify how passenger costs vary with the production mix. Production histories distinguish a cost incurred when the outsourcing share changes from a return earned each quarter at the chosen share. Entry and continuation decisions identify the cost of entering a route and the return to remaining on it. The dynamic likelihood links these decisions through continuation values. Carrier-by-route effects absorb permanent differences in service profitability.

Demand estimation uses instruments for log fares and log within-route shares, both of which can respond to current demand innovations. The log number of active brands shifts within-route shares and identifies $\sigma$ \citep{Berry1994}. Participation is chosen before the current demand innovation is realized, making the number of brands predetermined relative to that innovation. The log Hausman fare is the same carrier's passenger-weighted fare on routes in the same quarter and distance decile that share neither endpoint with the focal route. This fare identifies $\alpha$ from quarterly cost variation, such as fuel and aircraft costs, shared by the carrier's routes of similar stage length \citep{Hausman1996,Nevo2001}. Both instruments are assumed orthogonal to the focal product's current demand innovation conditional on carrier-by-route and carrier-by-quarter effects, current and lagged outsourcing, lagged log seats per departure and log departures, and Southwest and other low-cost carrier presence. Carrier-by-quarter effects absorb network-wide demand shocks.\footnote{The estimating sample contains the 75 percent of products with departures on the route in the preceding quarter, which the lagged seat and departure controls require. Appendix Table~\ref{tab:app_price_slope} reports the first stage. Effects for each distance decile in each quarter of the year absorb seasonal demand differences between short and long routes. With these effects, the estimates of $\alpha$ and $\sigma$ are 3.08 (0.55) and 0.41 (0.06). Replacing the current number of brands with its value two quarters earlier gives 2.65 (0.46) and 0.44 (0.07) on products with that lag available. Appendix~\ref{app:robustness_demand} reports the estimates with the seasonal effects.}

Given the demand parameters, the demand derivatives and ownership structure determine markups and map the pricing conditions into effective passenger costs. Cost moments identify the cost-minimizing share and the curvature of the cost schedule. Instruments are lagged outsourcing, its square, inherited passenger scale, the products of lagged outsourcing with passenger scale, distance, hub status, and weather, and the product of squared lagged outsourcing with weather. These instruments are assumed orthogonal to current idiosyncratic cost innovations conditional on the included effects and cost shifters.

Production histories identify fixed and variable adjustment costs through different margins. Comparing adjustment rates across previous outsourcing shares identifies the fixed component, including the shift $\theta_I$ for carriers already combining internal and independent production. Carrier shifts $\theta_{R,i}$ absorb persistent differences in adjustment propensity. Identification of $\theta_D$ uses the decline in adjustment as arrangements age, given the inherited allocation and the carrier shifts. The same decline appears within carrier-route pairs that pass through several arrangements, so the duration pattern remains after removing persistent differences across pairs.\footnote{After the split-panel correction, the within-pair slopes of adjustment on log duration are $-0.053$ under exclusive production and $-0.147$ under mixed production, compared with pooled slopes of $-0.050$ and $-0.180$ (Appendix Table~\ref{tab:state_duration}). Allowing the outsourcing slope and the fixed adjustment cost to differ with the production regime inherited in each pair's first observed quarter gives a duration coefficient of 0.75, compared with 0.79 in the benchmark (Appendix~\ref{app:measurement_robustness}).}

Conditional on adjustment, the distribution of new shares identifies the recurring outsourcing return and the quadratic adjustment cost. Differences in chosen shares across carriers and eras identify shifts in the outsourcing return. The quadratic adjustment cost depends on how far the chosen share is from the previous share. The recurring payoff depends on the chosen share itself. Carriers starting from different outsourcing shares therefore face different adjustment costs for reaching the same new share, which separates adjustment costs from recurring returns. Entry decisions supply a further comparison: the production choices of entrants identify $\kappa_O$, $\kappa_M$, and $\gamma_E$, with the frequency of mixed production among incumbents relative to its rarity at entry identifying $\kappa_M$. The likelihood uses the recorded shares after adjustment as well as the probability masses at retention and the corners.\footnote{Appendix Table~\ref{tab:measurement} compares production measures.}

Participation histories identify entry costs and returns to continued service. Within carrier-route pairs observed both with and without service, the difference between starting service and continuing it identifies entry costs conditional on calendar effects. Calendar effects shift the payoffs of entrants and incumbents equally, so the year shifts are identified by entry relative to continuation across years. Entry by acquiring brands on the routes of acquired brands identifies $\kappa_T$. The comparison contrasts entry during integration with entry by other carriers in the same year and by the same carrier on the same route in other quarters. The contrast between first entry and reentry by the same carrier on the same route identifies $\kappa_P$. Continuation decisions of the same carrier on the same route at different inherited scales identify $b_z$, and consecutive active quarters identify the transition of scale. The unrestricted carrier-by-route effects absorb a common fixed cost of activity. Exits by the last active legacy brand identify the return to continued service without legacy rivals.\footnote{Appendix~\ref{app:payoff_normalization} derives the normalization of the activity cost and the treatment of constant participation histories.}

Strategic interactions are identified from variation in predicted rival behavior conditional on the brand's own state and payoff controls, as in two-step estimators of entry games \citep{AguirregabiriaMira2007}. The activity payoff already includes the preceding quarter's aggregate rival count. Rival-specific histories, inherited allocations, and route profitability supply additional variation in predicted current activity and outsourced activity. Identification requires these rival characteristics to affect the focal brand's current payoff only through the included controls and rivals' current choices, and requires the predicted rival terms to retain variation conditional on those controls. The discount factor of 0.98 fixes the intertemporal valuation of payoffs \citep{MagnacThesmar2002}.\footnote{Appendix Table~\ref{tab:robustness_game} evaluates the estimates and counterfactual responses at a discount factor of 0.95.}

The counterfactuals use the payoff differences identified under the model's normalizations. A minimum internal share rules out production arrangements with too much outsourcing and leaves the payoffs to the remaining choices unchanged; higher outsourcing costs reduce the payoff to purchasing independent regional service. Both retain the terminal values and sequences of public states. The inactive flow payoff is fixed at zero, defining the comparison under which entry costs are estimated. Shifting all carrier-by-route effects and a common activity cost by the same amount preserves flow payoffs and counterfactual values \citep{AguirregabiriaSuzuki2014,KalouptsidiScottSouzaRodrigues2017,KalouptsidiScottSouzaRodrigues2021}.\footnote{Appendix~\ref{app:payoff_normalization} proves the location invariance. Appendix~\ref{app:policy_merger_design} describes the counterfactual choice sets and simulation.}

\subsection{Estimation}
\label{sec:estimation_procedure}

We estimate demand, pricing, and cost parameters separately from the game, using each product's available contemporaneous and lagged moments. The set of potential service decisions includes route-quarters with legacy service in the preceding quarter and brands available in both quarters. Inclusion depends on the lagged state that enters the likelihood. The game has 149 common parameters covering payoffs, inherited scale, yearly entry costs, the entry cost during merger integration, carrier and era shifts of the outsourcing return, adjustment-cost shifters, calendar effects, and strategic interactions, plus 15,035 carrier-by-route effects.

We estimate the game in two stages. A static logit model with $\beta=0$, carrier-by-route effects, and no simultaneous rival terms provides first-stage choice probabilities. These probabilities give the expected number of rivals serving the route and the sum of their expected outsourcing shares in each observed game. The second stage estimates dynamic payoffs by solving each brand's problem conditional on the fitted rival policies \citep{AguirregabiriaMira2007,BajariBenkardLevin2007,PesendorferSchmidtDengler2008}. The benchmark counterfactuals in Section~\ref{sec:counterfactuals} solve for the equilibrium in which every brand re-optimizes.

To maximize the second-stage dynamic likelihood, we combine the value linearization underlying efficient pseudo-likelihood (EPL) with iteratively reweighted least squares (IRLS) \citep{DearingBlevins2025,KohYoon2026}. At each outer iteration, backward recursion evaluates choice values and their analytic derivatives at the current parameters. The derivatives propagate payoff changes through all continuation values. Linearizing the choice values gives an affine index for each choice and a logit likelihood on the measure $\nu_\ell$. IRLS maximizes this likelihood, after which we update the parameters and recompute values for the next outer iteration.

Joint estimation is feasible because the fixed effects enter separate route problems. An effect for one carrier and route changes choice values on that route but does not enter another route's likelihood. Each IRLS step therefore eliminates the local effects through blocks of at most $7\times7$ and solves a system of dimension 149 for the common parameters. This block elimination gives exactly the same parameter update as solving the full joint IRLS system. We apply the method of \citet{KohYoon2026} to a dynamic game whose production choices have both probability masses and continuous densities. The fixed effects enter all continuation values and remain in the counterfactual calculations.\footnote{Appendix~\ref{app:estimation_algorithm} gives the derivatives and the exact elimination formula.}

Continuation values are approximated as functions of inherited outsourcing, log duration, and inherited scale, with separate representations for the corners and interior allocations that preserve differences in fixed adjustment costs. The terms in inherited scale also differ across the three production regimes.\footnote{Appendix~\ref{app:estimation_algorithm} describes the value representation.}

\subsection{Bias Correction and Inference}
\label{sec:inference}

We use the split-panel jackknife of \citet{DhaeneJochmans2015} to correct the leading incidental parameter bias from estimating carrier-by-route effects \citep{HahnNewey2004,FernandezValWeidner2016}. We split each route's quarters at its median and hold calendar effects, the year shifts of entry costs, and the era shifts of outsourcing returns at their full-sample estimates in both half panels. Table~\ref{tab:game_estimates} reports corrected payoff coefficients and standard errors. The entry-cost estimate rises from 3.12 to 3.86 after correction. Counterfactuals retain the full-sample time effects, use the corrected payoff coefficients, and re-estimate the carrier-by-route effects conditional on these common parameters.

Standard errors allow dependence within airport pairs, which contain both directions of a route, and within calendar quarters \citep{CameronGelbachMiller2011}. For the corrected estimator, the influence function combines full-sample and half-panel scores and includes uncertainty in the calendar, year, and era effects held fixed in the half panels.\footnote{The calculation conditions on fitted first-stage rival probabilities and the fitted scale transition. Appendix~\ref{app:inference} derives the influence function.}

Counterfactual standard errors use 180 draws of the reported common parameter vector from a normal distribution centered at its estimate with the estimated covariance. For each draw, we re-estimate the carrier-by-route effects given the drawn parameters and solve every game counterfactual. These calculations cover outsourcing limits, cost increases, disruptions, reorganization costs, and route-size comparisons, both in equilibrium and with rivals held at their first-stage probabilities. Standard errors for fares, traffic, consumer valuation, and variable profit combine the delta method for demand and cost parameters with variation across these draws.

\section{Estimates}
\label{sec:estimates}

Tables~\ref{tab:current_point_estimates} and \ref{tab:game_estimates} report the demand and passenger cost estimates and the dynamic payoff estimates, respectively.

\subsection{Demand and Heterogeneity in Production Costs}

The estimated price coefficient is 2.74, with a standard error of 0.48, and the nesting parameter is 0.40, with a standard error of 0.05. Least squares gives 0.57 and 0.82, consistent with fares and within-route shares responding to unobserved demand. The first-stage $F$ statistics are 23.1 for log fares and 380.6 for log within-route shares, and the Kleibergen-Paap statistic for the joint first stage is 17.2 \citep{KleibergenPaap2006}.

A uniform one percent increase in fares reduces route traffic by 2.74 percent. In markets with several brands, the median own-price elasticity in absolute value is 3.8 and the median Lerner index is 0.27. The positive nesting parameter implies closer substitution among brands on a route than with the outside good. Moving from complete internal production to complete outsourcing lowers mean utility by 0.069, so reducing outsourcing raises demand at fixed fares.

\begin{table}[!tp]
\centering
\small
\singlespacing
\caption{Demand and Passenger Cost Estimates}
\label{tab:current_point_estimates}
\begin{tabularx}{\textwidth}{@{}Xrrr@{}}
\toprule
Parameter & Estimate & Standard error & $p$ value \\
\midrule
Price coefficient on log fare, $\alpha$ & 2.7431 & 0.4815 & $<0.001$ \\
Nesting parameter, $\sigma$ & 0.4004 & 0.0517 & $<0.001$ \\
Outsourcing share in mean utility, $\gamma_o$ & $-0.0689$ & 0.0222 & 0.0019 \\
Cost-minimizing share intercept & $-0.4943$ & 0.1529 & 0.0012 \\
Log cost curvature & 4.1686 & 0.1531 & $<0.001$ \\
Cost-minimizing share, log lagged passengers & 0.3652 & 0.0897 & $<0.001$ \\
Cost-minimizing share, distance & 0.2449 & 0.0842 & 0.0036 \\
Cost-minimizing share, hub & $-0.7682$ & 0.2075 & $<0.001$ \\
Cost-minimizing share, weather & 0.4225 & 0.0848 & $<0.001$ \\
Weather effect on log curvature & 0.4072 & 0.0829 & $<0.001$ \\
Baseline cost, log lagged passengers & $-0.0239$ & 0.0026 & $<0.001$ \\
\bottomrule
\end{tabularx}
\par\smallskip
\begin{minipage}{\textwidth}
\footnotesize\justifying
\noindent\textit{Notes.} The demand and passenger cost parameters are estimated separately from the participation and organization game. The demand parameters are two-stage least squares estimates with the log number of brands on the route and the log Hausman fare as instruments, on 263,912 products. The cost parameters are estimated by the generalized method of moments given the demand estimates. The coefficients for the cost-minimizing share enter its logistic index. The scale covariate is the standardized logarithm of inherited passengers in both the baseline cost and the equation for the cost-minimizing share. Standard errors are clustered by airport pair.
\end{minipage}
\end{table}

Passenger marginal cost has a median of \$182.06. The cost-minimizing outsourcing share has a median of 0.38 and 10th and 90th percentiles of 0.22 and 0.60. Most observed shares are zero or one, placing production on either side of these interior minima. Across feasible allocations, the largest increase in marginal cost above a product's minimum has a median of \$12.70 and a 90th percentile of \$21.72. Lowering the maximum permitted outsourcing share can therefore move some products toward their cost minima and others away. Higher lagged passenger volume for the brand, longer distances, and adverse weather shift the cost-minimizing share toward outsourcing. Hub status shifts it toward internal production, and adverse weather also makes passenger costs more sensitive to deviations from the minimizing share.

The fitted cost schedule and differences in markups account for the fare patterns in Table~\ref{tab:reduced_form_main}. With carrier-by-route and quarter effects, log marginal cost is U-shaped in outsourcing, with a minimum at 0.38, compared with 0.36 for log fares. Complete outsourcing raises log marginal cost by 0.043 relative to internal production, compared with 0.035 for log fares. Across carriers in the same route and quarter, log marginal cost has the same shape, with its minimum at 0.46, and the ratio of fare to marginal cost is 0.056 log points lower under complete outsourcing than under internal production. On routes with at least two modeled carriers, brands with mixed production carry an average of 60 percent of route traffic, compared with 41 percent under internal production and 40 percent under complete outsourcing, and these brands set the highest markups.

Across active products, revenue less modeled passenger variable cost averages 34.09 percent of revenue. The counterfactual price calculations combine changes in passenger costs with demand substitution and route traffic responses, holding each product's fitted dollar pricing residual fixed.\footnote{Appendix~\ref{app:static_quantities} reports the cost distribution and pricing residuals used in the counterfactuals.}

\subsection{Organizational Adjustment}

Adjustment costs depend on the size of the reallocation, the inherited production mix, and the duration of that mix. The quadratic adjustment coefficient is 4.52, with a standard error of 0.30, implying a cost of 0.41 payoff units for a change of 30 percentage points. At zero duration under complete internal production, the reference fixed component is 2.02. The fixed component includes a discount of 2.55 for inherited mixed production. The coefficient on $\log\left(1+d\right)$ is 0.79, so four quarters of retention add 1.27 units to the fixed component.\footnote{Delta is the reference carrier. At complete internal production, four quarters raise the reference fixed component from 2.02 to 3.29. The component also includes a carrier shifter and the term for the inherited share. At zero duration, the reference value under complete outsourcing is about 1.71 payoff units. The shift for mixed production is additional to these components.} For Delta, leaving complete outsourcing carries a fixed component equal to 44 percent of the 1998 entry cost at zero duration and 77 percent after four quarters.

\begin{table}[!tp]
\centering
\singlespacing
\caption{Estimates of the Participation and Organization Game}
\label{tab:game_estimates}
\small
\begin{tabularx}{\textwidth}{@{}X*{4}{>{\raggedleft\arraybackslash}p{1.9cm}}@{}}
\toprule
 & \multicolumn{2}{c}{Benchmark} & \multicolumn{2}{c}{No simultaneous rival terms} \\
\cmidrule(lr){2-3}\cmidrule(l){4-5}
Parameter & Estimate & SE & Estimate & SE \\
\midrule
\multicolumn{5}{@{}l}{\textit{Panel A. Production allocation}} \\
Outsourcing share, $\gamma_1$ & $-$2.689 & 0.128 & $-$2.813 & 0.125 \\
Outsourcing share squared, $\gamma_2$ & 2.339 & 0.116 & 2.472 & 0.114 \\
Share $\times$ lagged network size & 0.059 & 0.017 & 0.059 & 0.017 \\
Share $\times$ lagged route traffic & $-$0.090 & 0.011 & $-$0.099 & 0.010 \\
Share $\times$ lagged rivals & $-$0.016 & 0.024 & $-$0.002 & 0.006 \\
Quadratic adjustment cost, $\theta_A$ & 4.523 & 0.296 & 4.493 & 0.292 \\
Fixed cost intercept, $\theta_R$ & 2.023 & 0.115 & 2.093 & 0.115 \\
Fixed cost, log duration, $\theta_D$ & 0.789 & 0.044 & 0.706 & 0.040 \\
Fixed cost, inherited share, $\theta_\ell$ & $-$0.313 & 0.055 & $-$0.325 & 0.055 \\
Fixed cost shift, mixed production, $\theta_I$ & $-$2.552 & 0.096 & $-$2.645 & 0.096 \\
Setup cost of outsourcing at entry, $\kappa_O$ & 1.297 & 0.226 & 1.272 & 0.227 \\
Setup cost of mixed production at entry, $\kappa_M$ & 3.376 & 0.172 & 3.433 & 0.174 \\
Outsourcing share at entry, $\gamma_E$ & $-$0.104 & 0.162 & $-$0.162 & 0.160 \\
\midrule
\multicolumn{5}{@{}l}{\textit{Panel B. Participation}} \\
Entry cost in 1998, $\kappa_E$ & 3.858 & 0.212 & 4.012 & 0.207 \\
Entry cost shift, merger integration, $\kappa_T$ & $-$1.065 & 0.383 & $-$1.183 & 0.405 \\
Entry cost, previous service on route, $\kappa_P$ & $-$0.429 & 0.271 & $-$0.432 & 0.274 \\
Inherited scale (std.), $b_z$ & 0.616 & 0.042 & 0.706 & 0.040 \\
Single sampled passenger, $b_1$ & $-$0.010 & 0.058 & 0.093 & 0.058 \\
Lagged route traffic (std.) & 0.228 & 0.027 & 0.150 & 0.021 \\
Lagged network size (std.) & $-$0.085 & 0.022 & $-$0.069 & 0.021 \\
Lagged number of rivals (std.) & 0.110 & 0.028 & $-$0.157 & 0.016 \\
Lagged Southwest presence & $-$0.005 & 0.052 & $-$0.020 & 0.049 \\
Lagged other low-cost presence & $-$0.051 & 0.034 & $-$0.058 & 0.031 \\
\midrule
\multicolumn{5}{@{}l}{\textit{Panel C. Rival interaction}} \\
Active $\times$ rivals' activity, $\rho_{AA}$ & $-$0.557 & 0.077 &  &  \\
Active $\times$ rivals' outsourced activity, $\rho_{AO}$ & $-$0.204 & 0.034 &  &  \\
Share $\times$ rivals' activity, $\rho_{OA}$ & $-$0.042 & 0.044 &  &  \\
Share $\times$ rivals' outsourced activity, $\rho_{OO}$ & 0.204 & 0.034 &  &  \\
\midrule
Carrier-by-route effects & \multicolumn{2}{r}{15,035} & \multicolumn{2}{r}{15,035} \\
Calendar quarter effects & \multicolumn{2}{r}{86} & \multicolumn{2}{r}{86} \\
Decisions (route-quarter games) & \multicolumn{4}{r}{818,335 (260,940)} \\
\bottomrule
\end{tabularx}
\par\smallskip
\begin{minipage}{\textwidth}
\footnotesize
\justifying
\noindent\textit{Notes.} Point estimates use the split-panel jackknife, holding calendar effects, the year shifts of entry costs, and the era shifts of outsourcing returns at full-sample values in the half panels. Payoffs are in logit disturbance units, and costs are subtracted. Standard errors use the jackknife influence function, clustered by airport pair and calendar quarter. Appendix Table~\ref{tab:entry_years} reports the year shifts of entry costs relative to 1998. The final column reports analytical standard errors of the corrected estimator in the specification without simultaneous rival terms, which retains lagged rival counts. Fixed adjustment-cost shifts relative to Delta are American 0.02 (standard error 0.06), Alaska 0.27 (0.09), Continental 0.30 (0.08), Northwest 0.19 (0.06), United 0.14 (0.04), and US Airways 0.16 (0.04). Shifts in the linear outsourcing return relative to Delta are American $-0.06$ (0.03), Alaska 0.19 (0.05), Continental 0.28 (0.09), Northwest 0.10 (0.05), United 0.19 (0.03), and US Airways 0.11 (0.03). Era shifts in the linear outsourcing return relative to 1998 to 2004 are 0.18 (0.06) in 2005 to 2009, 0.21 (0.05) in 2010 to 2015, and 0.16 (0.05) in 2016 to 2019. Inherited scale enters the payoff of active incumbents. The quarterly discount factor is 0.98.
\end{minipage}

\end{table}

These estimates explain why carriers tend to keep fully internal or fully outsourced service but often change their outsourcing shares once they combine both forms of production. For Delta in 1998 to 2004, the recurring outsourcing component is $-2.689o+2.339o^2$, which is convex. Across carriers and eras, its value at complete outsourcing relative to internal production ranges from $-0.41$ to 0.14. Once an incumbent decides to adjust, however, the cost of moving far from its inherited allocation makes its current payoff concave in the new share: the coefficient on the share squared is $\gamma_2-\theta_A\simeq-2.18$. A carrier can therefore prefer a share between zero and one, depending on its previous share and the recurring return to outsourcing. For mixed producers, the discount can make the fixed adjustment component negative soon after an adjustment, encouraging another reallocation. The fixed cost rises as the arrangement persists, and these future adjustment costs enter the value of the current production choice.

Operating conditions also shape outsourcing returns. Outsourcing is less attractive on routes with more traffic and more attractive for carriers that sell nonstop service on a larger share of routes from their origin airports. Rivals' outsourced activity increases the return to a brand's own outsourcing, with $\rho_{OO}=0.20$ and a standard error of 0.03. Competition reduces the return to service, while rivals' use of independent production raises the relative return to outsourcing. The estimates of $\rho_{AO}$ and $\rho_{OO}$ have nearly equal magnitudes and opposite signs, so rivals' outsourced activity enters the payoff of an active brand approximately as $-\rho_{OO}\left(1-o\right)M^{AO}_{ie}$. Rivals' outsourced service therefore lowers the return to a brand's internally produced service and leaves the return to complete outsourcing nearly unchanged. The carrier shift in the outsourcing return is largest for Continental and smallest for American, and the era shifts raise the return after 2004 by 0.16 to 0.21.

The lower adjustment cost under mixed production applies after the carrier has established both its internal operations and its independent service. At entry, outside production has a setup cost of 1.30, with a standard error of 0.23, and mixed production costs an additional 3.38, with a standard error of 0.17. The recurring return to outsourcing is close to that of incumbents, with $\gamma_E=-0.10$ and a standard error of 0.16. The additional setup cost explains why 88 percent of entrants using outside production choose complete outsourcing (Table~\ref{tab:fit_regimes}), even though mixed production is common among incumbents that already operate both forms of production.

\subsection{Participation and Rival Interaction}

The entry cost is 3.86 payoff units in 1998, with a standard error of 0.21.\footnote{Appendix~\ref{app:robustness_profit} reports a participation payoff linked to predicted variable profit. A loading of 1.789 payoff units per \$1 million of quarterly variable profit converts that specification's entry-cost estimate of 3.942 into \$2.20 million of variable profit, incurred in the entry quarter.} All subsequent year shifts are positive, ranging from 0.19 to 0.82 through 2015 and from 1.36 to 2.86 in 2016 to 2019. The late increase coincides with consolidation among legacy carriers. Merger integration lowers the entry cost of the acquiring brand on the routes of the acquired brand by 1.06, with a standard error of 0.38. On these routes the acquiring brand takes over the airport facilities and customers of the acquired brand.

Within carrier-route pairs, higher lagged route traffic raises the return to service. A one-unit increase in the brand's standardized inherited scale raises the payoff from continuation by 0.62 payoff units, with a standard error of 0.04. Small incumbents therefore exit more often than large incumbents on the same route. A new entrant starts below the brand's typical scale on the route.

An additional expected active rival lowers the direct payoff from service by 0.56 payoff units, with a standard error of 0.08. Conditional on this expected competition and on the carrier-by-route effect, the positive coefficient on lagged rivals captures time-varying route attractiveness. The combined direct effect of a rival present in the current and preceding quarters is about $-0.39$, before interactions with outsourced activity and continuation effects.\footnote{The lagged rival count has a standard deviation of 0.671, so its standardized coefficient implies an increase of about 0.16 per additional lagged rival. The combined effect holds other payoff arguments fixed.} Omitting simultaneous rival terms makes the coefficient on lagged rivals negative, as lagged presence then absorbs the competitive effect. The organizational coefficients remain close to the benchmark estimates.

Permanent differences in route profitability matter for the counterfactual responses. Replacing carrier-by-route effects with one intercept per carrier gives an entry-cost estimate of 4.03, compared with 3.86 in the benchmark model. In the counterfactual that makes carriers operate at least 25 percent of service internally on each route (Section~\ref{sec:counterfactuals}), the predicted loss is 5.57 active brands per 100 route-quarter markets, compared with 3.99 in the benchmark, or 1.4 times the benchmark loss. The benchmark retains the estimated distribution of carrier-by-route profitability when calculating this response.\footnote{Appendix~\ref{app:no_route_effects} reports the estimates of both specifications.}

\subsection{Model Fit}
\label{sec:fit}

\begin{figure}[!tp]
\centering
\includegraphics[width=\textwidth]{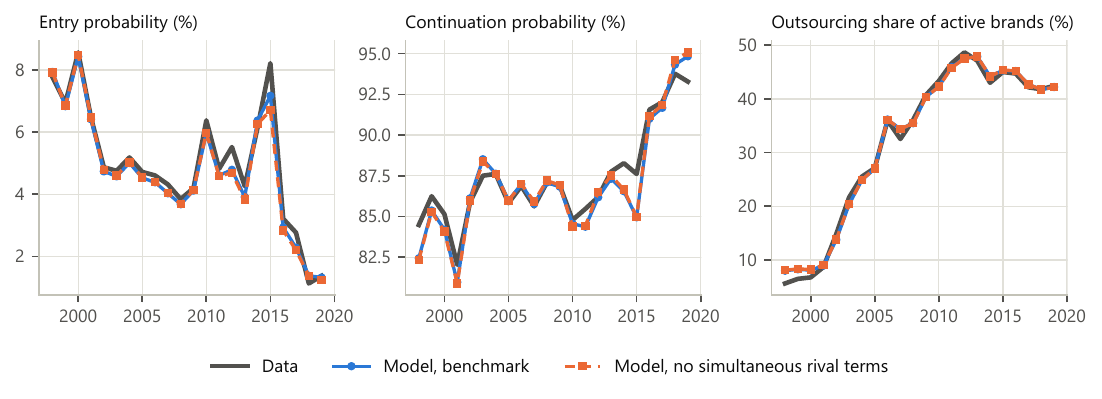}
\caption{Model Fit by Year}
\label{fig:fit}
\begin{minipage}{\textwidth}
\footnotesize\justifying
\noindent\textit{Notes.} The panels show the entry probability of inactive brands, continuation probability of active brands, and outsourcing share of active brands. Model predictions use the maximum likelihood estimates of the benchmark specification and the specification without simultaneous rival terms. The benchmark holds rivals at their first-stage choice probabilities.
\end{minipage}
\end{figure}

Figure~\ref{fig:fit} compares observed and fitted annual entry, continuation, and outsourcing. The benchmark model reproduces the rise in outsourcing through the middle of the sample and its subsequent decline. Entry-cost shifts reproduce falling entry after 2015 alongside more persistent incumbent service. In 2016 to 2019, fitted entry is 2.0 percent, compared with 2.1 percent in the data.

The model also reproduces the short tenure of many entrants. In the quarter after entry, 58.4 percent of entrants exit in the data and 54.3 percent exit in the model. Because entrants that exit quickly carry few passengers, passenger-weighted exit rates are lower, 11.7 percent in the data and 16.0 percent in the model. Inherited scale links these early exits to the small scale at which entrants start.

\begin{figure}[!tp]
\centering
\includegraphics[width=\textwidth]{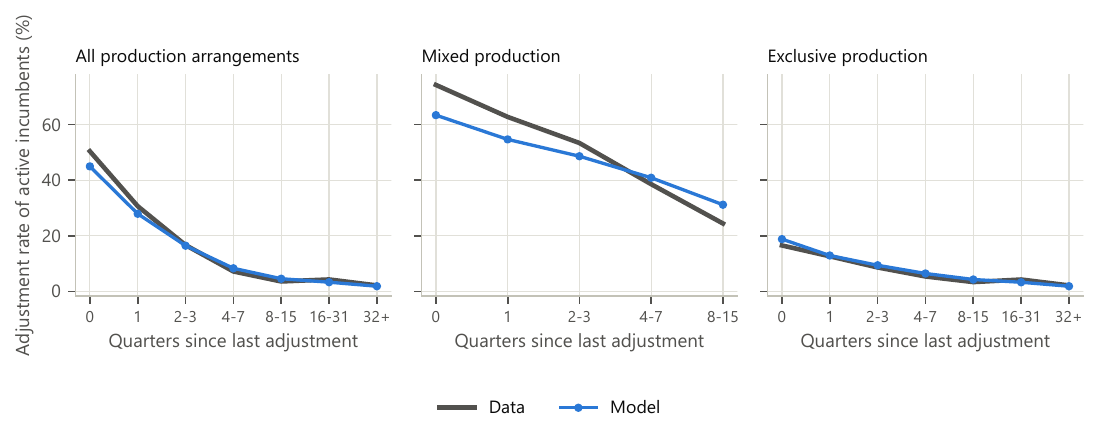}
\caption{Adjustments by Time since the Last Adjustment}
\label{fig:fit_duration}
\begin{minipage}{\textwidth}
\footnotesize\justifying
\noindent\textit{Notes.} Adjustment rates among active incumbents are grouped by the number of quarters since the last production adjustment. The panels show all inherited production arrangements, mixed production, and exclusive production with inherited shares of zero or one. Bins with fewer than 100 observations are omitted. Model rates use the maximum likelihood estimates of the benchmark specification.
\end{minipage}
\end{figure}

Figure~\ref{fig:fit_duration} compares observed and predicted adjustment rates by the duration of an arrangement. In the quarter following an adjustment, the model predicts adjustment rates of 63 percent under mixed production and 19 percent under fully internal or fully outsourced production, compared with 74 and 17 percent in the data. Both rates fall with duration.

Table~\ref{tab:fit_regimes} compares the distribution of production regimes and their quarterly transitions in the game sample. Predicted shares of complete internal production, mixed production, and complete outsourcing are 54.8, 25.2, and 20.0 percent, compared with 55.2, 24.0, and 20.8 percent in the data. Conditional on continued activity, the predicted persistence probabilities are 0.936, 0.827, and 0.871, compared with 0.945, 0.813, and 0.873. Among entrants, the model predicts complete outsourcing for 16.7 percent and mixed production for 2.3 percent, compared with 16.6 and 2.2 percent in the data. Direct switches between complete internal production and complete outsourcing are rare in both the model and the data. These patterns determine how carriers' existing production arrangements affect their responses to changes in the cost or availability of outside production.

\begin{table}[!tp]
\centering
\singlespacing
\small
\caption{Production Regimes of Active Brands}
\label{tab:fit_regimes}
\begin{tabularx}{\textwidth}{@{}Xrrrrrrr@{}}
\toprule
 & & \multicolumn{3}{c}{Data} & \multicolumn{3}{c}{Model} \\
\cmidrule(lr){3-5}\cmidrule(l){6-8}
Decisions & Count & Internal & Mixed & Outsourced & Internal & Mixed & Outsourced \\
\midrule
All active brands & 323,963 & 0.552 & 0.240 & 0.208 & 0.548 & 0.252 & 0.200 \\
\midrule
Previously internal only & 160,800 & 0.945 & 0.051 & 0.005 & 0.936 & 0.064 & 0.000 \\
Previously mixed & 75,789 & 0.088 & 0.813 & 0.099 & 0.097 & 0.827 & 0.076 \\
Previously outsourced only & 63,091 & 0.010 & 0.117 & 0.873 & 0.001 & 0.128 & 0.871 \\
\midrule
Entrants & 24,283 & 0.812 & 0.022 & 0.166 & 0.810 & 0.023 & 0.167 \\
\bottomrule
\end{tabularx}
\par\smallskip
\begin{minipage}{\textwidth}
\footnotesize
\justifying
\noindent\textit{Notes.} The columns report shares at zero outsourcing, a mixed allocation, and complete outsourcing in the game estimation sample. The first row covers all active decisions, including entrants. The next three rows cover active incumbents by the inherited production regime, giving quarterly transition distributions conditional on continued activity. The last row covers entrants. Model probabilities use the maximum likelihood estimates of the benchmark specification and first-stage rival probabilities, averaged over the same decisions as the data.
\end{minipage}

\end{table}

\section{Counterfactuals}
\label{sec:counterfactuals}
\label{sec:conditional_counterfactuals}

\subsection{Design of the Experiments}
\label{sec:policy_design}

When purchased regional service becomes more expensive, a carrier can continue buying that service, shift production to its own operations, or stop serving the route. The first experiment raises the recurring cost of purchasing service from independent regional providers relative to internal production. Republic's 2016 restructuring provides an example of such a change: amendments to its agreements with Delta and United increased reimbursement rates \citep{Republic2016QuarterlyReport}. We represent the increase by subtracting $\chi o$ from the payoff to serving a route in each quarter. A carrier can still outsource any share of production, including all of it, but pays the additional cost on whatever share it outsources. Internal production costs are held fixed.\footnote{The increase is common across carriers and periods and is measured in payoff units per unit of the outsourcing share. It changes the recurring return to operating a route with outside production, not the passenger marginal-cost equation. The experiment does not assign a dollar increase to a passenger, departure, or block hour.}

To distinguish higher outsourcing costs from the need to establish internal operations, we also impose a hypothetical minimum share of internal production. Each carrier must carry at least a share $\tau$ of its passengers on each route on flights operated by the carrier or a wholly owned regional subsidiary. Its outsourcing share must therefore satisfy $o\leq1-\tau$. We call $\tau$ the minimum internal share, or floor. Even a small positive value prevents a carrier from serving a route entirely through independent providers. The comparison leaves the payoff to each permitted production choice unchanged. It does not reproduce a particular supplier withdrawal or a scope clause: another independent provider may replace a withdrawing supplier, and scope clauses can apply to wholly owned regional airlines as well as independent providers.

We consider minimum internal shares of 1, 10, 25, and 50 percent, with 25 percent as the reference case. They reduce the mean outsourcing share among active carrier-route pairs by 17 to 25 percentage points.\footnote{For comparison, the decline among active pairs exposed to Chautauqua's withdrawal is 29.9 percentage points relative to the comparison routes in Appendix Table~\ref{tab:app_republic_did}. We do not use that decline to identify a cost shock.} We choose $\chi$ to approximately match the reductions under the 10 and 25 percent minimum shares. This gives two ways to obtain a similar average decline in outsourcing: make purchased service more expensive, or require some production to be internal. We then compare how many carriers continue to serve each route.

The cost increase and the floors apply from 1998Q2, the first decision quarter of the sample, and remain in place throughout the simulated horizon. All experiments, including the baseline, start each brand from the same observed production arrangement in its first game on a route. We condition on these initial arrangements and assume that the cost increase is unanticipated at the first simulated decision. Carriers then expect the higher cost to persist. The experiment does not reconstruct a historical cost increase in 1998. When a route's sample games resume after a gap, each brand continues from its last simulated state. A brand first observed on a route after 1998Q2 pays no fixed adjustment cost in its first game in any experiment, since its inherited arrangement was formed outside the simulated history.

In each quarter, we draw participation and production choices from the estimated choice distribution and update each brand's state. Inherited passenger volume follows the estimated transition. We use the same 200 draws for each carrier and route across experiments.\footnote{A route sequence is a run of consecutive retained games on a route. Brand availability, public states, and the twenty continuation quarters follow estimation. Appendix~\ref{app:policy_merger_design} defines the counterfactual choice sets and state transitions.} Under the cost increase, current and continuation payoffs incorporate the higher cost of outsourcing. Under a minimum internal share, a carrier whose existing outsourcing share exceeds $1-\tau$ must shift production internally or leave the route. The fixed cost of this initial reorganization depends on the existing production mix and how long it has been in place. Subsequent choices must also satisfy the minimum internal share. Any outsourcing share strictly between zero and one constitutes mixed production.

In the benchmark equilibrium, every brand re-optimizes. Each brand's expectations of rival activity and outsourced activity equal the averages generated by its rivals' choices over the simulation draws. Public states, including the lagged number of rivals, remain at their observed values; conditional on expected competition, the lagged count measures time-varying route attractiveness. We also report results with rivals' expectations held at their first-stage choice probabilities and from the specification without simultaneous rival terms. All experiments use the bias-corrected common parameters and the corresponding carrier-by-route effects.

We measure participation by the number of active carrier-route pairs, reported per 100 route-quarter games. A carrier's withdrawal counts as a loss even if other carriers continue serving that route. This measure records the availability of carrier services, rather than the number of flights or passengers. One brand per 100 games corresponds to approximately 30 active carrier-route pairs in an average quarter. Entry and continuation effects are percentage-point changes conditional on whether the carrier served the route in the preceding simulated quarter. The simulated baseline carries each brand's state forward from its first game, whereas the fit in Section~\ref{sec:fit} conditions on observed states in every quarter. The equilibrium baseline has 111.9 active brands per 100 games and a mean outsourcing share of 27.9 percent, compared with 124.3 and 31.8 percent in the data. With rivals held at their first-stage choice probabilities evaluated at observed states, baseline participation is 106.6 brands per 100 games. Parenthetical values below are standard errors in the same units as the estimates.

\subsection{Higher Outsourcing Costs and Route Participation}
\label{sec:network}

Higher outsourcing costs leave carriers free to decide how much regional service to purchase on each route. Nevertheless, the cost increase reduces participation more than the minimum internal share when the two generate approximately the same mean outsourcing share among active carrier-route pairs. Each active pair receives equal weight in this average, whose composition changes as carriers enter and exit. Cost increases of $\chi=0.35$ and $\chi=0.53$ reduce the mean share by 21.1 and 23.2 percentage points, approximately matching the reductions under the 10 and 25 percent minimum internal shares (Table~\ref{tab:network}). Participation falls by 4.65 and 5.21 brands per 100 games, compared with 3.30 and 3.99 under the corresponding minimum shares. The additional losses are 1.35 (0.17) and 1.22 (0.19). Relative to the 25 percent minimum internal share, higher outsourcing costs reduce the number of active carrier-route pairs by 31 percent (6.5) more.

The two experiments differ in what happens to the outsourcing that carriers continue to use. The minimum internal share rules out outsourcing shares above $1-\tau$ but leaves the return to every smaller share unchanged. The cost increase makes all positive outsourcing shares less profitable, in every quarter. Carriers therefore lose some of the value of purchasing regional service even when they reduce outsourcing enough to match the average response under the minimum share. The same cost increase also applies to outsourcing by future entrants.\footnote{Proposition~\ref{prop:menu_charge}(ii) separates these effects at fixed states, continuation values, and aggregation weights. For a given mean outsourcing share, the common cost increase changes the distribution of production choices least under the relative-entropy measure, but also lowers the payoff on the outsourcing that carriers retain.}

Most of the additional participation loss occurs among carriers that initially operated a route entirely internally or had not entered it. Relative to the minimum internal share, the cost increase reduces participation by 0.54 brands per 100 games (0.05) among initially internal producers and by 0.48 (0.16) among initially inactive brands. The corresponding losses for initially mixed and fully outsourced brands are 0.11 (0.01) and 0.08 (0.01). The first two groups account for 84 percent of the difference (Panel B). These carriers may purchase regional service in later quarters. Making that future option more costly reduces the value of entry and continued service today, even if the carrier currently operates all flights internally. Panel C shows a larger decline in active carrier-route pairs under the cost increase in every route-size group, with the largest difference in the middle third.

To assess how much of this difference comes from lower profitability, we compensate carriers for the average additional cost while retaining the incentive to reduce outsourcing. Every active carrier-route pair receives the same payment in each quarter, as in Proposition~\ref{prop:menu_charge}(iii). The payment equals $\chi$ times the resulting mean outsourcing share among active pairs and does not depend on the recipient's own share. Serving a route is therefore more profitable, but outsourcing remains more costly at the margin. This experiment requires outside funding; the comparison excludes its financing cost. At $\chi=0.53$, the payment is 0.025 payoff units per quarter. For a brand that remains active, the discounted value approaches $0.025/\left(1-0.98\right)=1.25$ units, about one third of the 1998 entry cost.

With the payment, the cost increase reduces participation by 2.47 brands per 100 games (0.69), or 1.52 (0.13) fewer than the 25 percent minimum internal share, while the mean outsourcing reduction remains 23.2 percentage points. The decline in active carrier-route pairs is now 38 percent (6.8) smaller than under the minimum share, compared with 31 percent (6.5) larger without the payment. In the comparison with the 10 percent minimum internal share, the payment yields a loss 1.31 brands per 100 games (0.15) smaller. Holding rivals at their first-stage choice probabilities, the losses with the payment are 2.59 (0.24) and 2.74 (0.19) brands per 100 games smaller than under the 10 and 25 percent minimum shares. The ranking reverses under both specifications of rival behavior. Compensating the average cost increase makes more routes worth serving while preserving carriers' ability to choose how much production to purchase.

\begin{table}[!tp]
\centering
\singlespacing
\footnotesize
\caption{Higher Outsourcing Costs and Route Participation}
\label{tab:network}
\renewcommand{\arraystretch}{0.92}
\begin{tabularx}{\textwidth}{@{}Xrrrr@{}}
\toprule
 & \multicolumn{2}{c}{Outsourcing decline of 21 pp} & \multicolumn{2}{c}{Outsourcing decline of 23 pp} \\
\cmidrule(lr){2-3}\cmidrule(l){4-5}
 & Floor & Cost increase & Floor & Cost increase \\
 & $\tau=0.10$ & $\chi=0.35$ & $\tau=0.25$ & $\chi=0.53$ \\
\midrule
\multicolumn{5}{@{}l}{\textit{Panel A. All route-quarter games}} \\
Outsourcing share of active brands (pp) & $-$21.02 & $-$21.05 & $-$23.19 & $-$23.22 \\
 & (1.71) & (1.61) & (1.83) & (1.80) \\
Active brands per 100 games & $-$3.30 & $-$4.65 & $-$3.99 & $-$5.21 \\
 & (0.64) & (0.68) & (0.72) & (0.78) \\
Difference from the floor & & $-$1.35 & & $-$1.22 \\
 & & (0.17) & & (0.19) \\
Active brands per 100 games, with payment & & $-$1.99 & & $-$2.47 \\
 & & (0.60) & & (0.69) \\
Difference from the floor, with payment & & 1.31 & & 1.52 \\
 & & (0.15) & & (0.13) \\
Entry probability (pp) & $-$0.44 & $-$0.05 & $-$0.42 & $-$0.12 \\
 & (0.08) & (0.05) & (0.08) & (0.06) \\
Continuation probability (pp) & 0.26 & $-$0.68 & 0.11 & $-$0.65 \\
 & (0.17) & (0.14) & (0.17) & (0.16) \\
\midrule
\multicolumn{5}{@{}l}{\textit{Panel B. Active brands per 100 games by initial production arrangement}} \\
Inactive & $-$2.30 & $-$2.90 & $-$2.78 & $-$3.26 \\
 & (0.48) & (0.53) & (0.54) & (0.60) \\
Internal production & $-$0.58 & $-$1.12 & $-$0.68 & $-$1.22 \\
 & (0.32) & (0.33) & (0.34) & (0.35) \\
Mixed production & $-$0.20 & $-$0.32 & $-$0.24 & $-$0.35 \\
 & (0.04) & (0.05) & (0.05) & (0.05) \\
Complete outsourcing & $-$0.23 & $-$0.32 & $-$0.29 & $-$0.37 \\
 & (0.05) & (0.05) & (0.05) & (0.06) \\
\midrule
\multicolumn{5}{@{}l}{\textit{Panel C. Active carrier-route pairs by route size (percent)}} \\
Thinnest third & $-$13.8 & $-$15.8 & $-$16.7 & $-$18.2 \\
 & (1.5) & (1.4) & (1.7) & (1.6) \\
Middle third & $-$6.6 & $-$8.8 & $-$7.9 & $-$9.7 \\
 & (1.0) & (1.1) & (1.2) & (1.2) \\
Densest third & $-$0.4 & $-$1.2 & $-$0.6 & $-$1.3 \\
 & (0.4) & (0.4) & (0.4) & (0.5) \\
\bottomrule
\end{tabularx}
\par\smallskip
\begin{minipage}{\textwidth}
\footnotesize\justifying
\noindent\textit{Notes.} Floor columns restrict the outsourcing share to at most $1-\tau$ on every route. Cost-increase columns leave all production choices available and subtract $\chi o$ from each active flow payoff. The value of $\chi$ approximately matches the unweighted mean outsourcing share among active carrier-route pairs under the corresponding floor. Payment rows add a common amount per active carrier-route pair and quarter, equal to $\chi$ times the resulting mean outsourcing share among active carrier-route pairs. The payment compensates the average additional cost; financing costs are not included. All brands re-optimize, and every column reports a change from the same baseline. Panel B groups brands by their production arrangement in 1998Q2; brands on routes entering the sample later count as inactive. Panel C reports percentage changes in active carrier-route pairs for thirds of routes by average legacy traffic. Parentheses report standard errors from 180 draws of the corrected game parameters, with experiments solved at each draw on the observed games and $\chi$ held fixed.
\end{minipage}
\end{table}

\subsection{Minimum Internal Production and Route Participation}
\label{sec:policy_results}

A minimum internal share reduces entry because a carrier can no longer begin serving a route entirely through independent providers. Even a 1 percent minimum share requires an entrant that would have outsourced completely to establish internal operations as well. Such an entrant must pay the additional setup cost $\kappa_M$ of mixed production or choose fully internal production. A fully outsourced incumbent must pay a fixed adjustment cost to keep serving the route. These fixed costs arise even when only a small fraction of production must be brought internally. Proposition~\ref{prop:menu_charge}(i) formalizes this discontinuity.

\begin{figure}[!tp]
\centering
\includegraphics[width=\textwidth]{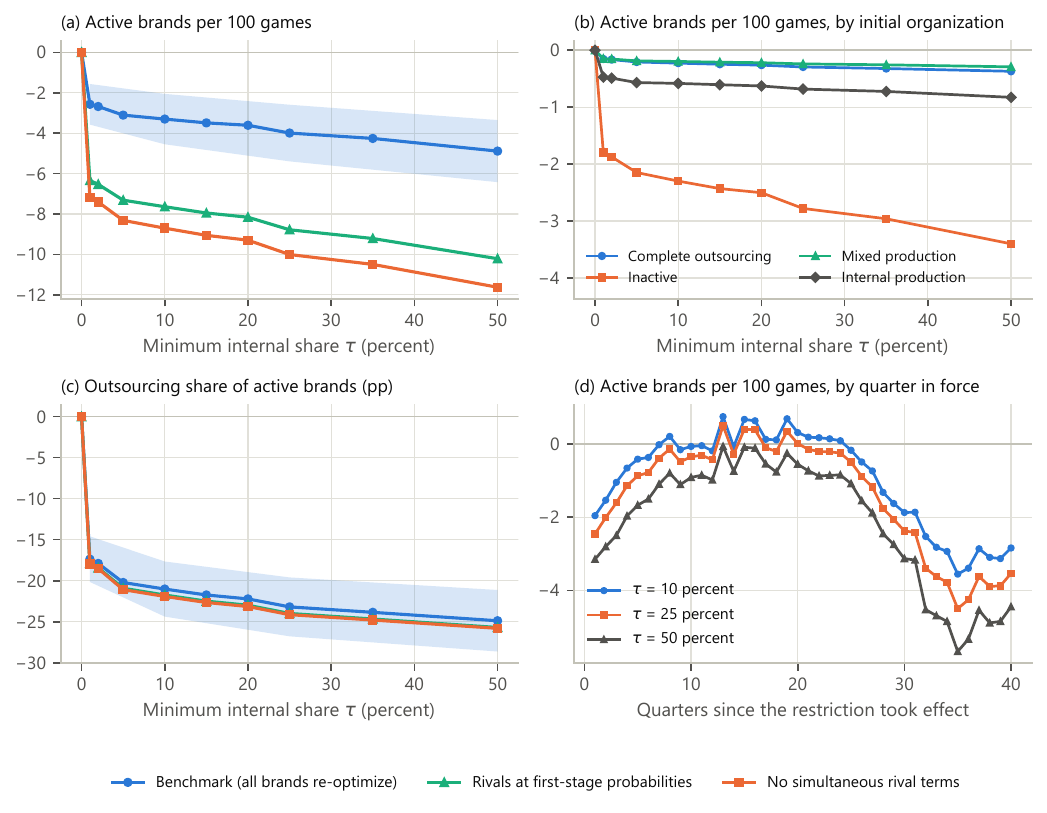}
\caption{Participation and Outsourcing under Minimum Internal Shares}
\label{fig:policy}
\begin{minipage}{\textwidth}
\footnotesize\justifying
\noindent\textit{Notes.} Panels (a) to (c) report changes from the simulated baseline. The horizontal axis is the minimum internal share $\tau$ in percent. Panels (a) and (c) compare the benchmark equilibrium, in which all brands re-optimize, with the experiment that holds rivals at their first-stage choice probabilities and with the specification without simultaneous rival terms. Shading shows the benchmark estimate plus or minus 1.96 standard errors, computed at floors of 1, 10, 25, and 50 percent and connected linearly. Panel (b) decomposes the benchmark activity effect by each brand's organization in 1998Q2, when the floor takes effect, with inactive brands shown separately. Brands on routes that enter the sample later count as inactive. The four components add up to the benchmark effect in panel (a). Panel (d) reports the benchmark effects of the 10, 25, and 50 percent floors by quarter since the floor took effect. Activity effects are brands per 100 route-quarter games, and share effects are percentage points.
\end{minipage}
\end{figure}

Table~\ref{tab:policy} and Figure~\ref{fig:policy} show the resulting participation response. A 25 percent minimum internal share lowers participation by 3.99 brands per 100 games (0.72), or 3.6 percent of active carrier-route pairs in the baseline and approximately 120 pairs in an average quarter. The decline ranges from 2.57 brands under the 1 percent minimum share to 4.89 under the 50 percent minimum share. The 1 percent experiment therefore produces 64 percent (2.7) of the loss under the 25 percent experiment. Raising the minimum internal share also removes mixed production choices with outsourcing above $1-\tau$, but those choices carry less probability than complete outsourcing. Participation consequently falls much less than proportionally with $\tau$.

\begin{table}[!tp]
\centering
\singlespacing
\footnotesize
\caption{Effects of Minimum Shares of Internal Production}
\label{tab:policy}
\renewcommand{\arraystretch}{0.97}
\begin{tabularx}{\textwidth}{@{}Xrrrrr@{}}
\toprule
 &  & \multicolumn{4}{c}{Minimum internal share $\tau$} \\
\cmidrule(l){3-6}
Outcome & Baseline & 1\% & 10\% & 25\% & 50\% \\
\midrule
\multicolumn{6}{@{}l}{\textit{Panel A. All brands re-optimize}} \\
Active brands per 100 games & 111.87 & $-$2.57 & $-$3.30 & $-$3.99 & $-$4.89 \\
 & & (0.51) & (0.64) & (0.72) & (0.78) \\
\quad Percent of baseline active pairs & & $-$2.3 & $-$3.0 & $-$3.6 & $-$4.4 \\
Entry probability (pp) & 4.91 & $-$0.48 & $-$0.44 & $-$0.42 & $-$0.41 \\
 & & (0.07) & (0.08) & (0.08) & (0.08) \\
Continuation probability (pp) & 90.52 & 0.45 & 0.26 & 0.11 & $-$0.04 \\
 & & (0.14) & (0.17) & (0.17) & (0.18) \\
Outsourcing share of active brands (pp) & 27.87 & $-$17.38 & $-$21.02 & $-$23.19 & $-$24.89 \\
 & & (1.42) & (1.71) & (1.83) & (1.91) \\
Adjustment probability of incumbents (pp) & 29.46 & $-$5.60 & $-$8.11 & $-$10.93 & $-$14.73 \\
 & & (0.53) & (0.71) & (0.88) & (1.08) \\
Active brands, first quarter in force & 131.61 & $-$1.66 & $-$1.96 & $-$2.46 & $-$3.15 \\
 & & (0.15) & (0.17) & (0.18) & (0.17) \\
Active brands, twenty-first quarter and later & 106.46 & $-$3.33 & $-$4.24 & $-$5.04 & $-$6.02 \\
 & & (0.60) & (0.75) & (0.84) & (0.92) \\
\midrule
\multicolumn{6}{@{}l}{\textit{Panel B. Rivals at first-stage choice probabilities}} \\
Active brands per 100 games & 106.59 & $-$6.34 & $-$7.64 & $-$8.78 & $-$10.21 \\
 & & (0.66) & (0.83) & (0.93) & (1.01) \\
\quad Percent of baseline active pairs & & $-$5.9 & $-$7.2 & $-$8.2 & $-$9.6 \\
Outsourcing share of active brands (pp) & 28.62 & $-$17.89 & $-$21.75 & $-$24.00 & $-$25.71 \\
 & & (1.30) & (1.59) & (1.69) & (1.74) \\
\midrule
\multicolumn{6}{@{}l}{\textit{Panel C. Fares, consumer valuation, and variable profit}} \\
Passenger-weighted fare (percent) & & $-$0.393 & $-$0.310 & $-$0.241 & $-$0.174 \\
 & & (0.125) & (0.141) & (0.156) & (0.173) \\
Traffic, active brands fixed (percent) & & 1.486 & 1.357 & 1.235 & 1.118 \\
 & & (0.389) & (0.436) & (0.474) & (0.517) \\
Passenger marginal cost (percent) & & $-$0.091 & 0.004 & 0.086 & 0.169 \\
 & & (0.071) & (0.084) & (0.096) & (0.109) \\
Consumer valuation, active brands fixed (\$ million per year) & & 74.7 & 67.2 & 60.0 & 53.1 \\
 & & (22.4) & (24.7) & (26.5) & (28.5) \\
Consumer valuation, participation (\$ million per year) & & $-$7.8 & $-$31.0 & $-$56.5 & $-$86.6 \\
 & & (4.1) & (11.7) & (19.5) & (28.2) \\
Consumer valuation (\$ million per year) & & 66.8 & 36.2 & 3.5 & $-$33.5 \\
 & & (20.9) & (22.8) & (26.7) & (32.9) \\
Variable profit, active brands fixed (\$ million per year) & & 57.2 & 54.7 & 51.9 & 49.4 \\
 & & (17.1) & (19.5) & (21.1) & (22.8) \\
Variable profit, participation (\$ million per year) & & $-$13.6 & $-$29.0 & $-$44.9 & $-$63.1 \\
 & & (3.5) & (7.1) & (10.7) & (14.4) \\
Variable profit (\$ million per year) & & 43.6 & 25.7 & 7.0 & $-$13.7 \\
 & & (16.5) & (18.4) & (20.1) & (22.1) \\
\bottomrule
\end{tabularx}
\par\smallskip
\begin{minipage}{\textwidth}
\footnotesize
\justifying
\noindent\textit{Notes.} A minimum internal share $\tau$ requires each carrier to operate at least that share internally on every route, so its outsourcing share cannot exceed $1-\tau$. Section~\ref{sec:policy_design} describes the simulations. Columns report simulated baseline levels and changes. In Panel A, each brand's expectations equal its rivals' simulated activity and outsourced activity. In Panel B, rivals' expected activity and outsourced activity remain at their first-stage choice probabilities. The conditional rows of Panel C hold observed active brands fixed, update outsourcing in demand and cost, and retain the pricing residuals. The participation rows weight each brand's contribution by its change in activity probability in Panel A. Profit contributions account for changes in rivals' profits following withdrawal. The totals add the participation and conditional components. Parentheses report standard errors from 180 draws of the corrected game parameters, with every experiment re-solved at each draw on the observed games. Panel C standard errors combine the delta method in the demand and cost parameters with the variation of the game across draws, which scales the conditional rows by each draw's outsourcing effect and the participation rows by its activity effect.
\end{minipage}
\end{table}

Most of the production response is a shift to internal operations. Under the 25 percent floor, the mean outsourcing share among active carrier-route pairs falls by 23.2 percentage points, from 27.9 to 4.7 percent. The estimated recurring payoff helps explain this response. For Delta in 1998 to 2004, its outsourcing component, $-2.689o+2.339o^2$, equals $-0.35$ at complete outsourcing and $-0.70$ at 0.75, the largest permitted share. This component is convex, with a minimum of $-0.77$ near a share of 0.57. The largest permitted share is therefore close to the least attractive allocation in recurring payoff terms, making internal production the principal alternative to complete outsourcing. Inherited shares and adjustment costs govern how carriers move between the two.

The decline in participation grows as complete outsourcing becomes more valuable in the baseline. In the first quarter, the 25 percent minimum internal share reduces participation by 2.5 brands per 100 games (0.2). The loss disappears in the thirteenth through twentieth quarters, after the initial reorganization and before the later increase in baseline outsourcing (Figure~\ref{fig:policy}, panel (d)). From the twenty-first quarter onward, the average loss is 5.04 brands per 100 games (0.84), or 4.7 percent of baseline active carrier-route pairs in those quarters. The outsourcing reduction is 9 percentage points in the first twenty quarters and 21 and 30 points in quarters 21 to 40 and 41 to 60. Thus, the later participation losses arise when more carriers would otherwise have used the production choices that the minimum share rules out.

Panel (b) of Figure~\ref{fig:policy} groups the loss by each brand's initial production arrangement. Brands inactive in 1998Q2, including every brand on a route that enters the sample later, account for 70 percent of the loss; brands initially using complete outsourcing account for 7 percent. Relative to each group's observed activity, however, the loss is largest for initially fully outsourced brands, at 12 percent. It is 5 percent for initially mixed or inactive brands and 1 percent for initially internal producers. The pooled entry probability falls by 0.42 percentage points (0.08). With complete outsourcing unavailable, an entrant can purchase outside production only as part of a mixed arrangement and must incur $\kappa_M$. Incumbent withdrawals make entry more attractive by reducing competition, but this response does not offset the loss of a production option.

The minimum internal share has its largest effect on routes with low passenger volume. We divide routes into thirds by average quarterly legacy traffic. In the baseline, the mean outsourcing share among active carrier-route pairs is 44 percent in the thinnest third and 22 percent in the densest third. The 25 percent minimum internal share reduces active pairs by 16.7 percent (1.7) in the thinnest third and 7.9 percent (1.2) in the middle third; the change in the densest third is $-$0.6 percent (0.4) (Table~\ref{tab:network}, Panel C). Thin routes combine greater reliance on independent providers with smaller inherited passenger volumes and a lower return to continued service. The ability to purchase all production from an independent provider is particularly valuable on these routes.

Rivals' responses offset part of the loss. Holding rivals at their first-stage choice probabilities, the 10 and 25 percent floors reduce activity by 7.64 (0.83) and 8.78 (0.93) brands per 100 games. The equilibrium losses are 3.30 (0.64) and 3.99 (0.72); without simultaneous rival terms, they are 8.70 and 10.01. Participation decisions are strategic substitutes \citep{BulowGeanakoplosKlemperer1985}: with $\rho_{AA}<0$, lower expected rival activity raises the return to service for the remaining brands. Under the 25 percent floor, this response reduces the participation loss by 55 percent (5.5) relative to the case with rivals' choices held fixed. Section~\ref{sec:policy_robustness} compares the experiments under both specifications of rival behavior.\footnote{Appendix~\ref{app:robustness} reports results across game specifications, samples, and production measures.}

\subsection{Fares and Consumer Valuation}
\label{sec:policy_welfare}

Under the minimum internal shares, fares change little when the observed set of products is held fixed. Panel C of Table~\ref{tab:policy} applies each brand's simulated outsourcing change to the demand and passenger-cost equations and solves the pricing conditions for that set. Under the 25 percent floor, the passenger-weighted fare changes by $-$0.24 percent (0.16) and passenger marginal cost by 0.09 percent (0.10). Traffic rises by 1.2 percent (0.5), reflecting both lower fares and the estimated demand response to internal production.

We compare consumer outcomes using a valuation index based on the uniform fare change that reproduces each counterfactual's route traffic. Let $\mathcal{P}^0$ denote the observed passenger-weighted fare and $Q^0$ observed route traffic. For counterfactual traffic $Q$, define the equivalent fare index $\mathcal{P}=\mathcal{P}^0\left(Q/Q^0\right)^{-1/\alpha}$. Applying this proportional fare change to the baseline set of products and production arrangements produces the same total traffic. The estimated price coefficient satisfies $\alpha>1$, so integrating the resulting scalar route demand curve gives the finite valuation index $\mathcal{V}=Q\mathcal{P}/\left(\alpha-1\right)$. Every comparison uses the same baseline normalization, which scales the index by observed route revenue.\footnote{Appendix~\ref{app:demand} derives the equivalent fare index and valuation integral. Appendix~\ref{app:counterfactual_valuation} describes aggregation across brands and routes. Dollar values use passenger counts expanded by the inverse sampling rate.} With the observed active set held fixed, this valuation rises by \$60 million per year (26) under the 25 percent floor. Passenger variable profit rises by \$52 million (21).

Reduced participation offsets most of the valuation gain under the 25 percent floor. We measure a brand's contribution by the decrease in route valuation when that brand withdraws and the remaining fares adjust. Weighting these contributions by the simulated changes in activity probabilities gives a participation component of $-$\$56.5 million per year (19.5) under the 25 percent floor. The combined valuation change is \$36 million (23) under the 10 percent floor and \$4 million (27) under the 25 percent floor.\footnote{The participation component is a first-order approximation in the activity probabilities. A brand inactive in a given quarter receives its mean contribution in active quarters on the same route.} As the floor increases, the participation component moves from a loss of \$8 million at 1 percent to a loss of \$87 million at 50 percent, while the valuation gain with the observed set of products held fixed remains between \$53 million and \$75 million. The combined change falls from \$67 million (21) under the 1 percent floor to $-$\$34 million (33) under the 50 percent floor.

Carriers withdraw service although total variable profit rises by \$7 million per year (20) under the 25 percent floor, the sum of a \$52 million gain with the observed set of products held fixed and a participation component of $-$\$45 million (11). The return to serving a route also includes the costs of maintaining a flight schedule and reorganizing production, which passenger variable profit excludes. Capacity purchase agreements compensate providers for aircraft, departures, and block hours, so many sourcing costs arise at the schedule level. A floor can therefore reduce the incentive to serve a route even when the change in production raises passenger variable profit with the observed products held fixed.

Table~\ref{tab:valuation_comparison} compares the consumer-valuation effects of higher outsourcing costs with those of the minimum internal shares. Without the payment, the cost increase at $\chi=0.53$ lowers the participation component by \$172.5 million per year (52.5), compared with \$56.5 million under the 25 percent minimum internal share. This loss is three times larger, although the decline in active carrier-route pairs is only 31 percent larger. Higher outsourcing costs lower continuation by 0.65 percentage points and approximately double the participation loss on the densest third of routes, where each brand carries more passengers. The minimum internal share acts mainly through reduced entry on thin routes. Which carrier services disappear therefore matters in addition to how many disappear.

Holding the observed set of products fixed, the consumer-valuation gain is also smaller under the cost increase: \$47 million, compared with \$60 million under the minimum internal share. Passenger-weighted outsourcing falls by 12.2 and 12.6 percentage points, respectively, and traffic rises by 1.0 and 1.2 percent. Passenger marginal cost rises by 0.15 percent under the cost increase and by 0.09 percent under the minimum share, while fares fall by 0.12 and 0.24 percent. These marginal-cost changes follow from the new production mix; $\chi$ does not enter passenger marginal cost directly. Under the minimum internal share, the maximum permitted outsourcing share remains above the share that minimizes passenger marginal cost for most products. The cost increase, in contrast, reduces the recurring return to every positive outsourcing share.

The combined annual valuation change is $-$\$126 million (50) under the cost increase and \$4 million (27) under the floor. The difference is $-$\$129 million per year (35); in the comparison with the 10 percent floor it is $-$\$118 million (33). These are consumer-valuation effects. They do not include the additional operating cost or the cost of financing the payment in the compensation experiment.

\begin{table}[!tp]
\centering
\singlespacing
\footnotesize
\caption{Higher Outsourcing Costs and Consumer Valuation}
\label{tab:valuation_comparison}
\renewcommand{\arraystretch}{0.97}
\begin{tabularx}{\textwidth}{@{}Xrrrr@{}}
\toprule
 & \multicolumn{2}{c}{Outsourcing decline of 21 pp} & \multicolumn{2}{c}{Outsourcing decline of 23 pp} \\
\cmidrule(lr){2-3}\cmidrule(l){4-5}
 & Floor & Cost increase & Floor & Cost increase \\
 & $\tau=0.10$ & $\chi=0.35$ & $\tau=0.25$ & $\chi=0.53$ \\
\midrule
Participation & $-$31.0 & $-$140.8 & $-$56.5 & $-$172.5 \\
 & (11.7) & (43.1) & (19.5) & (52.5) \\
Observed products fixed & 67.2 & 58.9 & 60.0 & 46.6 \\
 & (24.7) & (23.4) & (26.5) & (25.3) \\
Total & 36.2 & $-$81.9 & 3.5 & $-$125.9 \\
 & (22.8) & (40.3) & (26.7) & (50.3) \\
Total, difference from the floor & & $-$118.2 & & $-$129.5 \\
 & & (32.6) & & (34.7) \\
\bottomrule
\end{tabularx}
\par\smallskip
\begin{minipage}{\textwidth}
\footnotesize\justifying
\noindent\textit{Notes.} Entries are changes in consumer valuation in millions of dollars per year. Floor columns restrict the outsourcing share to at most $1-\tau$ on every route. Cost-increase columns leave all production choices available and subtract $\chi o$ from each active flow payoff. The value of $\chi$ approximately matches the unweighted mean outsourcing share among active carrier-route pairs under the corresponding floor. All brands re-optimize, and every column reports a change from the same baseline, without the common payment. The row for observed products holds the observed active set fixed and updates production shares in demand and passenger cost. The participation component weights each brand's contribution by its simulated change in activity probability. Baseline passenger-cost shifters are held fixed. Parentheses report standard errors from 180 draws of the corrected game parameters, with experiments solved at each draw on the observed games and $\chi$ held fixed. Standard errors also include the delta method in the demand and cost parameters, applied jointly to the two experiments in the difference row.
\end{minipage}
\end{table}

\subsection{Reorganization Costs and the Transition to Internal Production}
\label{sec:reorganization}

The estimates imply that changing production is less costly when a carrier already uses both internal operations and independent providers. The cost also rises with the time that the existing arrangement has remained unchanged. We examine the role of each feature by setting $\theta_I$ or $\theta_D$ to zero, holding the other parameters and initial states fixed. For each change, we compare equilibria with and without the 25 percent minimum internal share. Removing the lower fixed cost associated with mixed production raises the first-quarter participation loss by 0.30 brands per 100 games. It also changes the baseline: incumbents moving from internal production to complete outsourcing usually pass through mixed production, which has now become more costly to adjust. The baseline mean outsourcing share falls from 27.9 to 9.7 percent. The minimum internal share is then evaluated against a baseline with much less outsourcing, and its participation loss from the twenty-first quarter onward disappears. Removing duration dependence has smaller effects, reducing the loss from the twenty-first quarter onward by 0.02 brands per 100 games and the first-quarter loss by 0.16.

\begin{figure}[!tp]
\centering
\includegraphics[width=\textwidth]{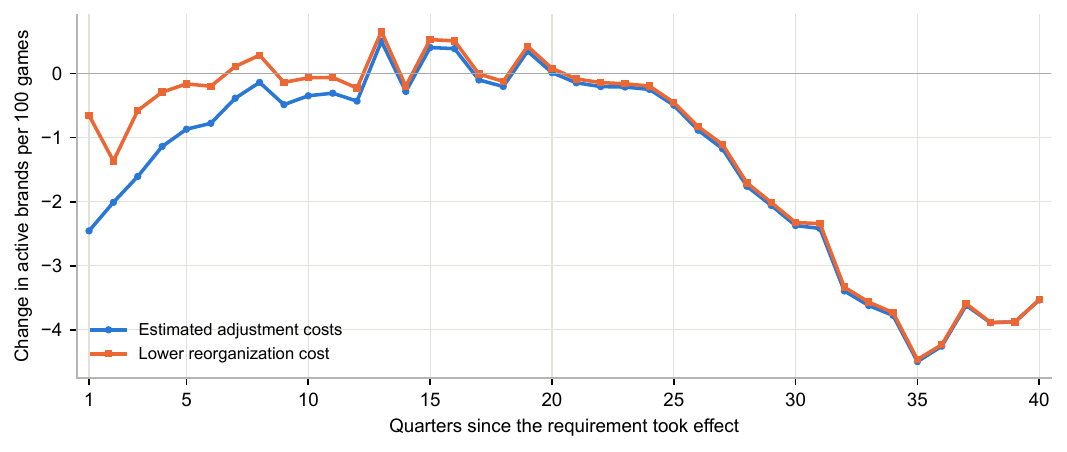}
\caption{Easier Reorganization and Route Participation}
\label{fig:reorganization_service}
\begin{minipage}{\textwidth}
\footnotesize\justifying
\noindent\textit{Notes.} Both curves report the effect of the 25 percent internal-share floor on active brands per 100 route-quarter games relative to the equilibrium without a floor. The lower-cost experiment grants an inherited fully outsourced arrangement the fixed-cost discount estimated for mixed production when the floor takes effect. Both curves use the estimated parameters and 200 common simulation draws, with all brands re-optimizing. The horizontal axis counts quarters from 1998Q2, when the floor takes effect on every route.
\end{minipage}
\end{figure}

A separate experiment makes it easier for fully outsourced incumbents to establish internal operations when the minimum share takes effect. We lower the fixed adjustment cost of their existing arrangements by the discount estimated for mixed production. Under the 25 percent minimum internal share, the first-quarter participation loss falls from 2.5 to 0.7 brands per 100 games, a reduction of nearly three quarters (Figure~\ref{fig:reorganization_service}). The gain is concentrated near the start: the first four quarters account for 44 percent of the total gain in active brand-quarters, while the average loss over all quarters falls only from 3.99 to 3.89 brands per 100 games. Lower adjustment costs help existing carriers remain on a route while bringing production internally. Future entrants still have to establish internal operations to serve the route. Easing the transition for incumbents therefore does little to offset the longer-run decline in participation.

\subsection{A Temporary Loss of Independent Regional Service}
\label{sec:disruption}

We next examine whether carriers that already operate part of a route internally are better able to keep serving it when independent regional service is unavailable. In the withdrawals described in Section~\ref{sec:republic}, carriers could often purchase replacement service from another independent provider. In this experiment, all independent production becomes unavailable for four quarters beginning in 2014Q2, the quarter of Chautauqua's withdrawal. The interruption is deliberately broader than the historical event. Carriers do not anticipate it and learn its duration when it begins. We compare the response with and without the minimum internal shares that have applied since 1998Q2. These histories differ both in production arrangements and in which carriers serve each route before the interruption. Every brand re-optimizes during and after the interruption.\footnote{Appendix~\ref{app:policy_merger_design} describes the implementation.}

When carriers have been free to outsource completely, the disruption reduces participation by 20.8 brands per 100 games (2.1) during the four disrupted quarters, equivalent to 19.3 percent of active carrier-route pairs (Table~\ref{tab:disruption}). Some carriers remain off the routes after independent production returns. Summing the differences in active carrier-route pairs between the simulations with and without the disruption gives a cumulative loss through 2019 of 161.6 brand-quarters per 100 games (13.7).

Carriers' ability to remain on a route depends strongly on whether they already operate some of its service internally. Brands using complete outsourcing when the disruption begins lose 3.29 quarters of activity each (0.22), compared with 1.34 (0.14) for mixed producers and 0.09 (0.04) for internal producers. A fully outsourced brand must establish internal operations and pay the cost of changing its existing production arrangement to remain on the route. Under the 25 percent minimum internal share, the mean outsourcing share among active carrier-route pairs is only 4.8 percent when the disruption begins. Relative to this experiment's own history without a disruption, the cumulative participation loss is 72 percent smaller (2.6) than when carriers have been free to outsource completely.

\begin{table}[!tp]
\centering
\small
\singlespacing
\caption{A Loss of Independent Regional Service with and without Minimum Internal Shares}
\label{tab:disruption}
\begin{tabularx}{\textwidth}{@{}X*{3}{>{\raggedleft\arraybackslash}p{1.5cm}}@{}}
\toprule
 & \multicolumn{3}{c}{Minimum internal share $\tau$} \\
\cmidrule(l){2-4}
 & 0\% & 10\% & 25\% \\
\midrule
\multicolumn{4}{@{}l}{\textit{Panel A. Four-quarter disruption beginning in 2014Q2}} \\
Outsourcing share of active brands before the disruption (percent) & 38.9 & 7.7 & 4.8 \\
 & (3.2) & (0.8) & (0.5) \\
Change in active brands per 100 games during the disruption & $-$20.84 & $-$5.61 & $-$4.94 \\
 & (2.10) & (0.58) & (0.51) \\
Cumulative loss through 2019 (brand-quarters per 100 games) & 161.6 & 52.3 & 45.2 \\
 & (13.7) & (5.6) & (4.9) \\
\midrule
\multicolumn{4}{@{}l}{\textit{Panel B. Average quarterly loss of active carrier-route pairs without a disruption}} \\
All quarters in force & & 3.30 & 3.99 \\
 & & (0.64) & (0.72) \\
Average from 2010Q2 to 2014Q1 & & 4.91 & 5.69 \\
 & & (0.97) & (1.07) \\
Twenty-first quarter in force and later & & 4.24 & 5.04 \\
 & & (0.75) & (0.84) \\
\midrule
\multicolumn{4}{@{}l}{\textit{Panel C. Years of lost carrier services in ordinary periods equal to the loss avoided in one disruption}} \\
Four quarters from 2014Q2, average loss in force & & 8.3 & 7.3 \\
 & & (1.4) & (1.1) \\
Four quarters from 2014Q2, loss from 2010Q2 to 2014Q1 & & 5.6 & 5.1 \\
 & & (0.9) & (0.8) \\
Two quarters from 2014Q2, average loss in force & & 6.7 & 5.9 \\
Eight quarters from 2014Q2, average loss in force & & 10.9 & 9.7 \\
Four quarters from 2009Q2, average loss in force & & 6.4 & 5.6 \\
\bottomrule
\end{tabularx}
\par\smallskip
\begin{minipage}{\textwidth}
\footnotesize\justifying
\noindent\textit{Notes.} A disruption makes independent production unavailable to every brand in the observed route-quarter decisions within the stated window. Brands do not anticipate the disruption and learn the end date when the disruption begins. Each floor has been in force on every route since 1998Q2, and all brands re-optimize before, during, and after the disruption. Panel A compares simulations with and without the disruption under the same floor and the same draws. The cumulative loss sums activity differences from the first disrupted quarter through 2019Q4 and multiplies by 100 divided by the number of games in the first disrupted quarter. Panel B measures the mean quarterly loss of active brands per 100 games relative to no floor, in the absence of a disruption. Its last row averages over games in which the floor has been in force for more than twenty quarters, as in Table~\ref{tab:policy}. If $L_\tau$ is the cumulative disruption loss under floor $\tau$ and $c_\tau$ is its mean quarterly participation loss in the stated comparison sample, Panel C reports $\left(L_0-L_\tau\right)/\left(4c_\tau\right)$. The comparison sums participation losses without discounting. Parentheses report standard errors from 180 draws of the corrected game parameters, which re-solve each experiment at each draw on the observed games.
\end{minipage}
\end{table}

Keeping more production internal reduces the loss during the interruption, but fewer carriers serve routes before it occurs. Over all quarters without a disruption, the 25 percent minimum internal share reduces participation by an average of 3.99 brands per 100 games. The avoided disruption loss, 116.4 brand-quarters per 100 games, equals 7.3 years (1.1) of the carrier services forgone at that quarterly rate. The corresponding comparisons are 5.9 years for a two-quarter disruption and 9.7 years for an eight-quarter disruption.\footnote{Using the average loss from 2010Q2 through 2014Q1 gives 5.1 years (0.8). A four-quarter disruption beginning in 2009Q2 gives 5.6 years.} This calculation compares the numbers of carrier-route pairs over time without discounting. It does not assign a probability to an interruption affecting all independent providers or establish an optimal internal share.

\subsection{Robustness of the Counterfactual Results}
\label{sec:policy_robustness}

The main comparisons also hold when rivals remain at their first-stage choice probabilities. Under this specification, the 25 percent floor reduces active carrier-route pairs by 8.2 percent, with the largest losses again on thin routes. A floor of 1 percent produces 72 percent of the loss under the 25 percent floor, consistent with the discontinuity in Proposition~\ref{prop:menu_charge}(i). At a comparable mean outsourcing share, the cost increase causes a larger participation loss than the 25 percent floor; with the common payment, the loss is 31 percent smaller. The floor reduces cumulative disruption losses by 78 percent. Across game specifications with carrier-by-route effects and simultaneous rival terms, it reduces equilibrium activity by 2.01 to 4.87 brands per 100 games.\footnote{Appendix Table~\ref{tab:policy_robust} reports both specifications of rival behavior. Appendix Table~\ref{tab:robustness_game} reports estimates under alternative specifications, and Appendix Table~\ref{tab:robustness_policy} compares the counterfactual results.}

\section{Conclusion}
\label{sec:conclusion}

When independent regional providers withdraw or become more expensive, the consequences for air service depend on what carriers can operate themselves and the cost of changing how their routes are served. Our estimates show that carriers already combining internal operations and independent providers face lower fixed costs when changing their outsourcing shares. Carriers that enter with both must pay additional setup costs. This helps explain why mixed production is frequently adjusted by incumbents while most entrants using independent providers outsource all service on the route.

In the counterfactual with persistently higher costs of purchased regional service, fewer carriers enter or continue serving routes, including carriers initially operating a route entirely internally. These carriers may purchase regional service on the same route in later quarters. Higher expected costs of doing so reduce the value of entering or remaining on a route today. A carrier's current outsourcing share therefore does not fully measure its exposure to higher supplier costs on that route. Evaluations of support for regional service should consider future sourcing opportunities as well as the flights currently purchased from independent regional carriers. In a separate experiment, each carrier must carry at least 25 percent of its passengers on a route itself or through a wholly owned regional subsidiary. The number of active carrier-route pairs falls by 3.6 percent, with the largest percentage decline on routes with low traffic. A carrier that would have entered using only independent providers must choose a different production arrangement or stay out. A minimum internal share of only 1 percent produces 64 percent of the loss under the 25 percent minimum. Making it cheaper for existing carriers to reorganize substantially reduces the initial loss, but does little to offset the average participation loss over the full experiment. At the same mean outsourcing share among active carrier-route pairs, higher outsourcing costs reduce participation more than this minimum because they lower the profits from purchased service that remains in use. Adding the same payment to each active carrier-route pair offsets the average additional cost and reverses that ranking. The calculation excludes the cost of financing the payment.

We also simulate a four-quarter interruption of all independent regional production. With the 25 percent minimum in place before the interruption, the cumulative loss of active carrier-route pairs through 2019 is 72 percent smaller than when carriers are free to outsource completely. Each loss is measured relative to the corresponding history without an interruption. Carriers enter the interruption with different production shares, and a different set of carriers serves the routes. The avoided loss equals about seven years of the carrier-route service forgone under the minimum when independent providers are available, without discounting. The minimum internal share reduces service losses during a broad supplier interruption but leaves fewer carriers serving routes when independent providers are available. An assessment based only on how existing airlines cope with interruptions would miss the loss of entry in ordinary periods.

\clearpage
{\interlinepenalty=10000
\small
\singlespacing

}

\clearpage
\appendix
\counterwithin{table}{section}
\counterwithin{figure}{section}
\numberwithin{equation}{section}
\section{Proofs and Model Implications}
\label{app:proofs}

\subsection{Demand Derivatives and the Price Policy}
\label{app:demand}

Suppress route and quarter subscripts, and define $a=\alpha/\left(1-\sigma\right)$ and $u_j=\delta_j/\left(1-\sigma\right)$. The inclusive value of the active brands is $V=\log\sum_k\exp\left(u_k\right)$. The demand equation in Section~\ref{sec:demand_model} implies $s_j=\exp\left(u_j\right)/\sum_k\exp\left(u_k\right)$ and $Q=N\exp\left\{\left(1-\sigma\right)V\right\}$. Because $\partial u_k/\partial p_i=-a\ind\left\{k=i\right\}/p_i$, differentiation gives
\[
\frac{\partial s_j}{\partial p_i}=-\frac{a}{p_i}s_j\left(\ind\left\{i=j\right\}-s_i\right),\quad\frac{\partial V}{\partial p_i}=-\frac{as_i}{p_i}.
\]
The route traffic response is therefore $\partial Q/\partial p_i=-\alpha Qs_i/p_i$. Differentiating $q_j=Qs_j$ gives equation~\eqref{eq:coherent_demand_jacobian}. With one active brand, the substitution term vanishes and the demand slope is $-\alpha q/p$.

The pricing system is $q+\Delta\left(p-mc\right)=r^p$. Multiplication by $\Delta^{-1}$ gives $p=mc-\Delta^{-1}q+\Delta^{-1}r^p$. Thus $\tilde r^p=\Delta^{-1}r^p$ is the residual expressed in fare units. Holding this residual fixed defines the local pricing map $F\left(p,o,\omega\right)=mc\left(o,\omega\right)-\Delta\left(p,o,\omega\right)^{-1}q\left(p,o,\omega\right)+\tilde r^p$.

Equivalently, define the effective passenger cost $\widetilde{mc}=mc+\tilde r^p$. The maintained pricing condition is $q+\Delta\left(p-\widetilde{mc}\right)=0$. Holding the dollar residual fixed therefore holds the product's pricing wedge fixed.

Counterfactual variable profit for brand $i$ is $q_i\left(p_i-\widetilde{mc}_i\right)$. For an owner $f$, let $\mathcal A_f$ contain its active brands. The maintained pricing conditions are the first-order conditions for the sum of these brand profits. For a twice differentiable local demand map with prices of other owners fixed and effective costs independent of price, the owner's profit Hessian has entries $D_{ik}+D_{ki}+\sum_{j\in\mathcal A_f}\left(p_j-\widetilde{mc}_j\right)\partial^2q_j/\partial p_i\partial p_k$ for $i,k\in\mathcal A_f$. Negative definiteness gives a strict local optimum. The price response derived below applies at a regular solution of these equations.

Define the route price index $\mathcal{P}$ by $\log \mathcal{P}=\log \mathcal{P}^0-\left(V-V^0\right)/a$, where $\mathcal{P}^0$ is the observed passenger-weighted fare, $Q^0$ is observed route traffic, and $V^0$ is the inclusive value at the observed allocation. The index expresses a change in the inclusive value as an equivalent proportional fare change at the observed dollar scale. Differentiation gives $\partial\log \mathcal{P}/\partial p_i=s_i/p_i$, and route traffic satisfies $Q=Q^0\left(\mathcal{P}/\mathcal{P}^0\right)^{-\alpha}$. For $\alpha>1$, define the consumer valuation index at the fixed baseline normalization by
\[
\mathcal{V}=\int_\mathcal{P}^\infty Q^0\left(\frac{x}{\mathcal{P}^0}\right)^{-\alpha}dx=\frac{Q\mathcal{P}}{\alpha-1}=\frac{Q^0\mathcal{P}^0}{\alpha-1}\left(\frac{Q}{Q^0}\right)^{\left(\alpha-1\right)/\alpha}.
\]
The index is finite, increases with route traffic, and satisfies $\partial\mathcal{V}/\partial \mathcal{P}=-Q$. The final expression shows how baseline revenue scales the valuation of a proportional change in traffic. Each counterfactual comparison uses the same $Q^0$, $\mathcal{P}^0$, and $V^0$. Removing a brand deletes its term from the inclusive value. Removing every brand sends the route index to infinity and traffic and valuation to zero. Outsourcing enters $u_j$ with coefficient $\gamma_o/\left(1-\sigma\right)$, so a change in outsourcing changes both brand shares and route traffic. The valuation integrates the scalar route demand relation at this fixed normalization.\footnote{For individual brand fares, the derivative is $\partial\mathcal{V}/\partial p_i=-q_i\mathcal{P}/p_i$. The reported object is therefore a valuation index defined from aggregate route demand.}

The cost response depends on the starting outsourcing share and the cost-minimizing allocation. Lowering the outsourcing share from $o$ to $o'<o$ in equation~\eqref{eq:marginal_cost} gives
\[
mc\left(o'\right)-mc\left(o\right)=-\left(c_2/2\right)\left(o-o'\right)\left(o+o'-2o^*\right).
\]
Costs fall when $o+o'>2o^*$ and rise when the inequality is reversed. A move from complete outsourcing to internal production lowers costs when $o^*<1/2$, raises costs when $o^*>1/2$, and leaves costs unchanged at $o^*=1/2$. A move from a mixed share to internal production raises costs whenever that share lies below $2o^*$. These inequalities determine which products experience higher or lower passenger costs when outsourcing falls.

\subsection{The Mixed Choice Measure}
\label{app:mixed_choice_proof}

Conditional on serving the route, the carrier can keep its previous outsourcing share, change to fully internal or fully outsourced service, or choose a mixed share. The active choice space consists of retention $\left(\ell,0\right)$, feasible adjustments to the corners, and feasible interior adjustments. The measure $\mu_\ell$ assigns unit mass to each discrete choice and Lebesgue measure to each feasible interior interval. Adding a unit atom for inactivity gives the finite measure $\nu_\ell$. Bounded choice values imply $0<Z=\int\exp\left\{W\right\}d\nu_\ell<\infty$. The density $f=\exp\left\{W\right\}/Z$ therefore integrates to one and gives the choice distribution in equation~\eqref{eq:dynamic_mixed_choice}.

Partition each feasible interior interval into cells $C_h$ of length $w_h>0$, and assign weight $w_h=1$ to each discrete choice. Let $H$ count the cells and atoms. For bounded measurable $W$, set $W_h=\log\left[w_h^{-1}\int_{C_h}\exp\left\{W\right\}d\nu_\ell\right]$ on each interior cell and retain the original value at each atom. Give alternative $h$ deterministic value $W_h+\log w_h$ and an independent type-I extreme value shock with unit scale and mean zero. Define $Z_H=\sum_{h=1}^Hw_h\exp\left\{W_h\right\}$. The finite model has probabilities $w_h\exp\left\{W_h\right\}/Z_H$ and expected maximum $\log Z_H$. This construction gives $Z_H=Z$ for every partition, and each cell receives its probability under $f$.

Let $f_H$ be the corresponding density, constant on each interior cell and unchanged at the atoms. Along refining interval partitions whose maximum cell length tends to zero, $f_H$ converges to $f$ in $L^1$. To establish convergence, approximate $f$ on the interior intervals by a continuous function $g$ and let $g_H$ contain its cell averages. Cell averaging gives $\left\|f_H-g_H\right\|_1\leq\left\|f-g\right\|_1$, so $\left\|f_H-f\right\|_1\leq2\left\|f-g\right\|_1+\left\|g_H-g\right\|_1$. The final term tends to zero, and the approximation error can be made arbitrarily small. Thus the finite probabilities converge to the mixed choice distribution. When $W$ is continuous on each feasible interval, values at representative points also give $Z_H\rightarrow Z$ and the same limiting probabilities and expected maximum. Polynomial choice values satisfy the continuity condition on each feasible interval.

For any density $g$ relative to $\nu_\ell$, the relative entropy is $\mathcal D\left(g\Vert f\right)=\int g\log\left(g/f\right)d\nu_\ell$. Substitution of $\log f=W-\log Z$ gives $\int gWd\nu_\ell-\int g\log gd\nu_\ell=\log Z-\mathcal D\left(g\Vert f\right)\leq\log Z$. Equality holds at $g=f$, so $\log Z$ is the maximum of the variational objective $\int gWd\nu_\ell-\int g\log gd\nu_\ell$ over densities $g$.

To justify the inequality, write $x=f/g$ on the set where $g>0$. The bound $-\log x\geq1-x$ gives $\mathcal D\left(g\Vert f\right)\geq1-\int_{g>0}f d\nu_\ell\geq0$. Equality requires $g=f$ almost everywhere. If the relative entropy is infinite, the variational objective is interpreted as minus infinity.

\subsection{Proof of Proposition~\ref{prop:revision_allocation}}

Let $Z_R=\int_{k=1}\exp\left\{W\left(o,1\right)\right\}d\mu_\ell\left(o,1\right)$ and $Z_0=\exp\left\{W^0\right\}$. Assumption~\ref{ass:org_regular}(i) and the boundedness in part (iii) give $Z_R,Z_0>0$. Conditional on activity, $P^R=Z_R/\left(Z_R+Z_0\right)=\Lambda\left(I^R-W^0\right)$. The mean allocation is $m^o=\left(1-P^R\right)\ell+P^R\bar o_R$.

Assumption~\ref{ass:org_regular}(ii) gives pointwise differentiability, and part (iii) provides an integrable bound for the derivative of $\exp\left\{W\right\}$ on the finite choice measure. Differentiation under the integral therefore gives $\dot Z_R=\int_{k=1}\exp\left\{W\right\}\dot Wd\mu_\ell$ and $\dot I^R=\E_R\left[\dot W\right]$. Differentiating the logistic formula for $P^R$ gives the first identity in equation~\eqref{eq:revision_decomposition}.

Conditional on adjustment, the density $f_R=\exp\left\{W\right\}/Z_R$ has derivative
\[
\dot f_R=f_R\left\{\dot W-\E_R\left[\dot W\right]\right\}.
\]
Since outsourcing is bounded, differentiating the mean conditional on adjustment gives $\dot{\bar o}_R=\operatorname{Cov}_R\left(o,\dot W\right)$. Differentiating $m^o=\ell+P^R\left(\bar o_R-\ell\right)$ then proves the second identity. When $\dot W$ is constant across adjustments, $\operatorname{Cov}_R\left(o,\dot W\right)=0$, so a common change in the payoff to adjustment moves $m^o$ only through the frequency term. The proof includes adjustments to the corners because the same measure represents atoms and the continuous intervals.

For an additional current outsourcing cost $\chi o$, set $\omega=\chi$ and hold the other components of choice values fixed, so $\dot W=-o$. Differentiation of the conditional active mean yields $\partial m^o/\partial\chi=\operatorname{Cov}_{f^A}\left(o,-o\right)=-\operatorname{Var}_{f^A}\left(o\right)$. The adjustment derivative is $\partial P^R/\partial\chi=-P^R\left(1-P^R\right)\left(\bar o_R-\ell\right)$. \hfill$\square$

\subsection{Proof of Proposition~\ref{prop:menu_charge}}
\label{app:menu_charge_proof}

For part (i), the outsourcing constraint imposes a minimum internal share on every carrier serving the route. Write $Z=\int\exp\left\{W\right\}d\mu_\ell$, so $I^A=\log Z$ and $f^A=\exp\left\{W\right\}/Z$. Assumption~\ref{ass:org_regular}(iii) bounds the choice values, and Assumption~\ref{ass:org_regular}(i) makes $\mu_\ell$ finite, so $Z$ is finite and positive. Let $\mathcal C_\tau$ denote the active choices with $o\leq1-\tau$. Complete internal production belongs to $\mathcal C_\tau$ and is always feasible, through retention when $\ell=0$ and through an adjustment to zero when $\ell>0$, because that adjustment changes the production regime. Hence $F^A\left(\mathcal C_\tau\right)>0$. Under the outsourcing limit, the active inclusive value is $\log\int_{\mathcal C_\tau}\exp\left\{W\right\}d\mu_\ell=I^A+\log F^A\left(\mathcal C_\tau\right)=I^A+\log\left(1-\Pi_\tau\right)$. The removed sets shrink as $\tau\downarrow0$, and their intersection is the set of choices with $o=1$. Continuity of the probability distribution $F^A$ from above therefore gives the limit of $\Pi_\tau$.

Under the cost increase, the active inclusive value is
\[
\log\int\exp\left\{W-\chi o\right\}d\mu_\ell=I^A+\log\E_{f^A}\left[\exp\left(-\chi o\right)\right].
\]
Since $0\leq o\leq1$, dominated convergence gives $\log\E_{f^A}\left[\exp\left(-\chi o\right)\right]\rightarrow0$ as $\chi\downarrow0$.

For part (ii), let $\Pi_{\tau j}$ denote the probability that brand $j$ assigns to choices removed by the outsourcing limit. The restricted distribution is $f_{\tau j}=f_j^A\ind\left\{\mathcal C_\tau\right\}/\left(1-\Pi_{\tau j}\right)$, so $\mathcal D\left(f_{\tau j}\Vert f_j^A\right)=-\log\left(1-\Pi_{\tau j}\right)$, the negative of the change in its active inclusive value. Let $E_{\chi j}=\E_{f_j^A}\left[\exp\left(-\chi o\right)\right]$ and let $m_{\chi j}$ be the mean share under $f_{\chi j}=f_j^A\exp\left(-\chi o\right)/E_{\chi j}$, the distribution after the cost increase. Then $\mathcal D\left(f_{\chi j}\Vert f_j^A\right)=-\chi m_{\chi j}-\log E_{\chi j}$. For $m'>0$, a finite common cost increase $\chi\geq0$ with $\sum_j\varpi_jm_{\chi j}=m'$ exists. Differentiating the tilted mean gives $\partial m_{\chi j}/\partial\chi=-\operatorname{Var}_{f_{\chi j}}\left(o\right)\leq0$, so the weighted mean under the cost increase is continuous and nonincreasing in $\chi$. This weighted mean equals the unrestricted weighted mean at $\chi=0$ and tends to zero as $\chi\rightarrow\infty$, because complete internal production has positive probability for every brand. Truncation above $1-\tau$ weakly lowers each brand's mean, so the positive target $m'$ lies in the range of the weighted mean under the cost increase. Weighting by $\varpi_j$ and using $\sum_j\varpi_jm_{\chi j}=m'$ gives the stated difference.

Consider any distributions $g_j$, absolutely continuous with respect to $f_j^A$, whose weighted mean share equals $m'$. The log ratios $\log\left(f_{\chi j}/f_j^A\right)=-\chi o-\log E_{\chi j}$ have the common slope $-\chi$ in $o$. The common slope and the weighted mean restriction imply that the weighted sum of these expected log ratios is the same under $g_j$ and $f_{\chi j}$. Therefore $\sum_j\varpi_j\mathcal D\left(g_j\Vert f_j^A\right)=\sum_j\varpi_j\mathcal D\left(g_j\Vert f_{\chi j}\right)+\sum_j\varpi_j\mathcal D\left(f_{\chi j}\Vert f_j^A\right)$, which is at least the last sum. The distributions under the outsourcing limit also have weighted mean $m'$, so setting $g_j=f_{\tau j}$ proves the inequality.

For part (iii), the payment adds $\chi m'$ to every active choice value of each brand under the cost increase and leaves each distribution $f_{\chi j}$ unchanged. The weighted change in active inclusive values under the cost increase therefore rises by $\chi m'$, and the difference in part (ii) becomes $\sum_j\varpi_j\left[\mathcal D\left(f_{\tau j}\Vert f_j^A\right)-\mathcal D\left(f_{\chi j}\Vert f_j^A\right)\right]\geq0$. With continuation values fixed, neither experiment changes the value of inactivity. Each brand's activity probability therefore moves in the same direction as the change in that brand's active inclusive value. \hfill$\square$

\subsection{Local Responses of Prices and Participation}
\label{app:transmission_proof}

Write the price condition as $\mathcal H\left(p,o,\omega\right)=p-F\left(p,o,\omega\right)=0$. When $F$ is continuously differentiable and $\mathcal H_p=\mathbf{I}-F_p$ is nonsingular at the solution, the implicit function theorem gives a differentiable local price function. Along a differentiable allocation path, differentiation yields $\left(\mathbf{I}-F_p\right)\dot p=F_o\dot o+F_\omega$, which proves the expression in Section~\ref{sec:transmission}.

Holding the inherited state and active choice set fixed, Assumption~\ref{ass:org_regular} permits differentiation under the mixed integral and gives $\dot I^A=\E_{f^A}\left[\dot W\right]$. Because $P=\Lambda\left(I^A-W\left(\varnothing\right)\right)$, $\dot P=P\left(1-P\right)\left\{\dot I^A-\dot W\left(\varnothing\right)\right\}$. Choice values contain both current payoffs and continuation values. The derivative therefore includes the change in the value of future production adjustments. A minimum internal share changes the choice set itself: a carrier can no longer serve the route entirely through independent providers. We evaluate the resulting participation and production responses by solving the value recursion with the corresponding choice measure.

\subsection{Payoff Normalizations and Participation Histories}
\label{app:payoff_normalization}

Suppose the activity component contains $\phi_{im}-\kappa_F$, where $\kappa_F$ is a common fixed cost. Replacing $\phi_{im}$ by $\phi_{im}+c$ and $\kappa_F$ by $\kappa_F+c$ leaves every flow payoff unchanged. The value recursion and the choice distribution consequently remain unchanged. Only the difference is identified. We therefore set the common activity cost to zero and leave the carrier-by-route effects unrestricted. Inactive flow payoff is zero, which fixes the payoff comparison governing entry and exit.

The calendar effects have a related location invariance. Adding $c$ to every carrier-by-route effect and subtracting $c$ from every calendar effect leaves all activity payoffs unchanged. Setting one calendar effect to zero removes that redundancy. The fixed adjustment cost and the outsourcing return are normalized by setting Delta's carrier shifts $\theta_{R,i}$ and $\gamma_{1,i}$ to zero. The year shifts $\kappa_{E,y\left(t\right)}$ are measured relative to 1998, and the era shifts $\gamma_{1,\mathit{era}\left(t\right)}$ relative to 1998 to 2004.

The invariance of values follows by backward induction through the extended sequence. Terminal values are the same under the two parameter vectors. If subsequent values agree, identical current flow payoffs give identical choice-specific values and log normalizing constants, completing the induction. The same invariance holds under each counterfactual. The production and ownership experiments leave $\phi_{im}-\kappa_F$ unchanged, so these normalizations also preserve each counterfactual flow payoff. Starting from the same counterfactual terminal values, backward induction gives identical counterfactual choice values and probabilities under the two parameterizations.

The set of players contains carrier-route pairs that serve the route in at least one quarter. The initial risk set includes 25,563 pairs that never operate and 1,118 pairs that always operate. Pairs that never operate are excluded. Always-active pairs choose from $\mu_\ell$ and contribute their production decisions. The pairs with changes in participation identify 15,035 activity effects. Participation is thus fixed for constant histories and estimated for histories with observed entry or exit.

For a finite horizon and bounded remaining payoffs, backward induction gives $V_{ie}\left(s\right)\rightarrow0$ and $P_{ie}\rightarrow0$ as $\phi_{im}\rightarrow-\infty$. The likelihood contribution of each inactive choice approaches one, which gives the boundary likelihood for a history without activity.

{\interlinepenalty=10000 For an always-active pair with activity effect $\phi=\phi_{im}$, the restriction to $\mu_\ell$ applies to current and future choices. Index decisions in the extended sequence, including continuation quarters, by $t=1,\ldots,T$. Define $K_{T+1}=0$ and $K_t=1+\beta K_{t+1}$, and write $V_t\left(s\right)=K_t\phi+\overline V_t\left(s\right)$. Because $K_{t+1}\phi$ is constant across next states, the conditional expectation preserves this decomposition. Substitution into the recursion restricted to activity factors $\exp\left\{K_t\phi\right\}$ out of the normalizer. The remaining recursion for $\overline V_t$ contains the organizational payoffs and no $\phi$. The same factor cancels from every conditional organizational probability.\par}

The same recursion restricted to active choices is the limit as $\phi\rightarrow\infty$ of the unrestricted finite-horizon problem with bounded remaining payoffs. Suppose $V_{t+1}-K_{t+1}\phi$ converges uniformly to $\overline V_{t+1}$. Conditional expectation under the fixed scale transition preserves this uniform convergence. After subtracting $K_t\phi$ from the current log normalizing constant, the inactive term is of order $\exp\left(-\phi\right)$ and the active terms converge to the integral over active choices. That integral is bounded away from zero on the bounded state set. Continuity of the logarithm proves the induction step, starting from the common zero terminal value. Activity probabilities converge to one, and conditional organizational densities converge to those from $\overline V_t$. This limiting argument characterizes the model with participation fixed.

\section{Estimation and Inference}
\label{app:estimation}

This appendix describes how we construct products, production histories, and the participation sample, and documents carriers' responses to supplier withdrawals. It then presents the likelihood, the joint estimation of payoff parameters and carrier-by-route effects, and inference for the counterfactual comparisons.

\subsection{Data and Measurement}
\label{app:data}

\subsubsection{Products and Decision Samples}

The source records distinguish the ticketing carrier, which markets the product and sets the fare, from the operating carriers that provide the flights. A product is the ticketing carrier's nonstop service on a directional route in a quarter. Flights in opposite directions constitute separate routes. We sum passengers within each product and calculate its fare as a passenger-weighted average of the mean ticket fares for flights operated by each carrier. Linked operating records identify the providers serving the product, and the outsourcing share is the fraction of its sampled tickets on flights operated by independent regional providers.

Flights operated by the ticketing carrier or a regional airline wholly owned by the carrier or its parent company count as internal production. Other codeshare operations are excluded from the regional outsourcing measure. We also measure outsourcing using available seats and departures in the operating records. The ticket measure covers all products, and the capacity measure covers 288,672 products.

The ownership classification requires full equity ownership by the ticketing carrier or its corporate parent. A branded service contract does not establish ownership. For example, GoJet belonged to Trans States Holdings, and Atlantic Coast Airlines belonged to Atlantic Coast Airlines Holdings. Both operators remained independent under their contracts with legacy carriers \citep{Delta2012Form10K,AtlanticCoast2003Form10K}.

An operator's flights count as internal production in a quarter when the ticketing carrier or its corporate parent wholly owns the operator for more than half of that quarter. Subsidiaries of the parent company, such as American Eagle and Executive Airlines under AMR, PSA and Piedmont under US Airways Group, and Horizon under Alaska Air Group, therefore count as internal production.

Corporate filings identify the ownership transitions. Delta completed full ownership of Atlantic Southeast Airlines on May 11, 1999, and sold the carrier to SkyWest on September 7, 2005 \citep{Delta1999Q1Form10Q,Delta2005Q3Form10Q}. Comair became wholly owned by Delta in January 2000, following a November 1999 tender that raised Delta's stake to about 90 percent \citep{Delta2000Q1Form10Q}. ExpressJet's April 23, 2002 public offering reduced Continental's ownership from 100 percent to about 53 percent \citep{ExpressJet2002Q2Form10Q}. Pinnacle was wholly owned by Northwest until January 15, 2003, when Northwest contributed the first Pinnacle shares to a pension plan. Pinnacle became wholly owned by Delta in May 2013 \citep{Northwest2004Form10K,Pinnacle2003Prospectus,Delta2013Form10K}. Mesaba became wholly owned by Northwest on April 24, 2007 \citep{MAIR2007Form10K}. Delta sold Mesaba to Pinnacle Airlines Corporation and Compass to Trans States Holdings in 2010, and both are external producers from 2010Q3.

Under this rule, Comair's flights count as outsourcing through 1999Q4 and internal production from 2000Q1. A change in ownership reclassifies the operator's flights and therefore changes the measured outsourcing share of the products it serves, even if the operator continues to provide the same flights.

Table~\ref{tab:app_sample} distinguishes the products used for demand and passenger costs from the service and production decisions used to estimate the dynamic game. The demand equations use the 263,912 products with departures in the preceding quarter. The contemporaneous pricing equations use all 351,811 products, and the lagged moments use the 299,696 products with the required preceding observations. The game contains 818,335 carrier-route-quarter decisions in 260,940 route-quarter markets on 6,159 routes, with 149 common parameters and 15,035 carrier-by-route effects.

\begin{table}[!tp]
\centering
\small
\caption{Sample Construction}
\label{tab:app_sample}
\begin{tabular*}{\textwidth}{@{\extracolsep{\fill}}lr@{}}
\toprule
Sample component & Count \\
\midrule
\multicolumn{2}{@{}l}{\textit{Panel A. Product and provider records}} \\
Ticketing carrier, operating carrier, route, and quarter records & 541,681 \\
Branded products & 351,811 \\
Route-quarter markets & 264,597 \\
Products in markets with at least two modeled carriers & 157,007 \\
Directional routes & 6,744 \\
Observed carrier-route pairs & 18,273 \\
Quarters & 88 \\
Modeled ticketing carriers & 7 \\
Operating carriers & 62 \\
Ticketing and operating carrier pairs & 142 \\
Independent regional providers & 23 \\
Candidate carrier and provider pairs & 154 \\
\midrule
\multicolumn{2}{@{}l}{\textit{Panel B. Estimation}} \\
Static pricing and cost products & 351,811 \\
Demand estimation products & 263,912 \\
Products with available lag moments & 299,696 \\
Adjacent organizational observations & 299,696 \\
Game decisions & 818,335 \\
Game route-quarters & 260,940 \\
Game routes & 6,159 \\
Game decision quarters & 87 \\
Estimated carrier-by-route effects & 15,035 \\
Always active pairs with conditional organizational choices & 1,118 \\
\bottomrule
\end{tabular*}
\par\smallskip
\begin{minipage}{\textwidth}
\footnotesize\justifying
\noindent\textit{Notes.} Product counts refer to the static sample. Adjacent organizational observations have an observed inherited allocation. The game applies its own eligibility requirements and retains organizational choices for always active pairs. The first decision requires a preceding quarter, so the dynamic period contains 87 quarters.
\end{minipage}
\end{table}

A brand is active on a route in a quarter if the ticket sample contains a product of that brand on the route. It is available in a quarter if the sample contains any of its products. A route-quarter enters the participation risk set when the route had legacy service in the preceding quarter. We therefore retain the quarter in which the last legacy brand leaves. Every brand available in the industry in both quarters is initially included as a potential entrant or continuing carrier. The risk set includes 19,251 route-quarter markets with no current legacy service and the 23,345 brand exits observed in those markets. Eligibility depends on the preceding quarter's state, which also provides the traffic, network, and rival covariates.\footnote{The entry margin is entry into routes with legacy service in the preceding quarter. Quarters immediately following an absence of legacy service have no observed lagged market covariates.}

A decision enters the game when we observe whether the brand served the route in the preceding quarter, its outsourcing share and duration, and the market covariates. We retain a route-quarter only when every decision in that game meets these conditions and exclude routes with fewer than two retained games. Of the candidate carrier-route pairs, 25,563 never operate and 1,118 always operate. Brands that never serve a route in a retained game are excluded, while always-active pairs contribute their production choices conditional on activity. We also exclude other brands' entry decisions in route-quarters in which an active brand retires, because the acquiring carrier's service continues the retired brand's operation. Duration is zero in the quarter following entry or a production adjustment and increases by one after each quarter without an adjustment. For arrangements already in place at the start of the sample, duration begins at zero in the first decision quarter.

The participation payoff uses market conditions in the preceding quarter. Lagged route traffic is the logarithm of passengers of the seven legacy brands on the route. The lagged number of rivals counts the other legacy brands active on the route. Among routes from the product's origin airport with sampled nonstop tickets, carrier network size is the share on which the ticketing carrier sells nonstop tickets in the quarter. The payoff uses the logarithm of average network size across the carrier's products. These three covariates enter in standard deviation units. Two indicators record whether Southwest served the route and whether other low-cost carriers served the route.

In the passenger-cost equation, hub status indicates whether either endpoint is a hub of the carrier. The weather index combines standardized precipitation, snowfall, and temperature extremes on the route into a single standardized measure. When weather data are missing, we use the route's mean for the same quarter of the year.

Table~\ref{tab:app_participation_support} reports descriptive participation statistics for all available brands and for brands ever active on a route, before applying the game eligibility requirements. A brand can enter a route when it did not serve it in the preceding quarter and can exit when it did. Retirement of an acquired brand changes the set of available players throughout the industry. Individual route entry and exit are measured among brands available in both adjacent quarters.

\begin{table}[!tp]
\centering
\small
\caption{Descriptive Participation Risk Sets}
\label{tab:app_participation_support}
\begin{tabular*}{\textwidth}{@{\extracolsep{\fill}}lrr@{}}
\toprule
& All available brands & Brands ever active on the route \\
\midrule
Carrier, route, and quarter alternatives & 1,564,115 & 843,014 \\
Alternatives at risk of entry & 1,217,662 & 496,561 \\
Alternatives at risk of exit & 346,453 & 346,453 \\
Quarterly entry probability & 0.0203 & 0.0498 \\
Quarterly continuation probability & 0.8650 & 0.8650 \\
Quarterly exit probability & 0.1350 & 0.1350 \\
Entries & 24,738 & 24,738 \\
Exits & 46,757 & 46,757 \\
\bottomrule
\end{tabular*}
\par\smallskip
\begin{minipage}{\textwidth}
\footnotesize\justifying
\noindent\textit{Notes.} The columns describe participation before applying the game eligibility requirements. The first column retains all brands available in adjacent quarters. The second retains only brands that serve the route at least once. An entry follows inactivity and an exit follows activity. Brand retirement changes the available set. Entry is measured on routes with legacy service in the preceding quarter, while exits include the disappearance of all legacy service from such routes.
\end{minipage}
\end{table}

\subsubsection{Organizational Adjustment}
\label{app:state_measurement}

The transition probabilities in Figure~\ref{fig:action_distribution} use adjacent observations of the same carrier on the same route. A production adjustment changes the outsourcing share by more than 0.05 or changes the regime among complete internal production, mixed production, and complete outsourcing. Small fluctuations within mixed production therefore count as retention. Table~\ref{tab:app_action_audit} summarizes the adjustment and measurement statistics. The duration comparisons in Figure~\ref{fig:fit_duration} use active incumbents in the game sample, distinguishing inherited mixed production from exclusive production, which combines fully internal and fully outsourced arrangements.

\begin{table}[!tp]
\centering
\small
\caption{Adjustment and Capacity Statistics}
\label{tab:app_action_audit}
\begin{tabular*}{\textwidth}{@{\extracolsep{\fill}}lr@{}}
\toprule
Statistic & Value \\
\midrule
\multicolumn{2}{@{}l}{\textit{Panel A. Adjacent organizational observations}} \\
Persistence, split-panel jackknife & 0.8681 \\
Airport-pair multiplier standard error & 0.0036 \\
Frequency of production adjustments & 0.2315 \\
Median absolute size of a production adjustment & 0.1263 \\
Adjustment probability at duration zero or one & 0.4482 \\
Adjustment probability at duration sixteen or more & 0.0337 \\
\midrule
\multicolumn{2}{@{}l}{\textit{Panel B. Capacity measure}} \\
Coverage of products & 0.8205 \\
Correlation with ticket outsourcing & 0.9522 \\
Fraction in the same production regime & 0.9216 \\
Coverage under complete internal production & 0.7286 \\
Coverage under mixed production & 0.9702 \\
Coverage under complete outsourcing & 0.8949 \\
\bottomrule
\end{tabular*}
\par\smallskip
\begin{minipage}{\textwidth}
\footnotesize\justifying
\noindent\textit{Notes.} Both panels define production regimes by exact shares of zero and one. Panel A uses the 299,696 adjacent organizational observations, and the persistence regression retains 297,904. Its duration groups and sample differ from those in Figure~\ref{fig:fit_duration}, which reports finer duration groups for active incumbents in the game. Panel B compares ticket and capacity outsourcing where both measures are available.
\end{minipage}
\end{table}

Table~\ref{tab:state_duration} asks whether the adjustment patterns remain within a carrier-route pair. Panel A regresses the production adjustment indicator on $\log\left(1+d\right)$ separately for inherited exclusive and mixed production. Quarter effects absorb common changes, and carrier-by-route effects account for persistent differences in the propensity to adjust. The split-panel correction combines estimates from the full panel and its two time halves. Panel B includes indicators for the inherited share and allows duration slopes to differ across these states. The state coefficients compare the indicated mixed allocations with complete outsourcing at zero duration. Mixed allocations remain more likely to adjust after controlling for carrier-by-route effects, and adjustment declines with duration.

\begin{table}[!tp]
\centering
\singlespacing
\small
\caption{Organizational State and Duration within Carrier and Route}
\label{tab:state_duration}
\begin{tabularx}{\textwidth}{@{}Xrrr@{}}
\toprule
Measure and inherited state & Pooled & Route effects & Corrected \\
\midrule
\multicolumn{4}{@{}l}{\textit{Panel A. Duration slopes}} \\
Ticket share, exclusive production & $-$0.0499 & $-$0.0390 & $-$0.0530 \\
 & (0.0013) & (0.0014) & \\
Ticket share, mixed production & $-$0.1797 & $-$0.0958 & $-$0.1465 \\
 & (0.0052) & (0.0049) & \\
Seat share, exclusive production & $-$0.0481 & $-$0.0351 & $-$0.0457 \\
 & (0.0015) & (0.0015) & \\
Seat share, mixed production & $-$0.2143 & $-$0.1237 & $-$0.1741 \\
 & (0.0044) & (0.0050) & \\
\midrule
\multicolumn{4}{@{}l}{\textit{Panel B. State differences from complete outsourcing}} \\
Ticket share, $\left[0.85,0.95\right)$ & 0.5816 & 0.5176 & 0.5427 \\
 & (0.0087) & (0.0092) & \\
Ticket share, $\left(0,0.8\right)$ & 0.5071 & 0.4625 & 0.4780 \\
 & (0.0059) & (0.0066) & \\
Seat share, $\left[0.85,0.95\right)$ & 0.6938 & 0.6145 & 0.6591 \\
 & (0.0116) & (0.0122) & \\
\bottomrule
\end{tabularx}
\par\smallskip
\begin{minipage}{\textwidth}
\footnotesize\justifying
\noindent\textit{Notes.} The dependent variable is a production adjustment. Panel A reports slopes on $\log\left(1+d\right)$ separately for inherited exclusive and mixed arrangements, using incumbent decisions for which duration since an adjustment is observed. The pooled regressions include carrier effects and quarter effects. The second column includes carrier-by-route effects and quarter effects, and the third applies the split-panel jackknife. Panel B reports state indicators from regressions that also allow duration slopes to differ across inherited states. The reference state is complete outsourcing. The coefficients on the state indicators give differences at zero duration. Ticket and seat regressions in Panel B use 223,583 and 174,660 decisions, respectively. Parentheses report standard errors clustered by airport pair for the pooled and uncorrected fixed-effects estimates.
\end{minipage}
\end{table}

\subsubsection{Production Measurement}

Table~\ref{tab:measurement} compares outsourcing shares measured by tickets and by seats. It also asks how often ticket sampling would generate an apparent adjustment when the underlying production mix is unchanged. For each pair of adjacent observations, we hold the true outsourcing share at the inherited value and calculate the adjustment probability under binomial sampling with the two observed ticket counts. Observed adjustment exceeds this benchmark in every ticket-count category. For groups with at least 500 sampled tickets in each quarter, observed adjustment probabilities are 0.524 and 0.568, compared with 0.031 and 0.011 under sampling alone.

\begin{table}[!tp]
\centering
\singlespacing
\small
\caption{Outsourcing Adjustments and Measurement}
\label{tab:measurement}
\begin{tabularx}{\textwidth}{@{}X*{5}{>{\raggedleft\arraybackslash}p{1.75cm}}@{}}
\toprule
 & \multicolumn{3}{c}{Remaining in the same regime} & \multicolumn{2}{c}{Adjusting next quarter} \\
\cmidrule(lr){2-4}\cmidrule(l){5-6}
Measure & Internal & Mixed & Outsourced & Mixed & Exclusive \\
\midrule
\multicolumn{6}{@{}l}{\textit{Panel A. Ticket and seat measures}} \\
Ticket share & 0.945 & 0.813 & 0.873 & 0.743 & 0.178 \\
Seat share & 0.919 & 0.824 & 0.926 & 0.722 & 0.176 \\
\end{tabularx}
\begin{tabularx}{\textwidth}{@{}X*{3}{>{\raggedleft\arraybackslash}p{3.2cm}}@{}}
\midrule
\multicolumn{4}{@{}l}{\textit{Panel B. Adjustments under mixed production by ticket count}} \\
Minimum tickets & Decisions & Observed adjustment & Sampling benchmark \\
1 to 49 & 16,081 & 0.828 & 0.587 \\
50 to 99 & 15,661 & 0.741 & 0.356 \\
100 to 199 & 20,369 & 0.667 & 0.235 \\
200 to 499 & 19,797 & 0.603 & 0.118 \\
500 to 999 & 3,308 & 0.524 & 0.031 \\
1,000 or more & 139 & 0.568 & 0.011 \\
\bottomrule
\end{tabularx}
\par\smallskip
\begin{minipage}{\textwidth}
\footnotesize\justifying
\noindent\textit{Notes.} Panel A compares transition and adjustment probabilities calculated from ticket shares in the ticket sample and seat shares in the operating capacity records, using the available sample for each measure. Internal, mixed, and outsourced production correspond to shares of zero, strictly between zero and one, and one. Exclusive production combines the two corners. The last two columns give adjustment probabilities in the quarter following entry or an adjustment. Figure~\ref{fig:fit_duration} excludes spells that begin at entry. Panel B groups adjacent observations by the smaller number of sampled tickets in the two quarters. The sampling benchmark holds the true share at the inherited share and calculates adjustments generated by two quarters of binomial sampling. Across the 174,657 decisions with both ticket and seat measures, the adjustment classifications agree in 86 percent of cases. The seat measure classifies 67 percent of adjustments of the ticket share as adjustments and 4.6 percent of retentions of the ticket share as adjustments.
\end{minipage}
\end{table}

Table~\ref{tab:app_near_one} follows carriers that outsource almost all, but not all, of their service on a route. Many adjustments from these allocations end at complete outsourcing. Such a move changes the production regime and therefore counts as an adjustment even when the change in the share is smaller than the adjustment threshold. Conditional on adjustment from an inherited share in $\left[0.98,1\right)$, the fraction ending at one is 0.779 under the ticket measure and 0.767 under the seat measure. The fractions changing the share by at least 0.2 are 0.117 and 0.127.

\begin{table}[!tp]
\centering
\small
\singlespacing
\caption{Adjustments of Arrangements Near Complete Outsourcing}
\label{tab:app_near_one}
\begin{tabularx}{\textwidth}{@{}Xrrrrr@{}}
\toprule
 & & & \multicolumn{2}{c}{Share of adjustments} & Large adjustments \\
\cmidrule(lr){4-5}
Inherited share & Decisions & Adjustment rate & Ending at one & Change $\geq0.2$ & per decision \\
\midrule
\multicolumn{6}{@{}l}{\textit{Panel A. Ticket share (Origin and Destination Survey)}} \\
$\left[0.85,0.9\right)$ & 2,104 & 0.818 & 0.300 & 0.236 & 0.193 \\
$\left[0.9,0.95\right)$ & 2,408 & 0.765 & 0.433 & 0.220 & 0.169 \\
$\left[0.95,0.98\right)$ & 3,215 & 0.688 & 0.682 & 0.156 & 0.108 \\
$\left[0.98,0.995\right)$ & 3,546 & 0.669 & 0.777 & 0.119 & 0.080 \\
$\left[0.995,1\right)$ & 424 & 0.618 & 0.798 & 0.095 & 0.059 \\
1 & 62,622 & 0.128 &  & 0.363 & 0.046 \\
\midrule
\multicolumn{6}{@{}l}{\textit{Panel B. Seat share (T100)}} \\
$\left[0.85,0.9\right)$ & 1,241 & 0.845 & 0.304 & 0.350 & 0.296 \\
$\left[0.9,0.95\right)$ & 1,399 & 0.844 & 0.498 & 0.238 & 0.201 \\
$\left[0.95,0.98\right)$ & 1,368 & 0.786 & 0.652 & 0.161 & 0.126 \\
$\left[0.98,0.995\right)$ & 1,584 & 0.797 & 0.760 & 0.134 & 0.107 \\
$\left[0.995,1\right)$ & 225 & 0.773 & 0.822 & 0.080 & 0.062 \\
1 & 66,005 & 0.074 &  & 0.424 & 0.032 \\
\bottomrule
\end{tabularx}
\par\smallskip
\begin{minipage}{\textwidth}
\footnotesize\justifying
\noindent\textit{Notes.} The sample consists of active incumbent legacy brands. Rows classify the inherited outsourcing share. An adjustment changes the share by more than five percentage points or changes the production regime, as defined in Section~\ref{sec:model}. Ending at one is the fraction of adjustments that reach complete outsourcing. A large adjustment changes the share by at least 0.2. The last column multiplies the adjustment probability by the fraction of adjustments that are large. Ticket and seat shares use their respective available samples.
\end{minipage}
\end{table}

Reallocating a single departure can change the outsourcing share substantially on routes with infrequent service. Table~\ref{tab:app_frequency} examines whether the adjustment patterns persist on routes where carriers operate more flights. It groups incumbent decisions by the brand's nonstop departures on the route in the preceding quarter. Complete outsourcing is more common below five daily departures, and mixed production measured by departure shares is rare below one daily departure. In every frequency group, complete outsourcing is persistent, carriers using both forms of production adjust more often than those using only one, and adjustment declines with duration. Mixed allocations near complete outsourcing also frequently move to complete outsourcing. The difference in adjustment rates between mixed and exclusive production narrows at higher frequencies.

\begin{table}[!tp]
\centering
\singlespacing
\small
\setlength{\tabcolsep}{4pt}
\caption{Organizational Patterns by Flight Frequency}
\label{tab:app_frequency}
\begin{tabularx}{\textwidth}{@{}Xrrrrrr@{}}
\toprule
 & & \multicolumn{4}{c}{Daily departures in the inherited quarter} & No nonstop \\
\cmidrule(lr){3-6}
 & All & Under 1 & 1 to under 2 & 2 to under 5 & 5 or more & record \\
\midrule
\multicolumn{7}{@{}l}{\textit{Panel A. Inherited production}} \\
Decisions & 299,696 & 32,891 & 43,416 & 123,011 & 65,680 & 34,698 \\
Routes & 5,866 & 3,877 & 3,597 & 3,660 & 1,656 & 3,552 \\
Median quarterly departures &  & 54 & 150 & 293 & 617 &  \\
Internal production & 0.537 & 0.481 & 0.457 & 0.461 & 0.602 & 0.832 \\
Mixed production & 0.253 & 0.220 & 0.239 & 0.291 & 0.312 & 0.055 \\
Complete outsourcing & 0.211 & 0.299 & 0.304 & 0.248 & 0.086 & 0.113 \\
Share in $\left[0.85,0.95\right)$ & 0.015 & 0.017 & 0.016 & 0.019 & 0.013 & 0.004 \\
\midrule
\multicolumn{7}{@{}l}{\textit{Panel B. Probability of remaining in the same regime}} \\
Internal production & 0.945 & 0.934 & 0.938 & 0.946 & 0.955 & 0.940 \\
Mixed production & 0.813 & 0.655 & 0.740 & 0.835 & 0.884 & 0.635 \\
Complete outsourcing & 0.873 & 0.862 & 0.881 & 0.884 & 0.861 & 0.803 \\
\midrule
\multicolumn{7}{@{}l}{\textit{Panel C. Adjustment probability}} \\
Mixed production, duration zero & 0.743 & 0.845 & 0.809 & 0.739 & 0.645 & 0.844 \\
Exclusive production, duration zero & 0.178 & 0.144 & 0.237 & 0.257 & 0.270 & 0.099 \\
Mixed production, duration four or more & 0.360 & 0.439 & 0.479 & 0.390 & 0.320 & 0.533 \\
Exclusive production, duration four or more & 0.054 & 0.056 & 0.050 & 0.052 & 0.052 & 0.066 \\
Adjustments from $\left[0.98,1\right)$ ending at one & 0.779 & 0.706 & 0.781 & 0.784 & 0.796 & 0.759 \\
\midrule
\multicolumn{7}{@{}l}{\textit{Panel D. Regimes of the departure share}} \\
Internal production & 0.523 & 0.524 & 0.489 & 0.482 & 0.621 &  \\
Mixed production & 0.202 & 0.056 & 0.127 & 0.225 & 0.280 &  \\
Complete outsourcing & 0.275 & 0.419 & 0.384 & 0.293 & 0.099 &  \\
\bottomrule
\end{tabularx}
\par\smallskip
\begin{minipage}{\textwidth}
\footnotesize
\justifying
\noindent\textit{Notes.} The sample contains decisions of active incumbent legacy brands with an observation in the preceding quarter, grouped by the brand's nonstop T100 departures on the route in that quarter. One daily departure corresponds to 90 departures per quarter. The last column contains decisions without a nonstop T100 departure record of the brand on the route in the inherited quarter. Panels A to C use the outsourcing share of sampled tickets and the definition of production adjustment of Section~\ref{sec:model}. Panel C uses observed duration, which begins at entry or a production adjustment. Duration zero denotes the following quarter. This sample includes spells initialized at entry and differs from the sample following adjustments in Figure~\ref{fig:fit_duration}. Panel D classifies the share of the brand's departures operated by independent regional carriers.
\end{minipage}

\end{table}

\subsubsection{Provider Relationships}
\label{app:provider_conditions}

We also examine changes in the set of independent providers that a carrier uses on a route. A change in the set of providers used on a route is associated with a 0.257 increase in the probability of an adjustment.\footnote{The adjustment indicator in these relationship regressions classifies shares at or below 0.02 and at or above 0.98 as exclusive production. An adjustment changes the share by more than five percentage points or crosses one of these category boundaries.} The ten largest independent providers account for 92.0 percent of outsourced passengers, and a legacy carrier may purchase service from the same provider on many routes.

We estimate an auxiliary logit equation for whether a carrier uses a provider in a quarter. To measure pressure on a provider's capacity, we allocate rivals' predicted outsourcing opportunities across providers using fixed historical relationship weights, then divide the demand assigned to each provider by its lagged capacity. The regressors include predicted operating opportunities, provider capacity pressure, lagged capacity, historical compatibility, the carrier's total passengers, lagged use of other providers, the lagged relationship indicator, seasonal indicators, and a quadratic time trend. The sample contains 5,943 observations on 154 candidate carrier-provider pairs in quarters when the provider operates independently for some carrier. Relationships are active in 28.1 percent of observations. Standard errors are clustered by carrier-provider pair.

Table~\ref{tab:app_provider_intensive} relates outsourcing intensity to capacity pressure on the primary provider among products with positive outsourcing. A one standard deviation increase in pressure is associated with a 0.0270 lower outsourcing share, controlling for carrier-by-route effects and quarter effects.

\begin{table}[!tp]
\centering
\small
\caption{Provider Capacity Pressure and Outsourcing Intensity}
\label{tab:app_provider_intensive}
\begin{tabularx}{\textwidth}{@{}Xrrr@{}}
\toprule
& Estimate & Standard error & $p$ value \\
\midrule
Lagged outsourcing share & 0.5813 & 0.0128 & $<0.001$ \\
Capacity pressure on the primary provider & $-0.0270$ & 0.0063 & $<0.001$ \\
Capacity of the primary provider & 0.0006 & 0.0016 & 0.7188 \\
Number of regional operators & 0.0323 & 0.0018 & $<0.001$ \\
\midrule
Observations & 129,130 & & \\
$R^2$ & 0.8298 & & \\
\bottomrule
\end{tabularx}
\par\smallskip
\begin{minipage}{\textwidth}
\footnotesize\justifying
\noindent\textit{Notes.} The sample contains products with positive outsourcing. The dependent variable is the outsourcing share. Capacity pressure and provider capacity are standardized. The specification includes carrier-by-route effects and quarter effects.
\end{minipage}
\end{table}

\subsection{Supplier Withdrawals}

\subsubsection{The 2014 Chautauqua Withdrawal}
\label{app:republic}

The regressions use an unbalanced panel of American, United, Alaska, and Delta on the routes in the game sample from 2012Q2 through 2015Q2. A pair is exposed if Chautauqua carried its passengers during 2013Q2 through 2014Q1. This definition identifies 53 American pairs and 66 United pairs. We measure outsourcing where the carrier continues to serve the route. Continuation records whether a carrier serving the route in the preceding quarter still serves it in the current quarter. An exit therefore lowers continuation and removes the pair from the observed outsourcing sample. Each regression excludes missing outcomes and carrier-by-route groups with only one usable observation. Standard errors are clustered by airport pair.

Let $Y_{imt}$ denote either outcome, $E_{im}$ indicate an exposed American, Delta, or United pair, $T_{im}$ indicate an exposed American or United pair, and $P_t$ indicate quarters from 2014Q2 onward. The triple-difference regression is $Y_{imt}=c_{im}+d_{it}+\delta^{D}T_{im}P_t+h_t E_{im}+u_{imt}$, where $c_{im}$ and $d_{it}$ are carrier-by-route and carrier-by-quarter effects. The interactions $h_t E_{im}$ absorb quarterly changes common to prior Chautauqua exposure, including Delta's exposed routes. Chautauqua continued flying for Delta in 2014. The difference-in-differences regression omits these interactions. Carrier-specific regressions define $T_{im}$ for the indicated carrier and exclude the other affected carrier's exposed pairs. The carrier-wide comparison assigns treatment to all American and United pairs and replaces $d_{it}$ with quarter effects.

The event studies replace the post-period treatment interaction with separate treatment interactions for quarters from eight before through four after 2014Q2, omitting 2014Q1. The event studies by route exposure include carrier-by-quarter effects, and the carrier-wide event study uses quarter effects. Both omit the additional $h_t E_{im}$ interactions used in the triple difference. The figures report pointwise intervals of 1.96 standard errors on either side of each coefficient.

Table~\ref{tab:app_republic_did} reports the production and continuation comparisons. Pooling American and United, outsourcing falls by 29.9 percentage points on exposed routes relative to the comparison routes. Continuation declines by an estimated 4.7 percentage points, with a standard error of 4.4. Accounting for changes on Delta's Chautauqua routes gives an outsourcing decline of 31.0 percentage points and a continuation estimate close to zero.

\begin{table}[!tp]
\centering
\small
\singlespacing
\caption{Outsourcing and Route Continuation after Chautauqua's Withdrawal}
\label{tab:app_republic_did}
\begin{tabularx}{\textwidth}{@{}Xrrr@{}}
\toprule
Comparison & Outsourcing share & Continuation & Treated pairs \\
\midrule
American and United compared with Alaska and Delta (quarter effects) & 0.119 & 0.029 & 2,837 \\
 & (0.010) & (0.008) & \\
Chautauqua routes, difference in differences & $-$0.299 & $-$0.047 & 118 \\
 & (0.078) & (0.044) & \\
Chautauqua routes, triple difference & $-$0.310 & $-$0.003 & 118 \\
 & (0.081) & (0.044) & \\
\quad American & $-$0.487 & 0.110 & 53 \\
 & (0.110) & (0.048) & \\
\quad United & $-$0.027 & $-$0.148 & 65 \\
 & (0.043) & (0.077) & \\
\quad American and United, excluding Cleveland routes & $-$0.402 & 0.139 & 77 \\
 & (0.093) & (0.041) & \\
\quad United, excluding Cleveland routes & $-$0.097 & 0.236 & 24 \\
 & (0.045) & (0.081) & \\
Delta's continuing Chautauqua routes & 0.008 & $-$0.043 & 252 \\
 & (0.022) & (0.013) & \\
\bottomrule
\end{tabularx}
\par\smallskip
\begin{minipage}{\textwidth}
\footnotesize\justifying
\noindent\textit{Notes.} The sample covers 2012Q2 through 2015Q2 on the routes of the game sample, with the post period beginning in 2014Q2, and includes American, United, Alaska, and Delta. The pooled outsourcing sample contains 34,914 active observations and the pooled continuation sample 37,435 incumbent observations. Counts fall in the indicated subsamples. The first row uses carrier-by-route and quarter effects. All remaining rows use carrier-by-route and carrier-by-quarter effects. The triple differences add interactions between prior Chautauqua exposure and quarter, using Delta's exposed routes as an additional comparison. Carrier-specific rows exclude the other affected carrier's exposed routes. Rows excluding Cleveland drop routes with an endpoint at Cleveland. The final row compares Delta's exposed routes with its other routes and excludes American's and United's exposed routes. Counts of treated pairs refer to the outsourcing sample, which contains 118 of the 119 exposed American and United pairs. Coefficients are in share or probability units. A coefficient of 0.10 corresponds to 10 percentage points. Parentheses report standard errors clustered by airport pair.
\end{minipage}
\end{table}

The carrier-specific comparisons show different production and participation patterns. On American's exposed routes, the triple-difference estimates give a 48.7 percentage point decline in outsourcing among active pairs and an 11.0 percentage point increase in continuation relative to the comparison routes. Operator records show that American expanded Envoy's service and purchased replacement service from other independent providers, as described in Section~\ref{sec:republic}.

American and US Airways entered common ownership in 2013Q4, within the exposure window, and both brands remained observed through 2015Q2. Carrier-by-quarter effects absorb quarterly changes common to American's network. The estimates compare American's exposed routes with its other routes within each quarter, after accounting for changes common to Chautauqua exposure.

United's outsourcing share changed little among active routes, while continuation was 14.8 percentage points lower relative to the comparison routes. United was also contracting its Cleveland hub during this period. Of the 56 exposed routes active in 2014Q1, 37 involved Cleveland, and 12 of those 37 remained active in 2014Q3. United attributed the contraction to persistent losses and reported that regional pilot shortages accelerated its timing.\footnote{United's 2013 Form 10-K describes the Cleveland reduction in Part II, Item 7, 2014 Outlook \citep{United2013Form10K}.} Excluding Cleveland routes gives United a continuation coefficient of 0.236, with a standard error of 0.081. These differences underscore the distinction between outsourcing among carriers that continue to serve a route and withdrawal from the route itself.

Figure~\ref{fig:republic_event_study} in Section~\ref{sec:republic} traces the carrier-specific comparisons by quarter. American's outsourcing decline develops over several quarters, while United's continuation decline is concentrated in 2014Q2 and 2014Q3. The post period begins in 2014Q2, when United's aircraft were scheduled to leave Chautauqua service. American's removals began earlier and continued over several months.

\begin{figure}[!tp]
\centering
\includegraphics[width=\textwidth]{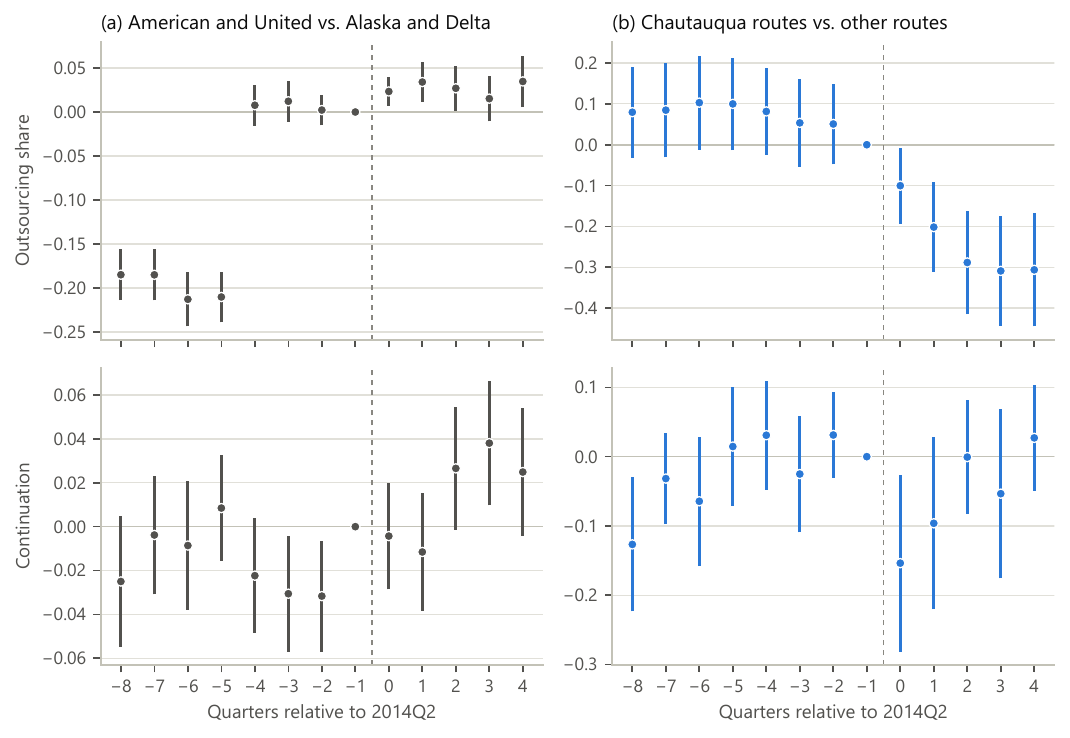}
\caption{Event Studies around Chautauqua's Withdrawal}
\label{fig:app_republic_event}
\par\smallskip
\begin{minipage}{\textwidth}
\footnotesize\justifying
\noindent\textit{Notes.} Event time is measured relative to 2014Q2, and 2014Q1 is the omitted quarter. Panel (a) compares all American and United routes with Alaska and Delta routes, with carrier-by-route and quarter effects. Panel (b) compares American and United routes served by Chautauqua during 2013Q2 through 2014Q1 with their other routes, with carrier-by-route and carrier-by-quarter effects. Both regressions omit the additional Chautauqua exposure-by-quarter effects used in the triple difference. Outsourcing is measured among active observations and continuation among incumbents. Coefficients are in share or probability units. Intervals are pointwise 95 percent confidence intervals based on standard errors clustered by airport pair.
\end{minipage}
\end{figure}

The first row of Table~\ref{tab:app_republic_did} compares all American and United routes with Alaska and Delta routes. Changes in ownership and in other provider relationships affect this carrier-wide comparison. Delta acquired Pinnacle in May 2013, when the provider emerged from bankruptcy as a wholly owned subsidiary.\footnote{Pinnacle was renamed Endeavor Air in 2013. Delta's 2013 Form 10-K reports the May acquisition in Item 1 and Note 18 \citep{Delta2013Form10K}.} The ownership classification treats Pinnacle as an independent provider through 2013Q1 and as internal production from 2013Q2, the quarter of the acquisition. American's outsourcing share rose from 0.03 in 2012 to 0.29 in 2015Q2 as other regional providers began flying for the carrier. Figure~\ref{fig:app_republic_event}(a) shows the resulting movement before 2014.

Panel (b) restricts exposure to Chautauqua routes and includes carrier-by-quarter effects, which absorb these carrier-wide changes. On Delta's Chautauqua routes, where Chautauqua continued flying, the outsourcing change is close to zero and continuation is 4.3 percentage points lower than on Delta's other routes. The triple difference uses these routes to account for changes common to Chautauqua exposure.

Table~\ref{tab:app_republic_retention} follows the fixed set of carrier-route pairs with service in 2014Q1, including entrants that quarter, through 2015Q2. We record whether each carrier serves the route in each subsequent quarter, allowing it to leave and later return. Among exposed American and United pairs with complete outsourcing, 31.6 percent are active in 2015Q2, compared with 73.2 percent on other routes. For mixed production the corresponding shares are 85.2 and 84.7 percent. The difference in subsequent activity is concentrated among pairs with complete outsourcing at baseline. Thirty-one of the 38 exposed pairs with complete outsourcing are United's routes at Cleveland.

\begin{table}[!tp]
\centering
\small
\singlespacing
\caption{Activity of the 2014Q1 Cohort by Inherited Organization}
\label{tab:app_republic_retention}
\begin{tabularx}{\textwidth}{@{}Xrrrrr@{}}
\toprule
Organization in 2014Q1 & Pairs & 2014Q2 & 2014Q3 & 2014Q4 & 2015Q2 \\
\midrule
\multicolumn{6}{@{}l}{\textit{American and United}} \\
Chautauqua routes, complete outsourcing & 38 & 60.5 & 36.8 & 36.8 & 31.6 \\
Chautauqua routes, mixed production & 61 & 88.5 & 86.9 & 86.9 & 85.2 \\
Other routes, complete outsourcing & 661 & 87.1 & 80.9 & 78.7 & 73.2 \\
Other routes, mixed production & 301 & 95.7 & 91.4 & 89.0 & 84.7 \\
\midrule
\multicolumn{6}{@{}l}{\textit{United at Cleveland}} \\
Chautauqua routes, complete outsourcing & 31 & 54.8 & 25.8 & 25.8 & 19.4 \\
Other routes, complete outsourcing & 17 & 76.5 & 29.4 & 29.4 & 17.6 \\
Chautauqua routes, mixed production & 6 & 83.3 & 66.7 & 66.7 & 66.7 \\
Other routes, mixed production & 15 & 100.0 & 80.0 & 80.0 & 80.0 \\
\bottomrule
\end{tabularx}
\par\smallskip
\begin{minipage}{\textwidth}
\footnotesize\justifying
\noindent\textit{Notes.} Entries are percentages of carrier-route pairs active in 2014Q1 that are active in the stated quarter. These pairs are incumbents at the 2014Q2 decision, including those that entered in 2014Q1. Missing subsequent records in the risk set are counted as inactive, including pairs on routes that lose all legacy service. Activity permits reentry and does not require uninterrupted service. Complete outsourcing denotes a 2014Q1 share of at least 0.99 and mixed production a share strictly between 0.01 and 0.99. Chautauqua exposure denotes positive passenger counts during 2013Q2 through 2014Q1. These groups summarize inherited organization near each corner. The structural model distinguishes exact corner allocations from interior allocations.
\end{minipage}
\end{table}

\subsubsection{Other Withdrawals of Independent Providers}
\label{app:supplier_events}

Operator records provide quarterly passenger counts for each independent provider and legacy carrier. We assign the ticketing codes of retired brands to the surviving carrier and use the ownership classification in Appendix~\ref{app:data} to identify independent operators. A withdrawal occurs when a provider's passengers for a carrier fall below 15 percent of their average over the preceding four quarters and remain below that threshold for four quarters. We exclude changes in provider ownership and mergers of operating certificates, since these can change the ownership classification or operator code without a withdrawal of service. We also exclude ExpressJet's flights for Continental in 2007Q2, when the operator field changes for most of those flights. Eleven withdrawals remain. We date Chautauqua's withdrawal to 2014Q2, as in Section~\ref{sec:republic}.

{\clubpenalty=10000\widowpenalty=10000 Table~\ref{tab:supplier_events} follows the carrier's pairs active in the quarter before each withdrawal. Exposed pairs are those on which the provider carried the carrier's passengers during the preceding four quarters. Carriers continued serving most exposed routes, usually purchasing replacement service from other independent providers. Among initially fully outsourced exposed pairs that remained active, the average outsourcing share three quarters after the withdrawal remained above 0.9. Service losses among initially fully outsourced pairs are concentrated in Colgan's and Mesaba's withdrawals from US Airways in 2006 and 2012 and Chautauqua's withdrawal from United in 2014.

The pooled comparison in Panel B includes fixed effects for each combination of withdrawal event and production arrangement. Each exposed pair is therefore compared with the same carrier's other routes that used the same production arrangement before the withdrawal. The placebo comparison records differences in activity for these relationships eight quarters before withdrawal, conditional on the same organization categories. For initially fully outsourced pairs, the placebo difference of $-7.4$ percentage points is close to the differences one and three quarters after withdrawal, so much of the lower activity on exposed routes predates the withdrawals. This pattern is consistent with carriers continuing to serve most exposed routes.\par}

\begin{table}[!tp]
\centering
\small
\singlespacing
\caption{Withdrawals of Independent Providers, 2004 to 2018}
\label{tab:supplier_events}
\begin{tabularx}{\textwidth}{@{}Xrrrrrr@{}}
\toprule
 & \multicolumn{3}{c}{Complete outsourcing} & \multicolumn{3}{c}{Mixed production} \\
\cmidrule(lr){2-4}\cmidrule(l){5-7}
Withdrawal & Exposed pairs & Exposed & Other & Exposed pairs & Exposed & Other \\
\midrule
\multicolumn{7}{@{}l}{\textit{Panel A. Percent of pairs active one quarter after the withdrawal}} \\
Atlantic Coast and United, 2004Q4 & 108 & 91.7 & 82.1 & 66 & 95.5 & 90.9 \\
Atlantic Coast and Delta, 2004Q4 & 0 &  & 100.0 & 56 & 96.4 & 82.0 \\
Air Wisconsin and United, 2006Q1 & 159 & 93.1 & 90.2 & 72 & 100.0 & 98.7 \\
Colgan and US Airways, 2006Q3 & 46 & 0.0 & 54.8 & 14 & 21.4 & 69.5 \\
Trans States and US Airways, 2009Q3 & 16 & 75.0 & 59.9 & 29 & 82.8 & 89.3 \\
Freedom and Delta, 2010Q4 & 3 & 66.7 & 69.8 & 116 & 94.0 & 91.2 \\
Mesaba and US Airways, 2012Q1 & 12 & 0.0 & 65.0 & 0 &  & 87.0 \\
Colgan and United, 2012Q4 & 43 & 90.7 & 70.2 & 52 & 84.6 & 84.4 \\
Chautauqua and United, 2014Q2 & 36 & 33.3 & 82.0 & 20 & 80.0 & 91.9 \\
Chautauqua and American, 2014Q2 & 10 & 60.0 & 48.6 & 59 & 83.1 & 83.9 \\
Air Wisconsin and American, 2018Q1 & 25 & 88.0 & 81.4 & 147 & 95.9 & 95.8 \\
\midrule
\multicolumn{7}{@{}l}{\textit{Panel B. Difference between exposed and other pairs (percentage points)}} \\
One quarter after & & $-$10.3 & & & 2.0 & \\
 & & (2.4) & & & (1.7) & \\
Three quarters after & & $-$9.6 & & & 2.9 & \\
 & & (2.7) & & & (2.0) & \\
Same relationships eight quarters earlier & & $-$7.4 & & & 1.9 & \\
 & & (2.9) & & & (2.0) & \\
\bottomrule
\end{tabularx}
\par\smallskip
\begin{minipage}{\textwidth}
\footnotesize\justifying
\noindent\textit{Notes.} A withdrawal is a quarter in which an independent provider's passengers for a legacy carrier fall below 15 percent of their average over the preceding four quarters and remain there for four quarters. Changes of provider ownership and mergers of operating certificates are excluded. The event cohorts use operator records with retired ticketing codes assigned to surviving carriers and follow the carrier's pairs active in the quarter before each withdrawal. The counts of exposed pairs refer to these cohorts, while Table~\ref{tab:app_republic_retention} follows the cohort in the game sample for the 2014 episode. Exposed pairs are those on which the provider carried the carrier's passengers during the preceding four quarters, and other pairs are the carrier's remaining pairs with the same organization in the quarter before the withdrawal. Complete outsourcing denotes a share of at least 0.99 and mixed production a share strictly between 0.01 and 0.99. Panel B reports coefficients from a linear probability model pooling the withdrawals, with fixed effects for each combination of withdrawal event and production arrangement and log passengers in the quarter before the withdrawal. The placebo row applies the same comparison to the relationships observed twelve to nine quarters before each withdrawal, with a hypothetical withdrawal eight quarters before the actual one. Parentheses report standard errors clustered by airport pair.
\end{minipage}
\end{table}

\FloatBarrier

\subsection{Estimation}
\label{app:estimation_algorithm}

\subsubsection{Choice Values and Natural Parameters}

The estimator must accommodate both discrete decisions, such as keeping an existing share, and the continuous choice of a new outsourcing share. We express the approximated choice values as an exponential family, which allows the score and information to be calculated from moments of these choices. At each observed state, subtract the value of inactivity. The resulting value of an interior adjustment is a polynomial of degree at most four in the chosen share. Retention has a separate value, and adjustments to the corners have separate continuation values. These values admit an eight-dimensional representation with statistics
\[
g\left(c\right)=\left(a,ao,ao^2,ak,ako^3,ako^4,ak\ind\left\{o=0\right\},ak\ind\left\{o=1\right\}\right)',
\]
where all coordinates are zero under inactivity. For natural parameters $\eta_{ie}\in\R^8$, the normalized choice value is $\eta_{ie}'g\left(c\right)$. The indicators for adjustments to the corners allow their continuation values to differ from those of interior allocations. The coefficient on $ak$ captures the difference between retention and adjustment, including the fixed adjustment cost $R_{ie}$.

Let $r$ be the normalized value of retention and let $\mathcal Q\left(o\right)=\sum_{h=0}^4\varrho_ho^h$ be the normalized value of an interior adjustment. Denote the values of feasible adjustments to zero and one by $r_0$ and $r_1$. Set $\eta_1=r-\varrho_1\ell-\varrho_2\ell^2$, $\eta_2=\varrho_1$, $\eta_3=\varrho_2$, $\eta_4=\varrho_0-\eta_1$, $\eta_5=\varrho_3$, and $\eta_6=\varrho_4$. If adjustment to zero is feasible, set $\eta_7=r_0-\varrho_0$, and otherwise set $\eta_7=0$. If adjustment to one is feasible, set $\eta_8=r_1-\sum_{h=0}^4\varrho_h$, and otherwise set $\eta_8=0$.

For retention, the inner product is $\eta_1+\varrho_1\ell+\varrho_2\ell^2=r$. For an interior adjustment, the first and fourth coordinates sum to $\varrho_0$, and the other polynomial coordinates give $\mathcal Q\left(o\right)$. The corner indicators recover $r_0$ and $r_1$ whenever those choices are feasible. When adjustment to a corner is infeasible, its adjustment indicator is zero throughout the support. The construction applies to every quartic interior value.

The quartic representation follows from the continuation basis and the state transition. After adjustment, the next inherited share is $o$ and duration is zero. Evaluating the tensor continuation basis at that duration leaves a polynomial of degree at most four in $o$. Retention leaves the share unchanged and advances duration by one, while scale is integrated out. Its normalized value remains an unrestricted scalar at the current state. Separate approximations at the two corners allow their values to differ from the interior polynomial.

Expected continuation values depend on inherited scale through the first and second conditional moments of next-quarter scale and the probability of a single sampled passenger, with separate coefficients for complete internal production, mixed production, and complete outsourcing. Within a production regime, these moments and the probability do not depend on the chosen share. Integration over scale therefore preserves the quartic representation. A higher polynomial degree adds the corresponding powers of the share chosen after adjustment to $g$, with the same construction for retention and the corners.

The log normalizing constant is $\mathcal{K}_{ie}\left(\eta\right)=\log\int\exp\left\{\eta'g\left(c\right)\right\}d\nu_\ell\left(c\right)$. For observed choice $c_{ie}$, the contribution is $\mathcal{L}_{ie}=\eta_{ie}'g\left(c_{ie}\right)-\mathcal{K}_{ie}\left(\eta_{ie}\right)$. The natural score and information of this contribution are
\begin{equation}
\mathcal{L}_{ie,\eta}=g\left(c_{ie}\right)-\E_f\left[g\right],\quad -\mathcal{L}_{ie,\eta\eta}=\operatorname{Var}_f\left(g\right).
\label{eq:natural_score}
\end{equation}
The moments integrate over the feasible intervals and sum over the atoms. Interior adjustments enter as densities evaluated at their recorded shares. For an always-active pair, the same expressions use $\mu_\ell$ and the unidentified common active level cancels.

Because the statistics are bounded and the reference measure is finite, differentiation under the integral is valid. If $Z\left(\eta\right)=\int\exp\left\{\eta'g\right\}d\nu_\ell$, then $Z_\eta=\int g\exp\left\{\eta'g\right\}d\nu_\ell$ and $Z_{\eta\eta}=\int gg'\exp\left\{\eta'g\right\}d\nu_\ell$. Differentiating $\log Z$ yields the mean and covariance of $g$, proving equation~\eqref{eq:natural_score}.

The variance in a given direction is zero if and only if the corresponding linear combination of statistics is constant almost everywhere on the choice support. Replacing each continuous integral by positive quadrature weights preserves these derivative identities for the approximated criterion.

\subsubsection{Backward Recursion and Value Derivatives}

Let $\zeta_m=\left(\vartheta',\phi_m'\right)'$ collect the common parameters and the activity effects on route $m$. Holding the first-stage rival probabilities and scale transition fixed, we solve equation~\eqref{eq:dynamic_mixed_choice} backward along the sequence of public states. Each observed sequence is extended by twenty quarters, keeping its final public covariates, calendar effects, year shifts of entry costs, era effects, rival policies, and brand availability. An added game belongs to an integration set $\mathcal{G}_i$ only when the final observed game of the sequence does. Own states continue to follow $T$, and the terminal value at the end of the extension is zero. These added quarters enter continuation values but make no contribution to the likelihood. Separate approximations at the corners preserve the discontinuity in fixed adjustment costs between fully internal or fully outsourced service and mixed production.

The derivative recursion follows directly from the log normalizing constant. For each parameter coordinate, $V_{ie,\zeta}=\E_f\left[W_{ie,\zeta}\right]$ and $W_{ie,\zeta}=u_{ie,\zeta}+\beta\E\left[V_{n\left(ie\right),\zeta}\left(T\left(s,c\right)\right)|s,c,X_e\right]$. The derivative of the terminal value is zero. Interpolation and integration over scale are linear in the fitted value coefficients, so the same backward pass propagates derivatives through the basis representation. Rival expectations and the first-stage scale transition remain fixed during this pass.

For every finite parameter vector, this recursion determines a unique conditional value function. The terminal value is fixed, and each preceding value is a finite log integral over feasible choices. Backward induction determines every earlier value, including those in the added quarters.

Inactivity sends the brand's state to $\left(0,0,0,0\right)$. Retention by an active incumbent advances duration, while entry or a production adjustment resets it to zero. The first-stage transition supplies the first and second conditional moments of next-quarter scale. We tabulate these moments on a grid of the predicted level separately for continuing brands, continuing brands with a single sampled passenger, and entrants. We apply the same recursion at the collocation states and interpolate after evaluating their log normalizing constants.

To express this calculation as an EPL update, stack choice-specific values at the fixed states and quadrature alternatives in $v_{mt}$. Let $\mathcal N_{m,t+1}$ map subsequent choice values to their log normalizing constants. Let $\mathcal M_{mt}$ be the fixed linear map from those normalizers to current expected continuation values, combining interpolation, evaluation at next states, and integration over the first-stage scale transition. Let $Z_{mt}$ contain the linear loadings of flow payoffs. The represented residual is $\mathcal R_{mt}=v_{mt}-Z_{mt}\zeta_m-\beta\mathcal M_{mt}\mathcal N_{m,t+1}\left(v_{m,t+1}\right)$, with terminal continuation zero. Holding $v_m$ fixed, this residual is affine in $\zeta_m$.

Stack these equations over the route's sequence to obtain $\mathcal R_m\left(\zeta_m,v_m\right)$. The Jacobian $\mathcal R_{m,v}$ is triangular in time with identity diagonal blocks and is therefore nonsingular. Differentiating $\mathcal R_m\left(\zeta_m,v_m\left(\zeta_m\right)\right)=0$ gives $v_{m,\zeta}=-\mathcal R_{m,v}^{-1}\mathcal R_{m,\zeta}$. Since the residual is affine in the trial payoff vector, at an anchor $\left(\zeta_{mk},v_{mk}\right)$ with $\mathcal R_m\left(\zeta_{mk},v_{mk}\right)=0$,
\begin{equation}
\widetilde v_m\left(\zeta_m\right)=v_{mk}-\mathcal R_{m,v}^{-1}\mathcal R_m\left(\zeta_m,v_{mk}\right)=v_{mk}+v_{m,\zeta}\left(\zeta_m-\zeta_{mk}\right).
\label{eq:epl_value_linearization}
\end{equation}
The last equality uses $v_{m,\zeta}=-\mathcal R_{m,v}^{-1}\mathcal R_{m,\zeta}$. The backward derivatives therefore give the pseudo-regressors required by the Newton transformation of the conditional value equations. As in \citet{KohYoon2026}, a route's values have zero derivatives with respect to another route's effects. Extending the sequence of public states and combining discrete and continuous production choices preserve this property.

A change in $\phi_m$ leaves terminal values and flow payoffs on every other route unchanged. Backward induction therefore leaves every other route's values unchanged. Equivalently, $\mathcal R_v$ and the columns of $\mathcal R_\zeta$ for local effects are block diagonal by route, and inversion of $\mathcal R_v$ preserves this structure. The identity in equation~\eqref{eq:epl_value_linearization} is exact for the fixed collocation and quadrature representation. Evaluation at an observed interior share uses the same continuation derivative and its own exact flow payoff loading.

At the anchor, the affine approximation matches the dynamic model's values and first derivatives and therefore gives the same likelihood score. Its conditional logit information follows from the curvature decomposition below. For the exact natural parameters $\eta\left(\zeta\right)$, let $J=\partial\eta/\partial\zeta'$ and $e=g\left(c\right)-\E_f\left[g\right]$. Twice differentiating a likelihood contribution gives
\begin{equation}
-\mathcal L_{\zeta\zeta}=J'\operatorname{Var}_f\left(g\right)J-\sum_{h=1}^8e_h\eta_{h,\zeta\zeta}.
\label{eq:structural_information}
\end{equation}
The first term is the IRLS information. The second captures the curvature of continuation values with respect to the parameters, including the fixed effects. Equation~\eqref{eq:structural_information} follows by differentiating $J'e$ and using $\partial e/\partial\zeta'=-\operatorname{Var}_f\left(g\right)J$. Under the specified conditional choice distribution, the curvature term has zero conditional mean.

\subsubsection{Iteratively Reweighted Least Squares and Fixed Effects}
\label{app:irls}

Within an outer iteration, the approximated natural parameters are affine in the trial common parameters and route effects. Stack their derivatives on route $m$ in matrices $X_m$ and $C_m$. Let $e_m$ stack the natural scores in equation~\eqref{eq:natural_score}, and let $\mathcal W_m$ contain the corresponding covariance blocks. Define $s_{\vartheta m}=X_m'e_m$, $s_{\phi m}=C_m'e_m$, $H_m=X_m'\mathcal W_mX_m$, $B_m=C_m'\mathcal W_mX_m$, and $A_m=C_m'\mathcal W_mC_m$. The IRLS Newton equations are $\sum_mH_m\Delta\vartheta+\sum_mB_m'\Delta\phi_m=\sum_ms_{\vartheta m}$ and $B_m\Delta\vartheta+A_m\Delta\phi_m=s_{\phi m}$.

When the local blocks and the profiled common information are nonsingular, the exact solution is
\begin{equation}
\Delta\vartheta=\mathcal{S}^{-1}\sum_m\left(s_{\vartheta m}-B_m'A_m^{-1}s_{\phi m}\right),\quad \Delta\phi_m=A_m^{-1}\left(s_{\phi m}-B_m\Delta\vartheta\right),
\label{eq:schur_update}
\end{equation}
where $\mathcal{S}=\sum_m\left(H_m-B_m'A_m^{-1}B_m\right)$. Solving the local equation for $\Delta\phi_m$ and substituting into the common equation proves equation~\eqref{eq:schur_update}. Block elimination therefore gives the same increments as solving the full joint IRLS system. Every local block has dimension at most seven, the number of legacy brands in the sample. We can therefore estimate the route effects without inverting a matrix whose dimension equals the total number of these effects. At a fixed common dimension and bounded route history, storage and arithmetic per iteration grow linearly with the number of routes.

The block decomposition also gives the rank condition for joint estimation. For a common direction $b$ and local directions $c_m$, completing the square gives
\[
\sum_m\left(X_mb+C_mc_m\right)'\mathcal W_m\left(X_mb+C_mc_m\right)=b'\mathcal S b+\sum_m\left(c_m+A_m^{-1}B_mb\right)'A_m\left(c_m+A_m^{-1}B_mb\right).
\]
If every $A_m$ and $\mathcal S$ is positive definite, the full information is positive definite. Conversely, full positive definiteness implies that every principal block $A_m$ is positive definite. Taking $c_m=-A_m^{-1}B_mb$ then proves positive definiteness of $\mathcal S$. This equivalence gives the rank condition after the calendar and carrier normalizations. Joint identification requires full rank of the combined parameter design, even when individual covariance blocks of the natural statistics are singular.

To obtain the IRLS update, expand the conditional criterion in parameter increments. Its quadratic approximation has linear term $\sum_ms_{\vartheta m}'b+\sum_ms_{\phi m}'c_m$ minus one half of the left side of the preceding identity. Minimizing the negative approximation first over $c_m$ gives $c_m=A_m^{-1}\left(s_{\phi m}-B_mb\right)$. Minimizing the remaining quadratic in $b$ gives equation~\eqref{eq:schur_update}. At the same accepted step length, block elimination and the full system produce the same iterate. This equivalence also holds for a damped system when the same damping is included in the blocks before elimination.

The working log likelihood is concave because each negative Hessian in equation~\eqref{eq:natural_score} is a covariance matrix and the working indices are affine. The working score and the dynamic likelihood score coincide at the anchor because equation~\eqref{eq:epl_value_linearization} matches both values and first derivatives there.

\subsection{Inference}
\label{app:inference}

Partition the common parameter vector as $\vartheta=\left(\theta',\upsilon'\right)'$, where $\theta$ contains the coefficients to be corrected and $\upsilon$ contains the calendar effects, the year shifts $\kappa_{E,y\left(t\right)}$, and the era shifts $\gamma_{1,\mathit{era}\left(t\right)}$. Let $\widehat\vartheta=\left(\widehat\theta',\widehat\upsilon'\right)'$ be the full-sample estimate. Each route's quarters are split at its median, and $\widehat\theta_h$ is estimated on half $h\in\left\{1,2\right\}$ with $\upsilon=\widehat\upsilon$. Each half re-estimates the first-stage choice probabilities with $\upsilon$ at its full-sample value, estimates carrier-by-route effects for pairs observed with and without service in that half, and evaluates continuation values along the full sequence of public states.

The split-panel jackknife is $\widehat\theta^{SPJ}=2\widehat\theta-\left(\widehat\theta_1+\widehat\theta_2\right)/2$ \citep{DhaeneJochmans2015}. We combine these corrected coefficients with the full-sample estimate $\widehat\upsilon$ of the calendar, year, and era effects. For the counterfactuals, we re-estimate the carrier-by-route effects conditional on the resulting coefficient vector.

The panel expansion shows how the correction removes the leading bias. Let $T$ index panel length, absorbing relative route lengths into the bias coefficients, and write $\widehat\theta-\theta_0=B/T+Z+r$ and $\widehat\theta_h-\theta_0=2B_h/T+Z_h+r_h$. The terms $Z$ and $Z_h$ are the leading centered sampling components, including the estimated time effects shared by the three fits. Substitution gives
\[
\widehat\theta^{SPJ}-\theta_0=\frac{2B-B_1-B_2}{T}+2Z-\frac{Z_1+Z_2}{2}+2r-\frac{r_1+r_2}{2}.
\]
If the leading half-panel biases satisfy $B_1+B_2=2B$, the term of order $1/T$ cancels. The covariance calculation below retains the combined sampling term $2Z-\left(Z_1+Z_2\right)/2$.

Let $\iota=\left(i,m,t\right)$ index an observed decision. For each fitted criterion, profile the carrier-by-route effects using the blocks in Appendix~\ref{app:irls}. At the full-sample fit, the common score contribution is $\widetilde s_\iota=s_{\vartheta,\iota}-B_m'A_m^{-1}s_{\phi,\iota}$, and its influence based on the conditional information is $\psi_\iota^F=\widehat{\mathcal S}^{-1}\widetilde s_\iota$. Partition this vector into $\psi_{\theta,\iota}^F$ and $\psi_{\upsilon,\iota}^F$.

At half-panel fit $h$, let $\widehat{\mathcal S}_{h,\theta\theta}$ and $\widehat{\mathcal S}_{h,\theta\upsilon}$ denote the corresponding profiled information blocks, and let $\widetilde s_{h,\theta,\iota}$ be the profiled score for $\theta$. This contribution is zero for decisions outside half $h$. Define
\[
\psi_{h,\iota}^0=\widehat{\mathcal S}_{h,\theta\theta}^{-1}\widetilde s_{h,\theta,\iota},\quad D_h=\widehat{\mathcal S}_{h,\theta\theta}^{-1}\widehat{\mathcal S}_{h,\theta\upsilon}.
\]
All blocks are evaluated at the corresponding full-sample or half-panel fit on the identified parameter coordinates.

The influence vector for the corrected coefficients is
\begin{equation}
\psi_{\theta,\iota}^{SPJ}=2\psi_{\theta,\iota}^F-\frac{\psi_{1,\iota}^0+\psi_{2,\iota}^0}{2}+\frac{D_1+D_2}{2}\psi_{\upsilon,\iota}^F.
\label{eq:spj_influence}
\end{equation}
To derive the exact local response, use the observed negative Hessian of the recomputed dynamic log likelihood. Hold the fitted first-stage rival probabilities and scale transition fixed. Let $A_m^O$, $B_m^O$, and $H_m^O$ denote the local, cross, and common blocks of that Hessian, and let $\mathcal S_F^O=\sum_m\left(H_m^O-B_m^{O\prime}\left(A_m^O\right)^{-1}B_m^O\right)$. Suppose the fitted score equations have a differentiable local solution and the local blocks and profiled common block are nonsingular on the identified coordinates. For observation weights $w_\iota$, define the full-sample influence response based on the observed Hessian by $d\widehat\vartheta=\sum_\iota\mathrm{IF}_{\iota}^{F,O}dw_\iota$. Differentiating the full-sample score equations at unit weights and eliminating the local effects gives
\[
\mathrm{IF}_{\iota}^{F,O}=\left(\mathcal S_F^O\right)^{-1}\left\{s_{\vartheta,\iota}-B_m^{O\prime}\left(A_m^O\right)^{-1}s_{\phi,\iota}\right\}.
\]
Partition $\mathrm{IF}_{\iota}^{F,O}$ into its $\theta$ and $\upsilon$ coordinates. In half $h$, the same Schur elimination defines the observed blocks $\mathcal S_{h,\theta\theta}^O$ and $\mathcal S_{h,\theta\upsilon}^O$. Apply the same nonsingularity conditions to the half-panel local blocks and profiled $\theta$ block. Define $\mathrm{IF}_{h,\iota}^{0,O}$ by this elimination with $\upsilon$ fixed, and define $D_h^O=\left(\mathcal S_{h,\theta\theta}^O\right)^{-1}\mathcal S_{h,\theta\upsilon}^O$.

Differentiation of the half-panel score equations gives $d\widehat\theta_h=\sum_\iota\mathrm{IF}_{h,\iota}^{0,O}dw_\iota-D_h^Od\widehat\upsilon$. The negative sign follows from the cross derivative of the score, $-\mathcal S_{h,\theta\upsilon}^O$.

Substituting these expressions into the differential of the split-panel estimator gives the exact local weight response $2\mathrm{IF}_{\theta,\iota}^{F,O}-\left(\mathrm{IF}_{1,\iota}^{0,O}+\mathrm{IF}_{2,\iota}^{0,O}\right)/2+\left(D_1^O+D_2^O\right)\mathrm{IF}_{\upsilon,\iota}^{F,O}/2$. Replacing the observed negative Hessian blocks by their conditional information counterparts yields equation~\eqref{eq:spj_influence}. The positive final term accounts for the full-sample time effects shared by both half-panel estimates.

Stack $\psi_{\theta,\iota}^{SPJ}$ with $\psi_{\upsilon,\iota}^F$ to obtain the influence vector $\psi_\iota^{SPJ}$ for the reported coefficients. Let $\bar m$ denote the airport pair that contains route $m$ and its reverse. Let $\Psi_{\bar m}=\sum_{\iota\in\bar m}\psi_\iota^{SPJ}$, $\Psi_t=\sum_{\iota\in t}\psi_\iota^{SPJ}$, and $\Psi_{\bar mt}=\sum_{\iota\in\left(\bar m,t\right)}\psi_\iota^{SPJ}$. The two-way covariance is
\begin{equation}
\widehat V^{SPJ}=\sum_{\bar m}\Psi_{\bar m}\Psi_{\bar m}'+\sum_t\Psi_t\Psi_t'-\sum_{\bar m,t}\Psi_{\bar mt}\Psi_{\bar mt}'.
\label{eq:spj_cluster_covariance}
\end{equation}
Equation~\eqref{eq:spj_cluster_covariance} allows serial dependence within an airport pair and common dependence within a quarter \citep{CameronGelbachMiller2011}. Subtracting the intersection term removes the duplicate pair-quarter contribution. Applying the same aggregation to $\psi_\iota^F$ gives the covariance of the uncorrected estimator.

The reported covariance uses conditional information in this score expansion. By equation~\eqref{eq:structural_information}, the observed negative Hessian differs from conditional information by a term linear in $e_\iota$. Under the specified conditional choice distribution, $\E\left[e_\iota|s_\iota,X_\iota\right]=0$, so this curvature term has zero conditional mean. Conditional information gives the Fisher scoring form of the first-order covariance calculation; the exact local response to observation weights uses the observed Hessian blocks defined above. Both calculations condition on fitted first-stage rival probabilities and the fitted scale transition. The standard errors in Table~\ref{tab:game_estimates} use conditional information for the corrected estimator. Counterfactual standard errors use the same covariance.

Table~\ref{tab:entry_years} reports the year shifts of entry costs in both specifications of Table~\ref{tab:game_estimates}. The half panels hold these shifts at their full-sample values, and their standard errors come from equation~\eqref{eq:spj_cluster_covariance}.

\begin{table}[!tp]
\centering
\singlespacing
\caption{Year Shifts of Entry Costs}
\label{tab:entry_years}
\small
\begin{tabularx}{\textwidth}{@{}X*{4}{>{\raggedleft\arraybackslash}p{1.9cm}}@{}}
\toprule
 & \multicolumn{2}{c}{Benchmark} & \multicolumn{2}{c}{No simultaneous rival terms} \\
\cmidrule(lr){2-3}\cmidrule(l){4-5}
Year & Estimate & SE & Estimate & SE \\
\midrule
1999 & 0.391 & 0.105 & 0.391 & 0.098 \\
2000 & 0.190 & 0.099 & 0.196 & 0.098 \\
2001 & 0.185 & 0.106 & 0.182 & 0.104 \\
2002 & 0.638 & 0.254 & 0.642 & 0.246 \\
2003 & 0.754 & 0.273 & 0.773 & 0.259 \\
2004 & 0.591 & 0.211 & 0.633 & 0.204 \\
2005 & 0.599 & 0.235 & 0.632 & 0.230 \\
2006 & 0.609 & 0.272 & 0.656 & 0.265 \\
2007 & 0.636 & 0.241 & 0.688 & 0.238 \\
2008 & 0.773 & 0.254 & 0.838 & 0.249 \\
2009 & 0.807 & 0.230 & 0.870 & 0.221 \\
2010 & 0.310 & 0.231 & 0.345 & 0.234 \\
2011 & 0.600 & 0.287 & 0.661 & 0.282 \\
2012 & 0.580 & 0.302 & 0.676 & 0.295 \\
2013 & 0.821 & 0.249 & 0.906 & 0.251 \\
2014 & 0.378 & 0.287 & 0.433 & 0.284 \\
2015 & 0.284 & 0.251 & 0.363 & 0.260 \\
2016 & 1.357 & 0.337 & 1.453 & 0.354 \\
2017 & 1.679 & 0.345 & 1.767 & 0.365 \\
2018 & 2.730 & 0.319 & 2.834 & 0.366 \\
2019 & 2.865 & 0.489 & 2.950 & 0.530 \\
\bottomrule
\end{tabularx}
\par\smallskip
\begin{minipage}{\textwidth}
\footnotesize
\justifying
\noindent\textit{Notes.} Entries are the year shifts $\kappa_{E,y\left(t\right)}$ of the entry cost relative to 1998, in logit disturbance units. Standard errors apply equation~\eqref{eq:spj_cluster_covariance} to the full-sample influence function, clustering by airport pair and calendar quarter. The specifications are those of Table~\ref{tab:game_estimates}.
\end{minipage}
\end{table}

\section{Counterfactuals and Robustness}
\label{app:counterfactuals_robustness}

\subsection{Counterfactual Construction}
\label{app:policy_merger_design}

\subsubsection{Counterfactual Aggregation and Consumer Valuation}
\label{app:counterfactual_valuation}

Participation counts each carrier separately on each route and in each quarter. A decline can therefore reflect a carrier's withdrawal from a route on which its rivals continue to operate. Let $\mathcal E$ contain the observed route-quarter games, $\mathcal I_e$ the eligible brands in game $e$, and $P_{ie}^c$ the probability that brand $i$ serves that route in experiment $c$. In the production experiments, $P_{ie}^c$ is the average of simulated activity indicators across draws. Expected participation per 100 games is $A_{100}^c=100\left|\mathcal E\right|^{-1}\sum_{e\in\mathcal E}\sum_{i\in\mathcal I_e}P_{ie}^c$. These experiments use the same observed games, eligible brands, and initial production arrangements, and allow subsequent states to evolve with simulated choices. The merger comparisons instead condition on observed state histories.

Entry and continuation in the production experiments depend on activity in the preceding quarter of each simulated path. Let $N_E^c$ and $N_C^c$ denote the numbers of inactive and active preceding states, averaged over draws, and let $p_E^c$ and $p_C^c$ denote the corresponding entry and continuation rates pooled across simulated decisions and draws, in probability units. Partitioning decisions by preceding activity gives $A_{100}^c=100\left(N_E^cp_E^c+N_C^cp_C^c\right)/\left|\mathcal E\right|$. An experiment can change both the entry and continuation rates and the number of decisions in each group.

In the merger comparisons, the conditioning histories are fixed, so $N_E$ and $N_C$ are common across experiments and $\Delta A_{100}=100\left(N_E\Delta p_E+N_C\Delta p_C\right)/\left|\mathcal E\right|$.

The mean outsourcing share among active carrier-route pairs weights each pair's conditional mean by the probability that the carrier serves the route. Writing $m_{ie}^{o,c}=\E\left[o_{ie}|a_{ie}=1,c\right]$, the mean is $\sum_{ie}P_{ie}^cm_{ie}^{o,c}/\sum_{ie}P_{ie}^c$. It therefore changes both when continuing carriers reallocate production and when carriers enter or leave routes. Active pairs receive equal weight regardless of their passenger volumes.

For the conditional price calculations, we multiply each observed product's outsourcing share by $m_{ie}^{o,c}/m_{ie}^{o,0}$. Under an outsourcing limit, we cap the result at $1-\tau$. Products outside the game sample, and products with $m_{ie}^{o,0}$ below 0.005, retain their observed shares subject to the same cap. We hold baseline demand and cost shifters and dollar pricing residuals fixed. The outsourcing cost increase changes the dynamic payoff but does not directly change passenger marginal cost; conditional fares respond through the resulting production mix. Quantities respond through both within-route shares and route traffic. Reported fare effects use passenger weights on the observed product set, with the active set held fixed.

To measure the participation component of consumer valuation, let $L_{ie}$ denote the decline in route valuation when brand $i$ withdraws in game $e$ and the remaining fares adjust. For a brand inactive in that game, we use its mean contribution in active quarters on the same route. The first-order participation component is the annualized sum of $\left(P_{ie}^c-P_{ie}^0\right)L_{ie}$, with each brand's contribution evaluated in the baseline environment. For the corresponding variable-profit component, we replace $L_{ie}$ by the decline in route variable profit, including the change in rivals' profits after fares adjust. Table~\ref{tab:policy} reports this participation component separately from the conditional price and demand component.

Each calculation solves the nested-logit pricing equations using the effective passenger costs $\widetilde{mc}$ of Appendix~\ref{app:demand}, which reproduce observed fares. A withdrawn brand is removed from the consumer's choice set. For $\alpha>1$, the consumer valuation index is $Q\mathcal{P}/\left(\alpha-1\right)$, where $\mathcal P$ is the route price index defined in Appendix~\ref{app:demand}. Changes in fares, outsourcing shares, and available brands all affect this valuation.

\subsubsection{Production Experiments and Equilibrium}

The hypothetical minimum internal share requires a carrier to carry at least share $\tau$ of its passengers on each route on flights that it or a wholly owned regional subsidiary operates. The outsourcing share is restricted to $\left[0,1-\tau\right]$, where $\tau\in\left[0,1\right]$. We also refer to $\tau$ as the internal-share floor. This experiment changes which production choices are available; it does not reproduce the loss of a particular supplier. The active choice measure $\mu_\ell^\tau$ is the restriction of $\mu_\ell$ to choices satisfying $o\leq1-\tau$. The unit atom at internal production represents retention when $\ell=0$ and adjustment otherwise. A brand can retain its inherited allocation only if $\ell\leq1-\tau$, and complete outsourcing remains available only if $\tau=0$. The continuous component retains Lebesgue measure on the feasible interior intervals in $\left(0,1-\tau\right]$. For $\tau=1$, the active measure reduces to the atom on internal production, reached by retention when $\ell=0$ and by adjustment otherwise.

Payoffs are those of equation~\eqref{eq:game_payoff} at the implemented shares, and the fixed adjustment cost depends on the inherited share and its duration exactly as in estimation. A brand with $\ell>1-\tau$ therefore pays $R_{ie}$ for leaving its inherited arrangement, plus the quadratic cost of the change, or becomes inactive. Changes of five percentage points or less within a regime count as retention. If an inherited mixed share exceeds $1-\tau$ by at most five percentage points, continued service requires either a reduction of more than five points within mixed production or a move to internal production. For $0<\tau<1$, an allocation of $1-\tau$ is mixed production in later quarters.

We solve values backward along each route's observed sequence of public states, using the restricted measure, given moments of rivals' choices, and the twenty continuation quarters used in estimation. Every outsourcing limit begins in 1998Q2 on every route, and elapsed quarters are counted from 1998Q2. The forward simulation starts each carrier-route pair from its observed state $\left(a_{i,e-1},\ell_{ie},d_{ie},z_{ie}\right)$ in its first game. When retained games resume after a gap, each brand continues from its last simulated state, and the duration of an active arrangement grows by the length of the gap. For a brand whose first game comes after 1998Q2, every experiment, including the baseline, removes the fixed adjustment cost from its adjustment choices in that game.

In each game, we draw an action from equation~\eqref{eq:dynamic_mixed_choice} at the brand's simulated state. The possible actions are inactivity, retention, adjustment to a corner, and an interior allocation drawn from the conditional continuous distribution. The state then evolves according to $T$. We use the same 200 uniform draws for each carrier and route in every experiment. The baseline and counterfactual therefore start from the same production arrangements, with differences arising from the altered choices. Reported outcomes average simulated activity, entry, continuation, outsourcing, and adjustment over draws and observed games. The continuation quarters enter the values but not the reported averages.

In the benchmark equilibrium, every brand re-optimizes. The moments of rival activity and outsourced activity used in each brand's decisions must equal the averages generated by its rivals' choices. We alternate between solving choices given these moments and simulating the resulting activity and outsourcing until the moments agree. The comparison with fixed rival behavior instead holds them at the values implied by the first-stage choice probabilities.

The reorganization exercise in Section~\ref{sec:reorganization} makes it less costly for a fully outsourced incumbent to begin operating internally when the minimum internal share takes effect. It adds $\theta_I=-2.552$ to the fixed adjustment cost of the incumbent's inherited arrangement. This gives the arrangement the fixed-cost discount estimated for mixed production while retaining its inherited share, duration, and quadratic adjustment cost. Complete outsourcing is unavailable under a positive internal-share floor. The cost reduction therefore applies to fully outsourced arrangements inherited in 1998Q2 and leaves subsequent production opportunities unchanged.

The outsourcing cost increase in Section~\ref{sec:network} replaces $\gamma_1$ by $\gamma_1-\chi$ in every active flow payoff, leaving all production arrangements available. It raises the recurring cost of independent production relative to internal production by $\chi o$ in the game's normalized payoff units. We choose $\chi$ to approximately match the unweighted mean outsourcing share among active carrier-route pairs under the corresponding internal-share floor. This mean is the ratio of expected outsourced activity to expected activity defined in Appendix~\ref{app:counterfactual_valuation}; the calibration is a comparison of experimental magnitudes, rather than an estimate of a historical supplier price increase. The experiment uses the same baseline, initial states, simulation design, and draws. Conditional on these initial arrangements, the cost increase is unanticipated at the first simulated decision. Carriers then expect the higher cost to persist throughout the simulated and continuation quarters. Backward solution incorporates the higher cost of future outsourcing as well as current outsourcing.

To separate the average cost burden from the incentive to outsource, a further experiment adds a common payment to every active flow payoff. The payment equals $\chi$ times the mean outsourcing share among active carrier-route pairs in the experiment with the payment. Each active carrier-route pair receives the same amount, regardless of its own share. This compensates the average modeled cost increase conditional on activity while preserving the additional cost of outsourcing at the margin. The payment requires outside funding, whose cost is not included in the model.

{\clubpenalty=10000\widowpenalty=10000 The disruption in Section~\ref{sec:disruption} begins in calendar quarter $t_s$ and lasts $H$ quarters. During the observed route-quarter decisions from $t_s$ through $t_s+H-1$, every active brand must use an outsourcing share of zero. Before $t_s$, values and choices are those of the experiment without the disruption, so brands do not anticipate it. At $t_s$, they learn when independent production will return. We solve values backward using the disrupted choice measure during the interruption and the pre-existing outsourcing limit afterward. The continuation quarters of each route sequence are not disrupted. In equilibrium, rivals' moments from $t_s$ on solve the fixed point under the disruption; before $t_s$, they remain at the equilibrium without it. Both simulations use the same draws and thus coincide before $t_s$. We group losses by each brand's simulated production arrangement at $t_s$, using route sequences that began before the disruption.\par}

Counterfactual standard errors use 180 draws $\vartheta^{\left(r\right)}$ from a normal distribution with mean equal to the reported coefficient estimate and covariance given by equation~\eqref{eq:spj_cluster_covariance}. For each draw, we re-estimate the carrier-by-route effects given $\vartheta^{\left(r\right)}$, holding the first-stage rival probabilities and scale transition at their estimates. We then resolve each counterfactual using the same simulation draws and average its outcomes over the observed games. This varies the parameters while holding the evaluation sample fixed. Standard errors are standard deviations across parameter draws. Statistics expressed relative to an estimated effect use the delta method, including the covariance of the numerator and denominator across draws.

\subsubsection{Static Quantities Used in the Counterfactuals}
\label{app:static_quantities}

The counterfactuals use the demand and cost estimates in Table~\ref{tab:current_point_estimates} and passenger counts expanded by the inverse sampling rate. At observed fares, the products covered by these calculations carry about 59 million passengers and generate \$15.7 billion of revenue per year. All reported dollar values refer to these passengers. Minimum and median passenger marginal costs are \$27.30 and \$182.06. Revenue less modeled passenger marginal cost is 34.09 percent of revenue. Using effective passenger cost $\widetilde{mc}$, which includes the dollar pricing residual, gives 33.19 percent. The counterfactual tables define passenger variable profit using this effective cost. Fixed operating, organizational, and entry costs are separate, as is the recurring outsourcing cost increase in the dynamic game.

The outsourcing share that minimizes passenger marginal cost depends on standardized log inherited passenger volume. Its median is 0.3787, with 10th and 90th percentiles of 0.2182 and 0.5963. All product states have cost-minimizing shares between 0.01 and 0.99. Counterfactual pricing retains product residuals to reproduce baseline fares. Their mean absolute value is \$34.90. Appendix~\ref{app:demand} gives the demand derivatives and the consumer valuation index used in these calculations.

\subsection{Common Ownership}
\label{app:common_ownership}

We use the three legacy mergers to estimate how common ownership changes the incentive to retain overlapping service. The conduct parameter places a weight on the bilateral payoff that one brand's choices generate for its merger partner. We first define this parameter in the estimated game, then apply the estimate to a hypothetical combination of American and United.

\subsubsection{Conduct and Service Incentives}
\label{app:merger_incentives}

For merger incentives, suppress the game index and define the activity vector $\mathbf a_i=\left(a_i,a_io_i\right)'$ and the interaction matrix $\boldsymbol\rho=\left[\begin{matrix}\rho_{AA}&\rho_{AO}\\\rho_{OA}&\rho_{OO}\end{matrix}\right]$. Brand $i$'s bilateral flow payoff from brand $j$ is $\mathbf a_i'\boldsymbol\rho\mathbf a_j$. Let $\lambda$ be the weight on brand $j$'s bilateral payoff in brand $i$'s decision. On the common payoff scale maintained in the game, the bilateral component of brand $i$'s objective is
\begin{equation}
\mathbf a_i'\boldsymbol\rho\mathbf a_j+\lambda\mathbf a_j'\boldsymbol\rho\mathbf a_i=\mathbf a_i'\left(\boldsymbol\rho+\lambda\boldsymbol\rho'\right)\mathbf a_j.
\label{eq:merger_internalization}
\end{equation}
The equality follows by transposing the second term, which is a scalar. At $\lambda=0$, the interaction is that estimated between independent brands; at $\lambda=1$, the objective includes both bilateral flow payoffs. The experiments use $\lambda=-0.537$, estimated from the legacy mergers.\footnote{Merger standard errors also account for uncertainty in $\lambda$. The calculation uses the standard error clustered by airport pair and scales each merger effect in proportion to $\lambda$.} Interactions with other brands retain $\boldsymbol\rho$. Write $\boldsymbol\rho^\lambda=\boldsymbol\rho+\lambda\boldsymbol\rho'$ and denote its entries by $\rho_{AA}^\lambda$, $\rho_{AO}^\lambda$, $\rho_{OA}^\lambda$, and $\rho_{OO}^\lambda$. Explicitly,
\[
\boldsymbol\rho^\lambda=\left[\begin{matrix}\left(1+\lambda\right)\rho_{AA}&\rho_{AO}+\lambda\rho_{OA}\\\rho_{OA}+\lambda\rho_{AO}&\left(1+\lambda\right)\rho_{OO}\end{matrix}\right].
\]
For given partner expectations $m_j^A=\E\left[a_j\right]$ and $m_j^{AO}=\E\left[a_jo_j\right]$, equation~\eqref{eq:merger_internalization} therefore contributes
\[
a_i\left\{\rho_{AA}^\lambda m_j^A+\rho_{AO}^\lambda m_j^{AO}+o_i\left[\rho_{OA}^\lambda m_j^A+\rho_{OO}^\lambda m_j^{AO}\right]\right\}
\]
to the incentive for brand $i$'s choice. Components of the partner's payoff that are constant across brand $i$'s choices cancel from the comparison. Both the reference experiment and the experiment at $\lambda\neq0$ recompute the two brands' policies under the same response rule for other brands. The effect at $\lambda=0$ is consequently zero.

For the historical comparison, let $C_e$ indicate common ownership of a merging pair in game $e$ before retirement of the acquired brand. The specification using only quarters of common ownership estimates a single coefficient $\lambda$, which measures conduct during common ownership. Let $\lambda_0$ denote a common baseline component for the three pairs. The specification using observations before and after closing replaces the bilateral internalization coefficient by $\lambda_0+\lambda C_e$, so $\lambda$ measures its change at closing. The post-closing coefficient is $\lambda_0+\lambda=-0.535$, close to the estimate of $-0.537$ from the quarters of common ownership alone.

We estimate each specification by maximizing the likelihood of the merging pairs' observed choices, conditional on the estimated payoff parameters, route effects, and other brands' first-stage policies. Standard errors are clustered by airport pair and condition on the fitted structural model.

\subsubsection{Historical Mergers}
\label{app:merger_data}

Closing is the first quarter in which the target brand is recorded under the acquirer's ownership. The closing quarters are 2008Q4 for Delta and Northwest, 2010Q4 for United and Continental, and 2013Q4 for American and US Airways. The last quarters with the acquired brands in the sample are 2010Q1, 2011Q4, and 2015Q2, respectively. Retirement is the first quarter after that last appearance.

The Federal Aviation Administration issued single operating certificates to the combined carriers in 2009Q4, 2011Q4, and 2015Q2. Integration, which defines the set $\mathcal{G}_i$ in Section~\ref{sec:long_run_mpe}, begins in the quarter of the operating certificate and ends eight quarters after retirement. These definitions follow the ownership and brand identifiers used in the panel.

Table~\ref{tab:past_mergers} follows routes served by both merging brands immediately before closing. About two years after brand retirement, the average number of active legacy brands ranges from 1.2 to 1.3, compared with 2.3 to 2.8 before closing. The merged group still serves 71 to 90 percent of these routes.

\begin{table}[!tp]
\centering
\caption{Route Consolidation after the Legacy Mergers}
\label{tab:past_mergers}
\small
\begin{tabularx}{\textwidth}{@{}Xrrrr@{}}
\toprule
Quarter & Legacy brands & Both merging & Merging group & Other legacy \\
& active & brands active & present & brands active \\
\midrule
\multicolumn{5}{@{}l}{\textit{Delta and Northwest. 47 overlapping routes}} \\
Before closing, 2008Q3 & 2.57 & 1.00 & 1.00 & 0.57 \\
Closing plus two quarters & 1.51 & 0.34 & 0.70 & 0.47 \\
Before retirement, 2010Q1 & 1.57 & 0.38 & 0.74 & 0.45 \\
Retirement plus three quarters & 1.23 & 0.00 & 0.79 & 0.45 \\
Retirement plus seven quarters & 1.26 & 0.00 & 0.77 & 0.49 \\
\midrule
\multicolumn{5}{@{}l}{\textit{United and Continental. 86 overlapping routes}} \\
Before closing, 2010Q3 & 2.26 & 1.00 & 1.00 & 0.26 \\
Closing plus two quarters & 1.66 & 0.66 & 0.76 & 0.24 \\
Before retirement, 2011Q4 & 1.22 & 0.51 & 0.56 & 0.15 \\
Retirement plus three quarters & 1.12 & 0.00 & 0.88 & 0.23 \\
Retirement plus seven quarters & 1.17 & 0.00 & 0.90 & 0.28 \\
\midrule
\multicolumn{5}{@{}l}{\textit{American and US Airways. 156 overlapping routes}} \\
Before closing, 2013Q3 & 2.78 & 1.00 & 1.00 & 0.78 \\
Closing plus two quarters & 2.09 & 0.56 & 0.85 & 0.69 \\
Before retirement, 2015Q2 & 1.76 & 0.38 & 0.78 & 0.60 \\
Retirement plus three quarters & 1.30 & 0.00 & 0.79 & 0.51 \\
Retirement plus seven quarters & 1.16 & 0.00 & 0.71 & 0.46 \\
\bottomrule
\end{tabularx}
\par\smallskip
\begin{minipage}{\textwidth}
\footnotesize\justifying
\noindent\textit{Notes.} The sample for each merger contains routes served by both merging brands in the quarter before closing. Closing is the first quarter in which the target is recorded under the acquirer's ownership. Retirement is the first quarter after the target brand's last appearance in the data. Entries are route averages. The first and last numeric columns are numbers of brands. The middle columns are probabilities of joint service and group presence.
\end{minipage}
\end{table}

The overlapping samples contain 47, 86, and 156 routes, respectively, with overlap fixed by service in the quarter before closing. A separate regression compares joint operation by the merging pair with joint operation by other legacy pairs on the same routes. It includes pair-by-route and merger-by-quarter effects. Using 115,799 pair-quarter observations on 576 routes, the coefficient on common ownership is 0.013, with a standard error of 0.019 clustered by airport pair.

\begin{figure}[!tp]
\centering
\includegraphics[width=\textwidth]{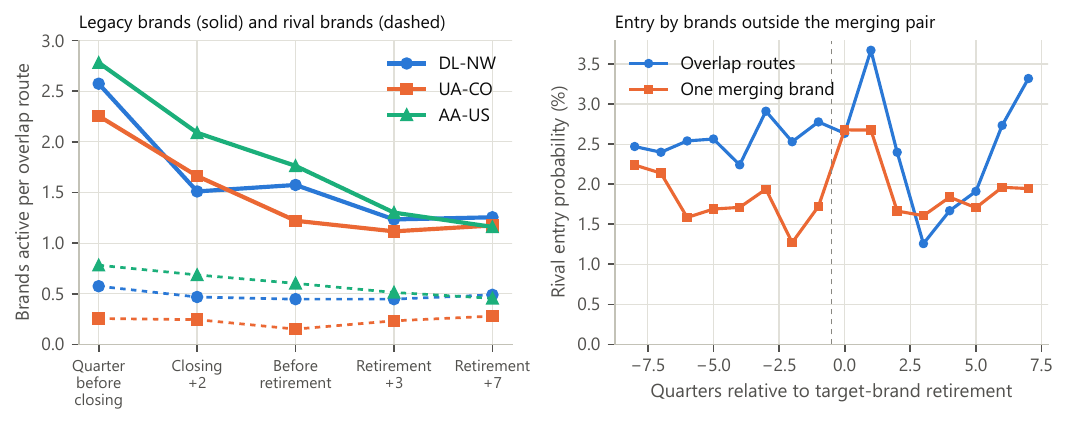}
\caption{Service and Rival Entry around the Legacy Mergers}
\label{fig:past_mergers}
\par\smallskip
\begin{minipage}{\textwidth}
\footnotesize\justifying
\noindent\textit{Notes.} The left panel reports the number of active legacy brands and the number outside the merging pair on routes served by both merging brands before closing. The right panel compares entry by other legacy brands on overlapping routes with entry on routes served by only one merging brand. Event time is measured relative to retirement of the acquired brand.
\end{minipage}
\end{figure}

Figure~\ref{fig:past_mergers} traces route consolidation and rival entry around retirement of the acquired brand. Rival entry regressions compare overlapping routes with routes served by only one merging brand before closing. They use the participation risk set of Appendix~\ref{app:data}, including the quarters in which the last legacy brand leaves a route. The regression around the ownership change uses 187,002 observations on 2,948 routes and yields a coefficient of $-$0.0021 with a standard error of 0.0029. The regression for brand retirement uses 178,389 observations on the same number of routes and yields $-$0.0032 with a standard error of 0.0031. Both include route-by-merger effects and merger-by-quarter effects, with standard errors clustered by airport pair. The pre-event entry probabilities are 0.030 and 0.025, respectively.

The traffic comparison in Table~\ref{tab:historical_traffic} defines overlap using service in the quarter immediately before closing. For each merger, the quarterly totals on overlapping routes and on routes served by only one merging brand are normalized by their respective means in the eight quarters before closing. Differences in log changes compare the two groups during common ownership and during the eight quarters following brand retirement. Airport-pair bootstrap draws provide the reported standard errors.

\begin{table}[!tp]
\centering
\singlespacing
\small
\caption{Traffic and Capacity after Historical Mergers}
\label{tab:historical_traffic}
\begin{tabularx}{\textwidth}{@{}Xrr@{}}
\toprule
Outcome & Common ownership & After retirement \\
\midrule
Route passengers & $-$0.060 & 0.133 \\
 & (0.096) & (0.122) \\
Merging group seats & $-$0.194 & 0.054 \\
 & (0.095) & (0.079) \\
Route seats & $-$0.130 & 0.049 \\
 & (0.087) & (0.096) \\
\bottomrule
\end{tabularx}
\par\smallskip
\begin{minipage}{\textwidth}
\footnotesize\justifying
\noindent\textit{Notes.} Entries compare routes served by both merging brands immediately before closing with routes served by one merging brand. Each group's quarterly aggregate is divided by its average during the eight quarters before closing, and the table reports the difference in log changes. The last column covers the eight quarters following retirement of the acquired brand. The three mergers are pooled. Parentheses report standard errors from 200 airport-pair bootstrap draws.
\end{minipage}
\end{table}

\subsubsection{Estimated Conduct}

For the conduct comparison in Table~\ref{tab:merger_conduct}, a route enters if both merging brands operated there during the eight quarters before closing. The reported observations are the route-quarter games during common ownership before the acquired brand disappears. This definition uses a longer pre-closing window than Table~\ref{tab:past_mergers}, whose overlap is measured in the immediately preceding quarter.

Table~\ref{tab:merger_conduct} reports conduct estimates from the three legacy mergers, conditional on estimated payoffs, route effects, and other brands' policies. The coefficient during common ownership is estimated at $-0.537$, with a standard error of 0.037 clustered by airport pair. A specification including the pairs' premerger choices estimates a change of $-0.782$ (0.043) at closing. The likelihood-ratio statistics against independent conduct and against full internalization are 611 and 3,579.

\begin{table}[!tp]
\centering
\small
\singlespacing
\caption{Conduct under Common Ownership}
\label{tab:merger_conduct}
\begin{tabularx}{\textwidth}{@{}Xrr@{}}
\toprule
\multicolumn{3}{@{}l}{\textit{Panel A. Conduct estimates from the legacy mergers}} \\
Parameter & Common ownership & Before and after closing \\
\midrule
Conduct coefficient, $\lambda$ & $-0.537$ & $-0.782$ \\
 & (0.037) & (0.043) \\
Baseline pair component, $\lambda_0$ & & 0.247 \\
 & & (0.017) \\
Conduct during common ownership, $\lambda_0+\lambda$ & $-0.537$ & $-0.535$ \\
Twice log likelihood difference for $\lambda=0$ & 611 & \\
Twice log likelihood difference for $\lambda=1$ & 3,579 & 4,900 \\
\end{tabularx}
\begin{tabularx}{\textwidth}{@{}X*{5}{>{\raggedleft\arraybackslash}p{1.75cm}}@{}}
\midrule
\multicolumn{6}{@{}l}{\textit{Panel B. Joint service during common ownership}} \\
Historical pair & Data & $\lambda=0$ & $\lambda=-0.537$ & $\lambda=1$ & Games \\
\midrule
Delta and Northwest & 0.176 & 0.161 & 0.196 & 0.114 & 983 \\
United and Continental & 0.714 & 0.455 & 0.576 & 0.289 & 462 \\
American and US Airways & 0.404 & 0.324 & 0.389 & 0.225 & 2,008 \\
\bottomrule
\end{tabularx}
\par\smallskip
\begin{minipage}{\textwidth}
\footnotesize\justifying
\noindent\textit{Notes.} Conduct weights the bilateral payoff generated by a brand's choices for its partner. In Panel A, the first specification estimates this coefficient during common ownership. In the second, let $\lambda_0$ denote the common baseline component for the three pairs and let $C$ indicate common ownership before retirement. The coefficient is $\lambda_0+\lambda C$, so $\lambda$ is the change at closing. The likelihood comparisons impose $\lambda=0$ or $\lambda=1$ in the stated specification. Inference clusters by airport pair and conditions on estimated payoffs, route effects, and other brands' policies. Panel B reports the probability that both brands are active during quarters of common ownership on routes where both historical brands operated during the eight quarters before closing, in the data and in the model at three conduct values.
\end{minipage}
\end{table}

The negative estimate of $\lambda$ makes overlapping service more attractive because the participation interaction between independent brands is negative. At $\lambda=-0.537$, the participation interaction within the merging pair is $\left(1+\lambda\right)\rho_{AA}=-0.26$, compared with $\rho_{AA}=-0.56$ between independent brands. This coefficient summarizes the incentive to retain both services during common ownership. Coordinated schedules, connecting traffic, shared airport facilities, and regional operations could contribute to that incentive, but the conduct estimate does not distinguish among them.

Full internalization of the estimated interactions between independent brands predicts joint service in only 11 to 29 percent of the historical overlapping route-quarters. Observed joint service ranges from 18 to 71 percent across the mergers, close to or above the predictions under independent conduct. The estimated conduct therefore describes service incentives during the period when both brands remain available; it does not impose joint maximization of the payoffs estimated between independent brands.

\subsubsection{American and United under Estimated Conduct}
\label{sec:merger}

We apply the estimated conduct to a hypothetical combination of American and United from 2016Q1, keeping both brands available and evaluating the observed route states through 2019. As in the estimation of $\lambda$, the comparison conditions on the observed state histories. Table~\ref{tab:merger} reports the effects.

\begin{table}[!tp]
\centering
\small
\singlespacing
\caption{American and United under Estimated Conduct, 2016 to 2019}
\label{tab:merger}
\begin{tabularx}{\textwidth}{@{}Xrrrr@{}}
\toprule
 & \multicolumn{2}{c}{All brands re-optimize} & \multicolumn{2}{c}{Merged brands move first} \\
\cmidrule(lr){2-3}\cmidrule(l){4-5}
Outcome & Baseline & Effect & Baseline & Effect \\
\midrule
\multicolumn{5}{@{}l}{\textit{Panel A. Participation, all route-quarter games}} \\
Merging brands active per 100 games & 67.31 & 1.55 & 67.32 & 1.53 \\
 & & (0.23) & & (0.22) \\
Merging brands' entry probability (pp) & 1.44 & 1.17 & 1.46 & 1.16 \\
 & & (0.25) & & (0.25) \\
Merging brands' continuation probability (pp) & 90.90 & 1.08 & 90.89 & 1.08 \\
 & & (0.12) & & (0.12) \\
Merging brands' outsourcing share (pp) & 46.54 & 0.15 & 46.52 & 0.15 \\
 & & (0.10) & & (0.10) \\
Rival brands active per 100 games & 42.11 & $-$0.08 & 42.09 & $-$0.08 \\
 & & (0.02) & & (0.02) \\
Rival brands' entry probability (pp) & 1.48 & $-$0.11 & 1.47 & $-$0.11 \\
 & & (0.03) & & (0.03) \\
All brands active per 100 games & 109.42 & 1.46 & 109.41 & 1.45 \\
 & & (0.21) & & (0.20) \\
\midrule
\multicolumn{5}{@{}l}{\textit{Panel B. Routes on which both brands are available}} \\
Both merging brands active (percent) & 11.35 & 2.81 & 11.35 & 2.79 \\
 & & (0.43) & & (0.42) \\
Neither merging brand active (percent) & 25.32 & $-$0.23 & 25.31 & $-$0.23 \\
 & & (0.03) & & (0.03) \\
All brands active per 100 games & 122.99 & 2.88 & 122.99 & 2.85 \\
 & & (0.41) & & (0.40) \\
\bottomrule
\end{tabularx}
\par\smallskip
\begin{minipage}{\textwidth}
\footnotesize\justifying
\noindent\textit{Notes.} American and United remain separate brands from 2016Q1 under a common owner with the conduct coefficient estimated from the legacy mergers, $\lambda=-0.537$. In the first pair of columns, all brands re-optimize. In the second pair, the merged brands re-optimize given rivals' first-stage policies, and rivals then respond to the pair. Each experiment is compared with its own baseline at $\lambda=0$. Panel B covers the 24,461 games in which both brands are available. Parentheses report standard errors from 180 draws of the corrected game parameters, which re-solve the experiment at each draw on the observed games, combined with the standard error of $\lambda$ clustered by airport pair.
\end{minipage}
\end{table}

When all brands re-optimize, the merging brands' activity rises by 1.55 brands per 100 games (0.23), with entry increasing by 1.17 percentage points (0.25) and continuation by 1.08 (0.12). Rival activity declines by 0.08 (0.02). On routes where both merging brands are available, the probability of joint service rises from 11.35 to 14.16 percent, an increase of 2.81 percentage points (0.43).

Under the alternative response rule, the merged brands first re-optimize given rivals' first-stage policies, and rivals then respond to the pair. The merging brands' activity increases by 1.53 and rival activity falls by 0.08. The estimated conduct thus preserves and expands overlapping service in this experiment.\footnote{Table~\ref{tab:robustness_profit} reports the corresponding experiment with the participation payoff linked to predicted variable profit.}

Common ownership changes the return to outsourcing through two opposing terms. With $\lambda<0$, the interaction between a brand's outsourcing share and its partner's activity, $\rho_{OA}^\lambda=\rho_{OA}+\lambda\rho_{AO}$, rises from $-0.04$ under independent conduct to 0.07. This increases the return to outsourcing when the partner is active. At the same time, the complementarity with the partner's outsourced activity, $\rho_{OO}^\lambda=\left(1+\lambda\right)\rho_{OO}$, falls from 0.20 to 0.09. The two changes largely offset one another: the merging brands' outsourcing share rises by only 0.15 percentage points (0.10), and the main effect is on participation.

\FloatBarrier

\subsection{Robustness}
\label{app:robustness}

\subsubsection{Rival Responses}
\label{app:policy_robust}

Table~\ref{tab:policy_robust} compares the main counterfactual results in the benchmark equilibrium, in which all brands re-optimize, with the experiments that hold rivals at their first-stage choice probabilities.

\begin{table}[!tp]
\centering
\singlespacing
\small
\caption{Counterfactual Results under Alternative Rival Responses}
\label{tab:policy_robust}
\begin{tabularx}{\textwidth}{@{}Xrr@{}}
\toprule
 & \shortstack{All brands\\re-optimize} & \shortstack{Rivals at first-stage\\choice probabilities} \\
\midrule
\multicolumn{3}{@{}l}{\textit{Panel A. Service under the 25 percent internal-share floor}} \\
Active carrier-route pairs (percent) & $-$3.6 & $-$8.2 \\
 & (0.6) & (0.9) \\
Thinnest third of routes (percent) & $-$16.7 & $-$21.1 \\
 & (1.7) & (1.9) \\
Densest third of routes (percent) & $-$0.6 & $-$4.8 \\
 & (0.4) & (0.6) \\
\midrule
\multicolumn{3}{@{}l}{\textit{Panel B. Counterfactual comparisons}} \\
Loss under the 1 percent internal-share floor relative to the 25 percent loss (percent) & 64.4 & 72.2 \\
 & (2.7) & (1.9) \\
Reduction of the 25 percent loss from easier reorganization (percent) & 2.6 & 1.2 \\
 & (0.5) & (0.2) \\
Additional participation loss under the cost increase (percent) & 30.5 & 15.5 \\
 & (6.5) & (2.8) \\
Reduction of the 25 percent loss with a common payment (percent) & 38.1 & 31.1 \\
 & (6.8) & (2.8) \\
Disruption loss avoided by the 25 percent internal-share floor (percent) & 72.0 & 77.5 \\
 & (2.6) & (2.1) \\
Years of the average participation loss equal to the avoided disruption loss & 7.3 & 4.6 \\
 & (1.1) & (0.5) \\
\bottomrule
\end{tabularx}
\par\smallskip
\begin{minipage}{\textwidth}
\footnotesize\justifying
\noindent\textit{Notes.} The first column repeats the benchmark equilibrium results of Section~\ref{sec:counterfactuals}. The second solves the same experiments with rivals' expected activity and outsourced activity fixed at the values implied by their first-stage choice probabilities. The cost increase is recalibrated to the outsourcing reduction under the 25 percent internal-share floor in that setting. Its additional participation loss is the difference between its loss and the floor's loss, divided by the latter and expressed as a percentage. The experiment with a common payment uses the payment rule in Table~\ref{tab:network}, recalculating the common amount from the resulting mean outsourcing share. The payment compensates the average modeled cost increase and requires outside funding. The disruption rows refer to the four-quarter interruption beginning in 2014Q2, with the participation loss in ordinary periods averaged over all quarters of the experiment. Parentheses report standard errors from 180 draws of the corrected game parameters. Rows expressed relative to an estimated effect use the delta method, including the covariance of the numerator and denominator across draws.
\end{minipage}
\end{table}

\subsubsection{Game Specification}
\label{app:robustness_game}

Table~\ref{tab:robustness_game} reports game estimates and counterfactuals under alternative payoff specifications, continuation values, and production measures. All experiments with outsourcing limits use the design of Section~\ref{sec:policy_design}. Across the other specifications with carrier-by-route effects and simultaneous rival terms, the 25 percent internal-share floor reduces activity by 2.01 to 4.87 brands per 100 games and the 10 percent internal-share floor by 1.57 to 4.00, compared with 3.99 and 3.30 in the benchmark specification. The American and United experiment raises the merging brands' activity by 0.79 to 1.80 brands per 100 games in every specification with simultaneous rival terms.

\begin{table}[!tp]
\centering
\singlespacing
\footnotesize
\setlength{\tabcolsep}{3pt}
\caption{Robustness of the Game Estimates and Counterfactuals}
\label{tab:robustness_game}
\begin{tabular*}{\textwidth}{@{\extracolsep{\fill}}lrrrrrrrrr@{}}
\toprule
 & \multicolumn{4}{c}{Estimates} & \multicolumn{2}{c}{25\%} & 10\% & \multicolumn{2}{c}{Merger} \\
\cmidrule(lr){2-5}\cmidrule(lr){6-7}\cmidrule(lr){8-8}\cmidrule(l){9-10}
Specification & $\kappa_E$ & $\theta_A$ & $\theta_D$ & $\rho_{AA}$ & Active & Entry & Active & Merging & Rivals \\
\midrule
Benchmark specification & 3.86 & 4.52 & 0.79 & $-$0.56 & $-$3.99 & $-$0.42 & $-$3.30 & 1.55 & $-$0.08 \\
No simultaneous rival terms & 4.01 & 4.49 & 0.71 & & $-$10.01 & & $-$8.70 & & \\
Common adjustment cost intercept & 3.93 & 4.97 & 1.42 & $-$0.59 & $-$2.28 & $-$0.54 & $-$1.89 & 1.53 & $-$0.08 \\
Common entry cost over time & 4.30 & 4.47 & 0.71 & $-$0.53 & $-$4.03 & $-$0.34 & $-$3.30 & 1.80 & $-$0.11 \\
Carrier-by-year effects & 3.81 & 4.53 & 0.81 & $-$0.64 & $-$2.01 & $-$0.34 & $-$1.57 & 1.11 & $-$0.04 \\
Last eight quarters of the sample dropped & 3.92 & 4.39 & 0.70 & $-$0.44 & $-$3.91 & $-$0.42 & $-$3.25 & 1.12 & $-$0.04 \\
Discount factor 0.95 & 3.96 & 4.58 & 0.72 & $-$0.56 & $-$4.03 & $-$0.42 & $-$3.21 & 1.45 & $-$0.07 \\
Activity payoff from variable profit & 3.94 & 4.49 & 0.74 & $-$0.19 & $-$4.26 & $-$0.42 & $-$3.49 & 0.79 & $-$0.02 \\
Forty continuation quarters & 3.81 & 4.53 & 0.84 & $-$0.60 & $-$3.63 & $-$0.45 & $-$2.98 & 1.65 & $-$0.10 \\
Adjustment threshold of 10 percentage points & 3.94 & 4.26 & 0.79 & $-$0.50 & $-$3.94 & $-$0.46 & $-$3.23 & 1.42 & $-$0.07 \\
Adjustment threshold of 15 percentage points & 3.99 & 4.05 & 0.82 & $-$0.47 & $-$3.69 & $-$0.49 & $-$2.97 & 1.34 & $-$0.06 \\
Corner tolerance of 2 percentage points & 3.83 & 4.36 & 1.05 & $-$0.59 & $-$3.65 & $-$0.47 & $-$2.90 & 1.65 & $-$0.09 \\
Routes with at least median sampled tickets & 3.91 & 5.09 & 0.74 & $-$0.57 & $-$3.44 & $-$0.38 & $-$2.94 & 1.56 & $-$0.08 \\
Routes with at least three daily departures & 4.13 & 6.72 & 0.76 & $-$0.54 & $-$3.45 & $-$0.34 & $-$2.87 & 1.79 & $-$0.11 \\
Seat outsourcing share & 3.96 & 2.41 & 0.63 & $-$0.51 & $-$4.87 & $-$0.51 & $-$4.00 & 1.41 & $-$0.07 \\
Types from the initial inherited share & 3.88 & 4.50 & 0.75 & $-$0.55 & $-$4.09 & $-$0.42 & $-$3.44 & 1.53 & $-$0.08 \\
Carrier intercepts only & 4.03 & 4.66 & 0.88 & $-$0.39 & $-$5.57 & $-$0.51 & $-$4.40 & 1.00 & $-$0.04 \\
\bottomrule
\end{tabular*}
\par\smallskip
\begin{minipage}{\textwidth}
\footnotesize\justifying
\noindent\textit{Notes.} Common parameters use the split-panel jackknife, and the carrier-by-route effects are re-estimated given the corrected parameters. The entry cost is $\kappa_E$, the quadratic adjustment cost is $\theta_A$, duration dependence is $\theta_D$, and the activity interaction is $\rho_{AA}$. The 25\% columns report the simulated effects of the 25 percent internal-share floor of Section~\ref{sec:policy_design} on active brands per 100 games and on the entry probability (percentage points), and the 10\% column reports the effect of the 10 percent internal-share floor on active brands per 100 games, with all brands re-optimizing. Merger columns report the change in active merging and rival brands per 100 games of 2016 to 2019, or of 2016 and 2017 when the last eight quarters are dropped, at the benchmark conduct estimate, $\lambda=-0.537$, with all brands re-optimizing. The specification without simultaneous rival terms has no rival responses. Other rows change the indicated component of the benchmark specification. Adjustment thresholds, the corner tolerance, seat measurement, sample selection, and initial types are defined in Appendix~\ref{app:measurement_robustness}. The last row replaces the carrier-by-route effects with one intercept per carrier (Appendix~\ref{app:no_route_effects}).
\end{minipage}
\end{table}

A common adjustment cost intercept removes the carrier and mixed-production shifts from the fixed adjustment cost. The 10 and 25 percent internal-share floors reduce activity by 1.89 and 2.28 brands per 100 games in this specification.

The organizational estimates remain close to the benchmark under alternative time controls and discounting, although the participation effects vary. Removing the year shifts of entry costs gives losses of 3.30 and 4.03 brands per 100 games under the 10 and 25 percent internal-share floors. Adding 105 carrier-by-year effects reduces these losses to 1.57 and 2.01. Re-estimation at a quarterly discount factor of 0.95 gives losses of 3.21 and 4.03, while extending the continuation horizon to forty quarters gives 2.98 and 3.63. Ending the sample in 2017 gives a quadratic adjustment coefficient of 4.39, duration dependence of 0.70, and an activity interaction of $-0.44$.

Table~\ref{tab:robustness_policy} reports the counterfactual comparisons in each specification. Across the other specifications with carrier-by-route effects and the carrier and mixed-production shifts of fixed adjustment costs, the 1 percent internal-share floor produces 60 to 71 percent of the participation loss under the 25 percent floor. The common cost increase reduces active carrier-route pairs by 15 to 43 percent more than the 25 percent floor, while the floor avoids 69 to 78 percent of the cumulative participation loss during and after the disruption. When the cost increase is accompanied by the common payment to active carrier-route pairs, the participation loss is 29 to 63 percent smaller than under the 25 percent floor in each of these specifications.

\begin{table}[!tp]
\centering
\singlespacing
\footnotesize
\caption{Counterfactual Comparisons across Specifications}
\label{tab:robustness_policy}
\begin{tabularx}{\textwidth}{@{}Xrrrr@{}}
\toprule
 & \shortstack{Loss at 1\%\\relative to 25\%} & \shortstack{Additional loss\\under higher costs} & \shortstack{Loss reduction with\\a common payment} & \shortstack{Disruption loss\\avoided at 25\%} \\
\midrule
Benchmark specification & 64.4 & 30.5 & 38.1 & 72.0 \\
Common adjustment cost intercept & 78.0 & $-$18.4 & 112.0 & 61.9 \\
Common entry cost over time & 61.9 & 35.3 & 35.0 & 75.0 \\
Carrier-by-year effects & 61.7 & 43.2 & 63.2 & 76.4 \\
Last eight quarters of the sample dropped & 64.0 & 33.6 & 40.9 & 75.9 \\
Discount factor 0.95 & 59.5 & 38.7 & 29.8 & 73.5 \\
Activity payoff from variable profit & 64.4 & 38.2 & 50.9 & 74.0 \\
Forty continuation quarters & 63.0 & 30.0 & 40.4 & 71.2 \\
Adjustment threshold of 10 percentage points & 67.8 & 20.7 & 48.6 & 76.1 \\
Adjustment threshold of 15 percentage points & 70.5 & 14.7 & 56.5 & 77.1 \\
Corner tolerance of 2 percentage points & 62.0 & 15.0 & 54.5 & 72.1 \\
Routes with at least median sampled tickets & 70.1 & 35.2 & 37.5 & 71.9 \\
Routes with at least three daily departures & 60.3 & 38.7 & 42.4 & 69.3 \\
Seat outsourcing share & 61.7 & 27.1 & 28.5 & 77.6 \\
Types from the initial inherited share & 65.9 & 31.3 & 35.9 & 73.9 \\
Carrier intercepts only & 60.4 & 43.9 & 12.2 & 71.4 \\
\bottomrule
\end{tabularx}
\par\smallskip
\begin{minipage}{\textwidth}
\footnotesize\justifying
\noindent\textit{Notes.} Entries are percentages computed in the equilibrium of each specification, with all brands re-optimizing. The first column divides the activity loss under the 1 percent internal-share floor by the loss under the 25 percent floor. The second reports the additional activity loss under a common cost increase, calibrated within each specification to the outsourcing reduction of the 25 percent floor, relative to that floor's loss. The third reports the reduction in the floor's loss when the cost increase is accompanied by a common payment to active carrier-route pairs, as in Table~\ref{tab:network}. This payment requires outside funding. The last column reports the share of the cumulative activity loss from the four-quarter disruption beginning in 2014Q2 that the 25 percent floor avoids. The specifications are those of Table~\ref{tab:robustness_game}.
\end{minipage}
\end{table}

\subsubsection{Carrier-by-Route Effects}
\label{app:no_route_effects}

Table~\ref{tab:app_no_route_effects} compares the benchmark model with a specification that replaces its 15,035 carrier-by-route activity effects with seven carrier intercepts. Both use the same players and observed quarters, payoff covariates, calendar effects, year shifts of entry costs, the entry cost during merger integration, and carrier-specific fixed adjustment costs. The specification with carrier intercepts also makes participation endogenous for the 1,118 pairs that are active in every observed quarter.

\begin{table}[!tp]
\centering
\footnotesize
\singlespacing
\caption{Estimates and Counterfactuals with and without Carrier-by-Route Effects}
\label{tab:app_no_route_effects}
\renewcommand{\arraystretch}{0.97}
\begin{tabularx}{\textwidth}{@{}Xrr@{}}
\toprule
 & With carrier-by-route effects & Carrier intercepts only \\
\midrule
\multicolumn{3}{@{}l}{\textit{Panel A. Participation}} \\
Entry cost in 1998, $\kappa_E$ & 3.858 & 4.035 \\
 & (0.212) & (0.093) \\
Entry cost shift, merger integration, $\kappa_T$ & $-$1.065 & $-$0.890 \\
 & (0.383) & (0.365) \\
Entry cost, previous service on route, $\kappa_P$ & $-$0.429 & $-$0.729 \\
 & (0.271) & (0.169) \\
Lagged route traffic (std.) & 0.228 & 0.154 \\
 & (0.027) & (0.010) \\
Lagged number of rivals (std.) & 0.110 & 0.116 \\
 & (0.028) & (0.022) \\
Active $\times$ rivals' activity, $\rho_{AA}$ & $-$0.557 & $-$0.390 \\
 & (0.077) & (0.045) \\
\midrule
\multicolumn{3}{@{}l}{\textit{Panel B. Organization}} \\
Outsourcing payoff, linear, $\gamma_1$ & $-$2.689 & $-$2.729 \\
 & (0.128) & (0.139) \\
Outsourcing payoff, squared, $\gamma_2$ & 2.339 & 2.356 \\
 & (0.116) & (0.121) \\
Quadratic adjustment cost, $\theta_A$ & 4.523 & 4.662 \\
 & (0.296) & (0.305) \\
Fixed cost intercept, $\theta_R$ & 2.023 & 1.902 \\
 & (0.115) & (0.106) \\
Fixed cost, inherited share, $\theta_\ell$ & $-$0.313 & $-$0.279 \\
 & (0.055) & (0.058) \\
Fixed cost shift, mixed production, $\theta_I$ & $-$2.552 & $-$2.490 \\
 & (0.096) & (0.086) \\
Fixed cost, log duration, $\theta_D$ & 0.789 & 0.885 \\
 & (0.044) & (0.045) \\
\midrule
\multicolumn{3}{@{}l}{\textit{Panel C. Fit}} \\
Carrier-by-route effects & 15,035 & 0 \\
Common parameters & 149 & 156 \\
Log likelihood at the maximum & $-$294,367.3 & $-$318,027.9 \\
Comparable log likelihood at the maximum & $-$294,367.3 & $-$314,650.5 \\
Yearly entry rate, root mean squared error (pp) & 0.35 & 0.34 \\
Yearly continuation rate, root mean squared error (pp) & 1.02 & 2.12 \\
Yearly outsourcing share, root mean squared error (pp) & 1.03 & 0.99 \\
\midrule
\multicolumn{3}{@{}l}{\textit{Panel D. Counterfactuals (brands active per 100 games unless noted)}} \\
25\% internal-share floor & $-$3.99 & $-$5.57 \\
10\% internal-share floor & $-$3.30 & $-$4.40 \\
25\% internal-share floor, entry probability (pp) & $-$0.42 & $-$0.51 \\
25\% internal-share floor, continuation probability (pp) & 0.11 & 0.02 \\
Merger at the benchmark conduct estimate, merging brands & 1.55 & 1.00 \\
Merger at the benchmark conduct estimate, rival brands & $-$0.08 & $-$0.04 \\
\bottomrule
\end{tabularx}
\par\smallskip
\begin{minipage}{\textwidth}
\footnotesize\justifying
\noindent\textit{Notes.} Both columns report split-panel corrected estimates with standard errors from the jackknife influence function, clustered by airport pair and calendar quarter. The carrier-intercept model also makes participation endogenous for the 1,118 always-active pairs. Its log likelihood includes their participation terms. The comparable likelihood removes these terms for their 65,615 observations and retains their organizational contributions in both columns. Root mean squared errors compare annual rates fitted at the maximum likelihood estimates with the data. Effects of outsourcing limits use the simulated design of Section~\ref{sec:policy_design} with all brands re-optimizing, and the merger uses $\lambda=-0.537$.
\end{minipage}
\end{table}

The entry cost rises from 3.86 to 4.03, and the activity interaction changes from $-0.56$ to $-0.39$. Duration dependence of fixed adjustment costs rises from 0.79 to 0.88, while the reference fixed adjustment cost changes from 2.02 to 1.90. The benchmark model's carrier-by-route effects have a standard deviation of 0.81. Panel C reports the likelihoods and annual fit of both specifications.

Replacing carrier-by-route effects with carrier intercepts increases the activity loss under the 25 percent internal-share floor from 3.99 to 5.57 brands per 100 games. The entry probability falls by 0.51 instead of 0.42 percentage points, and the continuation probability rises by 0.02 instead of 0.11. Under the 10 percent floor, the activity loss rises from 3.30 to 4.40. Allowing persistent returns to differ across a carrier's routes therefore matters for the predicted participation response.

\subsubsection{Measurement and Inherited Organization}
\label{app:measurement_robustness}

We first raise the threshold for adjustments within a regime to 10 or 15 percentage points, continuing to count every regime change as an adjustment. The corresponding losses are 3.23 and 2.97 brands per 100 games under the 10 percent internal-share floor and 3.94 and 3.69 under the 25 percent floor. Larger thresholds lower the quadratic adjustment cost from 4.52 to 4.26 and 4.05, while duration dependence remains between 0.79 and 0.82. A separate specification recodes outsourcing shares below 0.02 as internal production and shares above 0.98 as complete outsourcing. It gives a duration coefficient of 1.05 and losses of 2.90 and 3.65 brands per 100 games under the 10 and 25 percent floors.

On the 3,081 routes with at least the median number of sampled tickets, re-estimation preserves substantial fixed adjustment costs and duration dependence. The 10 and 25 percent internal-share floors reduce activity by 2.94 and 3.44 brands per 100 games. Measuring outsourcing by independent providers' share of T100 seats gives corresponding losses of 4.00 and 4.87.

Table~\ref{tab:app_frequency} reports adjustment patterns by flight frequency. Restricting estimation to the 1,754 routes averaging at least three daily departures per brand gives activity losses of 2.87 and 3.45 brands per 100 games under the 10 and 25 percent internal-share floors. The outsourcing payoff remains convex, and the quadratic adjustment coefficient rises from 4.52 to 6.72.

Finally, we allow the linear return to outsourcing and the fixed adjustment cost to differ with the carrier's inherited production regime in its first observed quarter on each route. These predetermined types add four parameters. The duration coefficient is 0.75, and the 10 and 25 percent internal-share floors reduce activity by 3.44 and 4.09 brands per 100 games.

\subsubsection{Samples}
\label{app:robustness_samples}

Table~\ref{tab:robustness_samples} reports the benchmark specification in six nested samples. The samples successively trim routes with extreme initial fares and distances and range from 6,159 to 994 routes. The corrected entry cost ranges from 3.80 to 4.05, the quadratic adjustment coefficient from 4.52 to 4.72, and the duration coefficient from 0.79 to 0.87. The fixed-cost shift for mixed production remains between $-2.55$ and $-2.37$, and the shift of entry costs during merger integration ranges from $-1.06$ to $-0.90$. The difference in fixed adjustment costs between mixed and exclusive production is therefore substantial throughout these samples.

\begin{table}[!tp]
\centering
\small
\singlespacing
\caption{Estimates across Samples}
\label{tab:robustness_samples}
\begin{tabular*}{\textwidth}{@{\extracolsep{\fill}}lrrrrrrr@{}}
\toprule
Sample (routes) & $\kappa_E$ & $\kappa_T$ & $\theta_A$ & $\theta_R$ & $\theta_I$ & $\theta_D$ & $\rho_{AA}$ \\
\midrule
6,159 & 3.86 & $-$1.06 & 4.52 & 2.02 & $-$2.55 & 0.79 & $-$0.56 \\
 & (0.13) & (0.39) & (0.29) & (0.09) & (0.08) & (0.04) & (0.06) \\
3,987 & 3.80 & $-$1.04 & 4.72 & 1.95 & $-$2.50 & 0.83 & $-$0.69 \\
 & (0.12) & (0.39) & (0.33) & (0.09) & (0.08) & (0.04) & (0.08) \\
2,988 & 3.87 & $-$0.93 & 4.60 & 1.91 & $-$2.45 & 0.80 & $-$0.70 \\
 & (0.13) & (0.40) & (0.33) & (0.09) & (0.09) & (0.04) & (0.09) \\
2,493 & 3.97 & $-$0.97 & 4.62 & 1.90 & $-$2.45 & 0.80 & $-$0.64 \\
 & (0.14) & (0.41) & (0.33) & (0.09) & (0.09) & (0.04) & (0.10) \\
1,986 & 4.00 & $-$0.90 & 4.72 & 1.86 & $-$2.42 & 0.85 & $-$0.67 \\
 & (0.13) & (0.42) & (0.35) & (0.09) & (0.09) & (0.04) & (0.11) \\
994 & 4.05 & $-$0.90 & 4.60 & 1.84 & $-$2.37 & 0.87 & $-$0.77 \\
 & (0.20) & (0.43) & (0.37) & (0.10) & (0.10) & (0.04) & (0.15) \\
\bottomrule
\end{tabular*}

\par\smallskip
\begin{minipage}{\textwidth}
\footnotesize\justifying
\noindent\textit{Notes.} Point estimates use the split-panel jackknife. Standard errors refer to the uncorrected estimator and are clustered by airport pair and calendar quarter. Table~\ref{tab:game_estimates} reports the standard errors of the corrected full-sample estimates. The samples successively trim routes by initial fares and distances. Within each sample, the half panels hold calendar effects, the year shifts of entry costs, and the era shifts of outsourcing returns at that sample's full-panel estimates. The reference carrier for fixed adjustment cost is Delta.
\end{minipage}
\end{table}

\begin{figure}[!tp]
\centering
\includegraphics[width=\textwidth]{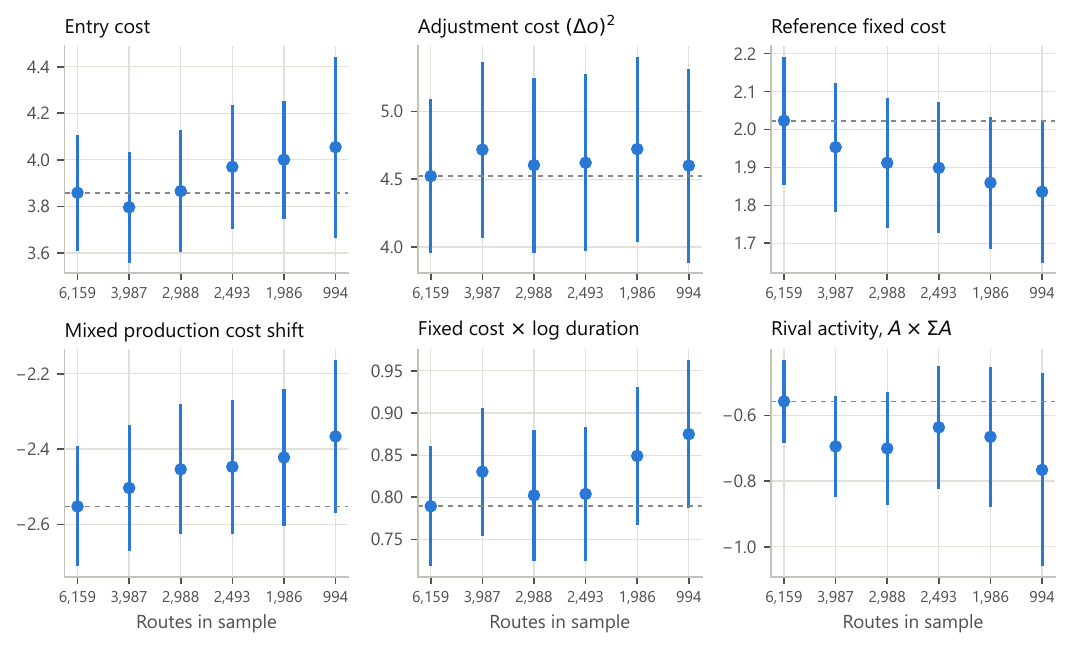}
\caption{Estimates across Samples}
\label{fig:estimates_samples}
\begin{minipage}{\textwidth}
\footnotesize\justifying
\noindent\textit{Notes.} Points are split-panel corrected estimates from the six nested samples, ordered by the number of routes. The intervals use the reported standard errors clustered by airport pair and quarter. Those standard errors refer to the uncorrected estimator. The dashed line marks the full-sample estimate. The panel for the reference fixed cost reports $\theta_R$ before adjustments for inherited shares, duration, and carriers.
\end{minipage}
\end{figure}

The activity interaction is more negative in the trimmed samples, ranging from $-0.77$ to $-0.64$, compared with $-0.56$ in the full sample. The main production and merger experiments use the full sample. Figure~\ref{fig:estimates_samples} plots the estimates and their reported intervals across the six samples.

\subsubsection{Demand Estimates}
\label{app:robustness_demand}

The demand system is the nested logit of Section~\ref{sec:demand_model}, estimated with the log number of brands on the route and the log Hausman fare as instruments (Section~\ref{sec:identification}). Table~\ref{tab:app_price_slope} reports the first stage and estimates that control for seasonal demand by route length.

\begin{table}[!tp]
\centering
\singlespacing
\small
\caption{Demand Instruments and Estimates}
\label{tab:app_price_slope}
\begin{tabularx}{\textwidth}{@{}X*{3}{>{\raggedleft\arraybackslash}p{2.7cm}}@{}}
\toprule
\multicolumn{4}{@{}l}{\textit{Panel A. First stage}} \\
 & Log fare & \multicolumn{1}{r}{Log within-route share} & \\
\cmidrule(lr){2-2}\cmidrule(lr){3-3}
Log number of brands on the route & $-$0.015 & $-$0.302 & \\
 & (0.004) & (0.011) & \\
Log Hausman fare & 0.130 & 0.050 & \\
 & (0.022) & (0.030) & \\
$F$ statistic, excluded instruments & 23.1 & 380.6 & \\
\midrule
\multicolumn{4}{@{}l}{\textit{Panel B. Demand parameters}} \\
 & (1) & (2) & (3) \\
 & Two-stage & Two-stage & Least \\
 & least squares & least squares & squares \\
\cmidrule(lr){2-2}\cmidrule(lr){3-3}\cmidrule(l){4-4}
Price coefficient on log fare, $\alpha$ & 2.743 & 3.077 & 0.574 \\
 & (0.482) & (0.555) & (0.021) \\
Nesting parameter, $\sigma$ & 0.400 & 0.414 & 0.819 \\
 & (0.052) & (0.056) & (0.010) \\
Outsourcing share in mean utility, $\gamma_o$ & $-$0.069 & $-$0.065 & $-$0.101 \\
 & (0.022) & (0.023) & (0.017) \\
Distance decile $\times$ quarter of the year effects & No & Yes & No \\
Kleibergen-Paap statistic & 17.2 & 15.1 & \\
Observations & 263,912 & 263,912 & 263,912 \\
\bottomrule
\end{tabularx}
\par\smallskip
\begin{minipage}{\textwidth}
\footnotesize
\justifying
\noindent\textit{Notes.} The Hausman fare is the carrier's passenger-weighted fare in the same quarter on routes in the same distance decile that share neither endpoint with the focal route. Every regression includes carrier-by-route and carrier-by-quarter effects, the brand's current and lagged outsourcing shares, its log seats per departure and log departures in the preceding quarter, and indicators for Southwest and other low-cost carriers on the route. Panel A reports the first stage of column (1). Column (2) adds effects for each distance decile in each quarter of the year. The sample contains products with departures on the route in the preceding quarter. The Kleibergen-Paap statistic tests the joint first stage of the log fare and the log within-route share \citep{KleibergenPaap2006}. Standard errors are clustered by airport pair.
\end{minipage}
\end{table}

\begin{table}[!tp]
\centering
\singlespacing
\caption{Participation Payoffs Linked to Variable Profit}
\label{tab:robustness_profit}
\small
\begin{tabularx}{\textwidth}{@{}Xrrr@{}}
\toprule
 & Benchmark & \multicolumn{2}{c}{Variable profit} \\
\cmidrule(l){3-4}
Parameter or outcome & Estimate & Estimate & \$ million \\
\midrule
\multicolumn{4}{@{}l}{\textit{Panel A. Estimates}} \\
Variable profit loading, $\theta_\pi$ (per \$ million per quarter) &  & 1.789 & \\
 & & (0.193) & \\
Entry cost in 1998, $\kappa_E$ & 3.858 & 3.942 & 2.20 \\
 & (0.212) & (0.202) & \\
Entry cost shift, merger integration, $\kappa_T$ & $-$1.065 & $-$1.079 & $-$0.60 \\
 & (0.383) & (0.357) & \\
Quadratic adjustment cost, $\theta_A$ & 4.523 & 4.493 & 2.51 \\
 & (0.296) & (0.295) & \\
Fixed cost intercept, $\theta_R$ & 2.023 & 2.076 & 1.16 \\
 & (0.115) & (0.114) & \\
Fixed cost shift, mixed production, $\theta_I$ & $-$2.552 & $-$2.612 & $-$1.46 \\
 & (0.096) & (0.095) & \\
Fixed cost, log duration, $\theta_D$ & 0.789 & 0.736 & 0.41 \\
 & (0.044) & (0.041) & \\
Active $\times$ rivals' activity, $\rho_{AA}$ & $-$0.557 & $-$0.191 &  \\
 & (0.077) & (0.033) & \\
Market covariates in the activity payoff & 5 free & \multicolumn{2}{r}{through $\theta_\pi$} \\
Log likelihood at the maximum & $-$294,367.3 & $-$294,875.9 & \\
\midrule
\multicolumn{4}{@{}l}{\textit{Panel B. Counterfactuals (all brands re-optimize)}} \\
25\% internal-share floor, active brands per 100 games & $-$3.99 & $-$4.26 & \\
25\% internal-share floor, entry probability (pp) & $-$0.42 & $-$0.42 & \\
25\% internal-share floor, continuation probability (pp) & 0.11 & 0.10 & \\
10\% internal-share floor, active brands per 100 games & $-$3.30 & $-$3.49 & \\
Merger, merging brands active per 100 games & 1.55 & 0.79 & \\
Merger, rival brands active per 100 games & $-$0.08 & $-$0.02 & \\
Merger, both merging brands active (pp) & 2.81 & 1.41 & \\
\bottomrule
\end{tabularx}
\par\smallskip
\begin{minipage}{\textwidth}
\footnotesize
\justifying
\noindent\textit{Notes.} The specification replaces the five market covariates in the activity payoff by $\theta_\pi\E\left[\pi_{ie}\right]$. Here $\pi_{ie}=10^{-6}q_{ie}\left(p_{ie}-mc_{ie}\right)$ is passenger variable profit in millions of dollars per quarter. The predicted value comes from a regression on the market covariates with carrier-by-route and carrier-by-quarter effects among active brands, evaluated for every eligible brand decision. Predictors exclude the brand's own lagged actions. The game's route effects absorb the carrier-by-route component, leaving the remaining variation to identify $\theta_\pi$. The last column converts costs to millions of dollars by dividing by $\theta_\pi$. Estimates use the split-panel jackknife, and standard errors use the jackknife influence function, clustered by airport pair and calendar quarter. Effects of outsourcing limits use the simulated design of Section~\ref{sec:policy_design}. The merger uses $\lambda=-0.537$, and the last merger row covers games in which both brands are available.
\end{minipage}
\end{table}

In the first stage, a one percent increase in the Hausman fare is associated with a 0.13 percent increase in the focal fare, and a larger number of brands is associated with lower within-route shares. The first-stage $F$ statistics are 23.1 for the log fare and 380.6 for the log within-route share. Least squares gives a smaller price coefficient and a larger nesting parameter than two-stage least squares, consistent with positive correlation of fares and within-route shares with unobserved demand.

Column (2) of Panel B adds effects for each distance decile in each quarter of the year, which absorb seasonal demand that differs between short and long routes. The resulting price coefficient is 3.08, with a standard error of 0.55, and the nesting parameter is 0.41, with a standard error of 0.06.\footnote{The Hausman fare moves with costs that grow with stage length, such as fuel and aircraft costs, and these costs change from quarter to quarter. Effects for each distance decile in each sample quarter would absorb this common cost variation, which is the variation that motivates the instrument.}

\subsubsection{Participation Payoffs and Variable Profit}
\label{app:robustness_profit}

Table~\ref{tab:robustness_profit} reports a specification that links participation to predicted passenger variable profit. Let $\pi_{ie}$ denote this profit in millions of dollars per quarter and $\theta_\pi$ its coefficient in the activity payoff. We replace the five market covariates in that payoff with $\theta_\pi\E\left[\pi_{ie}\right]$. To obtain the prediction, we regress variable profit on the market covariates with carrier-by-route and carrier-by-quarter effects among active brands, then evaluate the fitted equation for every eligible brand decision. Predictors exclude the brand's own lagged actions. The game's fixed effects absorb the carrier-by-route component, leaving the remaining variation in predicted profitability to identify $\theta_\pi$.

{\interlinepenalty=10000 \clubpenalty=10000 The estimated loading $\theta_\pi=1.789$, with a standard error of 0.193, implies an entry cost of \$2.20 million paid in the quarter of entry, a quadratic cost of \$2.51 million for a complete reallocation, and a reference fixed adjustment component of \$1.16 million.\par}

The organizational coefficients remain close to their benchmark estimates. The 10 and 25 percent internal-share floors reduce activity by 3.49 and 4.26 brands per 100 games. The activity interaction changes from $-0.56$ to $-0.19$. At the benchmark conduct estimate, the American and United merger raises the merging brands' activity by 0.79 brands per 100 games and raises the probability that both brands serve a route by 1.41 percentage points.

\end{document}